\pdfoutput=1
\documentclass[aps,showpacs,onecolumn,floatfix,amsmath,amssymb,nofootinbib]{revtex4}
\usepackage{amssymb}
\usepackage{amsmath}
\usepackage{epsfig}
\usepackage{graphicx}
\usepackage{color}
\usepackage{latexsym}
\usepackage{bm,latexsym}
\usepackage{mathrsfs}
\usepackage{float}
\usepackage{pbox}
\usepackage{pifont}

\def\Journal#1#2#3#4{{#1} {\bf #2}, #3 (#4)}

\def\CMM{{Commun. Math. Phys.}}

\def\CQG{{Class. Quantum Grav.}}
\def\EPJC{{Eur. Phys. J. C}}
\def\GRG{{Gen. Relativ. Gravit.}}
\def\IJMPD{{Int. Jour. Mod. Phys. D}}
\def\JCAP{{JCAP}}

\def\JMP{{J. Math. Phys.}}
\def\LRR{{Living Rev. Rel.}} 
 
\def\NAT{{Nature}}

\def\NPB{{\em Nucl. Phys.} B}

\def\PLB{{Phys. Lett.}  B}

\def\PRL{Phys. Rev. Lett.}

\def\PRD{{Phys. Rev.} D}
\def\PR{{Phys. Rep.}}

\def\RMP{{Rev. Mod. Phys.}}

\newcommand{\xmark}{\ding{55}}

\begin{document}


\title{Spherically symmetric black holes in $f(R)$ gravity: Is geometric scalar hair supported ?} 

\author{Pedro Ca\~nate$^1$}
\email{pedro.canate@ciencias.unam.mx}

\author{Luisa G. Jaime$^2$}
\email{jaime@thphys.uni-heidelberg.de}

\author{Marcelo Salgado$^1$}
\email{marcelo@nucleares.unam.mx}

\affiliation{$^1$Instituto de Ciencias Nucleares, Universidad Nacional
Aut\'onoma de M\'exico, A.P. 70-543, M\'exico D.F. 04510, M\'exico \\
$^2$ Institut f\"ur Theoretische Physik,\\ Ruprecht-Karls-Universit\"at Heidelberg,\\
Philosophenweg 16, 69120 Heidelberg, Germany}


\date{\today}

    
\begin{abstract}
We discuss with a rather critical eye the current situation of black hole (BH) solutions in 
$f(R)$ gravity and shed light about its geometrical and physical significance. 
We also argue about the meaning, existence or lack thereof of a Birkhoff's theorem in this kind of modified gravity. We focus then on 
the analysis and quest of {\it non-trivial} (i.e. hairy) {\it asymptotically flat} (AF) BH solutions in static and spherically symmetric (SSS) 
spacetimes in vacuum having the property that the Ricci scalar does {\it not} vanish identically in the domain of outer communication. 
To do so, we provide and enforce the {\it regularity conditions} at the horizon in order to prevent the presence  
of singular solutions there. Specifically, we consider several classes of $f(R)$ models 
like those proposed recently for explaining the accelerated expansion in the universe and which have been thoroughly 
tested in several physical scenarios. Finally, we report analytical and numerical evidence about the {\it absence} of 
{\it geometric hair} in AFSSSBH solutions in those $f(R)$ models. First, we submit the models to 
the available no-hair theorems, and in the cases where the theorems apply, the absence of hair is demonstrated analytically. In the cases where 
the theorems do not apply, we resort to a numerical analysis due to the complexity of the non-linear differential equations. Within that aim, a code 
to solve the equations numerically was built and tested using well know exact solutions.
In a future investigation we plan to analyze the problem of hair in De Sitter and Anti-De Sitter backgrounds. 
\end{abstract}



\maketitle
\newpage

\section{Introduction}
\label{Intro}
Modified $f(R)$ gravity has become one of the most popular mechanisms to generate a late accelerated expansion 
in the universe without the need of introducing new fields~\cite{f(R)}. 
It was also one of the first consistent models for early inflation~\cite{Starobinsky1980}. 
During the past fifteen years or so, several 
specific $f(R)$ models have been thoroughly analyzed in many scenarios, but only a few of them can survive the classical 
tests (e.g. solar system, binary pulsar) while predicting the correct accelerating expansion, and in general, a successful 
cosmological model, both at the background and at the perturbative level. Therefore, it is still unclear until what extent 
this kind of alternative theories of gravity can recover {\it all} the successes of general relativity (GR) while making 
new testable predictions.

As concern black hole solutions, the situation of $f(R)$ gravity can, in some sense, differ from GR, and in 
other sense be almost the same. The last statement is related with the content of Section~\ref{sec:theory} about the 
existence of the same kind of vacuum black-hole solutions found in GR, while the former concerns the existence of hairy solutions, or lack thereof, that we analyze in all the rest of the paper.

Perhaps the first and simplest theorem concerning BH solutions in GR was the Birkhoff's theorem (BT). Roughly speaking, this theorem establishes 
that in vacuum all spherically symmetric (SS) spacetimes are also static, and those that are asymptotically flat (AF) are 
represented by a one-parameter family of solutions, namely, the ubiquitous Schwarzschild solution, where the parameter is 
interpreted as the (ADM) {\it mass} $M$ of the spacetime (see~\cite{Schmidt2013} for a discussion). Remarkably, when including 
an electric field, the SS solution can be extended as to include the charge $Q$ of the BH, it is the well known two-parameter 
Reissner--Nordstr\"om (RN) solution. In the AF case, both the Schwarzschild and the RN solution are $R=0$ 
solutions of the Einstein's field equations (also termed {\it Ricci flat} solutions). 
When including a cosmological constant $\Lambda$, the Schwarzschild and RN black 
holes become a two and three-parameter family respectively, and the solutions are asymptotically de Sitter (ADS) or asymptotically 
anti de Sitter (AADS), depending if $\Lambda >0$ or $\Lambda <0$, respectively.

In the decade of 1960's, motivated by the discovery of the Kerr solution, several theorems (notably, the {\it uniqueness theorems}) 
were established for stationary and axisymmetric spacetimes both in vacuum and with an electromagnetic field (see~\cite{BHs,Wald1984} for details and reviews). 
One of the main consequences of those theorems is that in the AF case the solutions of the Einstein field equations under such symmetries 
are characterized {\it only} by three parameters, the mass $M$, the charge $Q$, and the angular momentum $J$ of the BH. These solutions 
are known as the Kerr--Newman family, which extends the Schwarzschild and RN black holes to 
more general spacetimes: stationary and axisymmetric. 
Due to the apparent simplicity of such solutions, Wheeler established the so called {\it no hair conjecture}, a statement that ``doomed'' 
all possible stationary AFBH solutions from having any other parameters than those three ($M,Q$ and $J$). This conjecture has been reinforced 
thereafter by the elaboration of several {\it no-hair} theorems (NHT's) that forbid the existence of BH solutions with more 
parameters associated with other kinds of matter fields (see~\cite{Bekenstein2000,Herdeiro2015} for a review). Among such theorems one can mention 
those that include several kinds of scalar fields. Eventually, this conjecture proved to be ``false'', for instance, within the 
Einstein--Yang--Mills system~\cite{Bizon1990}, and Einstein-scalar-field system with ``exotic'' potentials that can be negative~\cite{Nucamendi2003,Anabalon2012}. Or when including rotation like in the Einstein-boson-field system~\cite{Radu2014}. 
Nonetheless, since most, if not all of such {\it hairy} solutions are unstable~\footnote{In the sense that a perturbation can 
lead to an eventual loss of the hair.}, the community (or at least part of it) consider those solutions as {\it weak} 
counterexamples to the Wheeler's no-hair conjecture. Therefore, it has been tantalizing to extend the conjecture in the following 
more precise, although still informal, statement: {\it the only stable stationary AFBH's are within the Kerr--Newman family}~\cite{Bekenstein2000}.

The proposal of alternative theories of gravity as a possible solution to the dark-matter and dark-energy problems and to other 
theoretical problems (e.g. inflation, gravity renormalization) has motivated people to generalize several of the theorems and 
conjectures mentioned above, and which pertain to GR, to the realm of other modified-gravity proposals. While it is out of the scope of the present paper to 
review all such attempts, we shall simply focus on $f(R)$ metric gravity. Unless otherwise stated, by $f(R)$ gravity we mean a theory 
that departs from the GR function $f(R)= R - 2\Lambda$.

As concerns this kind of theory, a large amount of analysis has been devoted to establish an analogue of the BT for the SS 
situation~\cite{BTaltth,Nzioki2010,Carloni2013}. However, the reality is that no rigorous BT exists today in $f(R)$ gravity, as far as we are aware. 
In fact, if one such theorem were proved, certainly should be restricted to some specific $f(R)$ models. Moreover, the theorem should 
establish at least four things, upon fixing the boundary conditions (i.e. regularity and asymptotic conditions): 1) {\it Staticity:} the only spherically 
symmetric solutions in vacuum [i.e. without any matter field associated with the standard model of particle physics or any other field that 
is not associated with the Lagrangian $f(R)$] are necessarily static (i.e. the existence of a static Killing field should be proved from 
the spherically symmetric assumptions); 2) {\it Existence:} the existence of an exact static spherically symmetric (SSS) solution in 
vacuum; 3) {\it Uniqueness:} the SSS solution found in point `2)' is the only solution in vacuum (or prove otherwise); 4) The conditions 
under which the solution in point `2)' matches or not the exterior solution of an SSS extended body. 

So far, in vacuum only a few BH exact solutions exist in $f(R)$ gravity, and those that are genuine AF, ADS or AADS correspond simply 
to the same kind of solutions found in GR, where $R=R_c=const$ everywhere in the spacetime (with $R_c=0$, $R_c>0$ or $R_c<0$, respectively). 
It is unclear if other solutions exist with the same kind of asymptotics but with a varying $R$ in the domain of outer communication of the BH. 
We shall elaborate more about this point below to be more precise. Furthermore, in the presence of matter (i.e. a star-like object), it 
is possible to find SSS solutions where $R$ can vary in space~\cite{f(R)stars,Miranda2009,Jaime2011}, however, those solutions are not exact, 
but given only numerically, and it is unclear if the exterior part (i.e. the vacuum part) of those solutions is the same solution found when 
matter is totally absent in the spacetime, if it exists at all, as it happens in GR where the exterior solution of extended objects under such symmetries 
is always given by the vacuum Schwarzschild solution
~\footnote{In GR and in the stationary and axisymmetric case, the exterior 
solution of rotating extended bodies does not match exactly the Kerr solution. On the other hand, in scalar-tensor theories of gravity (STT)
without a potential, and due to the spontaneous scalarization phenomenon~\cite{spontscal}, there exist star-like SSS solutions where the exterior solution 
is not given by the Schwarzschild metric, since the scalar field there is not zero. It turns out, however, that in the absence of 
ordinary matter, the exterior solution within the same STT is given only by the Schwarzschild solution. These two scenarios show that the exterior 
solutions in presence of ordinary matter are not necessarily the same as in {\it complete} absence of it, and so this illustrates 
the difficulty of establishing a generalization of a BT when the hypothesis change.}.

Now, despite the absence of such BT, some NHT's have been proved in this kind of theories. 
In order to do so, people have resorted to the equivalence between a certain class of $f(R)$ models (notably, those where 
$f_R>0$ and $f_{RR}>0$, where the subindex indicates differentiation) with scalar-tensor theories (STT). The point is that 
one performs a conformal transformation from the original {\it Jordan frame} to the so-called {\it Einstein frame} where the 
conformal metric appears to be coupled minimally to gravity and a new scalar-field $\phi (\chi)$ emerges, where $\chi:= f_R$,  
which is also coupled minimally to the conformal metric but endowed with an ``exotic'' potential $\mathscr{U}(\phi)$. Thus, the 
available NHT's constructed for the Einstein-scalar-field system in GR can be applied 
for these theories as well (see Section~\ref{sec:STT}), notably in vacuum, and when the spacetime is AF and the potential satisfies 
the condition $\mathscr{U}(\phi)\geq 0$~\cite{Sudarsky1995,Bekenstein1995}\footnote{A more recent proof of the same result~\cite{Sotiriou2012} 
adopts a weaker convexity assumption $\mathscr{U}''(\phi)\geq 0$ for the potential. This proof is similar to 
Bekenstein's~\cite{Bekenstein1972} which assumes only {\it stationarity} as opposed to {\it staticity}.}.

We stress that the applicability of such NHT's is possible because the non-minimal coupling between the 
scalar field $\phi (\chi)$ and the matter fields that usually appears under the Einstein frame obviously vanishes in the absence of the matter. 
The only caveat of this method is that the potential $\mathscr{U}(\phi)$ is not given a priori but is the result of the specific $f(R)$ model considered 
ab initio, and thus, $\mathscr{U}(\phi)$ can be negative or even not well defined (i.e. it can be multivalued), which in turn can 
jeopardize the use of the NHT's. Consequently, the existing NHT's in $f(R)$ gravity can reduce the 
kind of AFSSSBH solutions that are available in some specific models, but do not rule out completely the absence of 
{\it geometric hair}. In this context, by (geometric) {\it hairy} 
solutions within $f(R)$ gravity we mean AFSSSBH solutions where the Ricci scalar is not {\it trivial} (i.e. constant), but 
rather a function that interpolates non-trivially between the horizon and spatial infinity. 

Thus, when the condition $\mathscr{U}(\phi)\geq 0$ fails and the NHT's are not applicable one can resort to a 
numerical analysis for evidence about the existence of such hair or its absence thereof. At this respect it is important to stress that 
regularity conditions have to be imposed at the inner boundary, namely, at the BH horizon $r_h$ in order to prevent the presence of singularities there.
In Section~\ref{sec:regcond} and Appendix~\ref{sec:regcond2} we obtain such regularity conditions and then in Sections~\ref{sec:expmodels} 
and~\ref{sec:numerical} we present analytical and numerical evidence, respectively, showing that hairy solutions 
are absent in several specific $f(R)$ models proposed as dark-energy alternatives in cosmology. In particular, the models considered in 
Section~\ref{sec:numerical} are precisely those for which the NHT's cannot be applied as the corresponding potential $\mathscr{U}(\phi)$ 
can be negative or is not even well defined. On the other hand, when such hairy solutions are absent, one may still find the trivial solution 
$R=R_1= const$ for which the field equations reduce to the Einstein field equations with an effective cosmological constant 
$\Lambda_{\rm eff}= R_1/4$, and an effective gravitational constant $G_{\rm eff}= G_0/f(R_1)$ where $R_1$ is a solution of an algebraic 
equation involving $f(R)$ and $f_R$. This includes the case where $R_1\equiv 0$. Therefore, in such circumstances, all the best known BH solutions 
found in GR exist also in $f(R)$ gravity simply by replacing the usual cosmological constant $\Lambda$ by $\Lambda_{\rm eff}$, and the 
Newton's gravitational constant $G_0$ by $G_{\rm eff}$. In view of this we shall argue in Section~\ref{sec:theory} that such solutions are so 
trivial (i.e. trivial in the context of $f(R)$ gravity) that almost nothing new arise from them. 

Finally, we mention that some ``non-trivial'' exact SSSBH solutions have been reported in the literature as a result of very 
ad hoc $f(R)$ models~\cite{Clifton,Nzioki2010,Sebastiani2011,nonvacuum}. Notwithstanding such solutions {\it cannot} be considered 
as hairy solutions because they have unusual asymptotics, and therefore, the corresponding ``hairless'' solution $R=const$ (including 
the Ricci-flat solution) does not even exist with the same kind of asymptotics. We shall discuss one such solution 
in Section~\ref{sec:exactsols}. 
\bigskip

The article is organized as follows: In Section~\ref{sec:theory} we discuss in a general setting the conditions for the existence of several 
trivial BH solutions. In Section~\ref{sec:SSS} we focus on SSS spacetimes and provide the corresponding differential equations to find 
BH solutions. We also discuss some exact solutions that will be used later to test a numerical code constructed to solve the equations. 
The boundary conditions appropriate to solve these equations with the presence of a BH are given in Section~\ref{sec:SSS} in form of 
{\it regularity conditions} at the horizon. No-hair theorems and the properties of $f(R)$ gravity formulated in the Einstein frame are analyzed 
in Section~\ref{sec:STT}. In that section we also provide strong numerical evidence about the absence of hair for several $f(R)$ models when the 
NHT's do not apply. Our conclusions and final remarks are presented in Section~\ref{sec:concl}. Several appendices at the end of the article 
complement the ideas of the main sections.


\section{$f(R)$ theory of gravity}
\label{sec:theory}

The general action for a $f(R)$ theory of gravity is given by 

\begin{equation}
\label{f(R)}
I[g_{ab},{\mbox{\boldmath{$\psi$}}}] =
\!\! \int \!\! \frac{f(R)}{2\kappa} \sqrt{-g} \: d^4 x 
+ I_{\rm matt}[g_{ab}, {\mbox{\boldmath{$\psi$}}}] \; ,
\end{equation}

where  $\kappa \equiv 8\pi G_0$ (we use units where $c=1$)~\footnote{In Section~\ref{sec:exactsols}, we extend our 
units so that $G_0=1$ as well.}, and $f(R)$ is a sufficiently smooth 
({\rm i.e.} $C^3$) but otherwise an a priori arbitrary function of the Ricci scalar $R$. 
The first term corresponds to the modified gravity action, while the second is the usual action for the matter, where 
${\mbox{\boldmath{$\psi$}}}$ represents schematically the matter fields.

The field equation arising from the action~(\ref{f(R)}) under the metric approach is
\begin{equation}
\label{fieldeq1}
f_R R_{ab} -\frac{1}{2}fg_{ab} - 
\left(\nabla_a \nabla_b - g_{ab}\Box\right)f_R= \kappa T_{ab}\,\,,
\end{equation}
where $f_R$ stands for $df/dR$ (we shall use similar notation for higher derivatives), $\Box= g^{cd}\nabla_c\nabla_d$ is the covariant D'Alambertian and $T_{ab}$ 
is the energy-momentum tensor of matter resulting from the variation of the matter action in~(\ref{f(R)}). It is straightforward, 
although a non-trivial result, to show that the conservation equation $\nabla^b T_{ab}=0$ holds also in this case (see Appendix~\ref{sec:consEMT} 
for a proof). In turn, this latter leads to the geodesic equation for free-fall particles $u^c\nabla_c u^a=0$. 
Therefore, the weak-equivalence principle (for point test particles) is also incorporated in this theory as well. Actually $f(R)$ metric gravity 
preserves all the axioms of GR but the one that assumes that the field equations for the metric $g_{ab}$ must be of 
second order. Clearly the only case where this happens is for $f(R)= R - 2\Lambda$, which leads to GR  plus a cosmological constant (hereafter GR$\Lambda$)
\footnote{A result that is also a corollary (when applied to a four dimensional spacetime) of a theorem known as {\it Lovelock's theorem}~\cite{Lovelock}.}. 

Now, taking the trace of Eq.~(\ref{fieldeq1}) yields
\begin{equation}
\label{traceR}
\Box R= \frac{1}{3 f_{RR}}\left[\rule{0mm}{0.4cm}\kappa T - 3 f_{RRR} (\nabla R)^2 + 2f- Rf_R \right]\,\,\,,
\end{equation}
where $T:= T^a_{\,\,a}$. When using~(\ref{traceR}) in~(\ref{fieldeq1}) and after some elementary manipulations we obtain~\cite{Jaime2011}
\begin{equation}
\label{fieldeq3}
G_{ab} = \frac{1}{f_R}\Bigl{[} f_{RR} \nabla_a \nabla_b R + f_{RRR} (\nabla_aR)(\nabla_b R) 
-\frac{g_{ab}}{6}\Big{(} Rf_R+ f + 2\kappa T \Big{)} + \kappa T_{ab} \Bigl{]} \; .
\end{equation}

Equations~(\ref{fieldeq3}) and~(\ref{traceR}) are the basic equations that 
we have used systematically in the past to tackle several problems in cosmology and 
astrophysics~\cite{Jaime2011,Jaime2012,Jaime2012e,Jaime2013,Jaime2014}, and that we plan 
to use in this article as well. 

Now, apart from the GR$\Lambda$ theory for which 
$f_R\equiv 1$, $f_{RR}\equiv 0$, and $R= 4\Lambda - \kappa T$, for more general models, one imposes the conditions $f_R >0$, for a positive $G_{\rm eff}$, 
and $f_{RR}>0$, for stability~\cite{Dolgov2003}. 
However, in this paper we shall sometimes relax these two assumptions in order to explore its consequences for the sake of finding BH solutions.

In vacuum, that is when $T_{ab}\equiv 0$, or more generally, in the presence of matter fields where $T\equiv 0$, like in electromagnetism 
or Yang--Mills theory, Eq.~(\ref{traceR}) admits in principle the trivial exact solution $R=R_c= const$ where $R_c$ is a solution of the algebraic 
equation $[2f(R) - Rf_{R}]/f_{RR}=0$. In particular, if $f_{RR}(R_c)\neq 0$ and $0<f_{RR}(R_c)< \infty$, as often happens in potentially viable $f(R)$ 
models, then $R_c$ is an algebraic solution of $2f(R) - Rf_{R}=0$ that we call $R_1$. 
For such kind of solutions the field Eq.~(\ref{fieldeq3}) reduces to
\begin{equation}
\label{Einsred}
G_{ab} + \Lambda_{\rm eff} g_{ab} = 8\pi G_{\rm eff} T_{ab} \,\,\,,
\end{equation}
where
\begin{eqnarray}
\label{Lambdaeff1}
\Lambda_{\rm eff} &=& \frac{R_1}{4} \,\,\,,\\
G_{\rm eff} &=& \frac{G_0}{f_R(R_1)} \,\,\,,
\end{eqnarray}
we assume $0<f_{R}(R_1)< \infty$. Moreover, the condition $0<f_{RR}(R_1)< \infty$ is an 
{\it stability} condition that is usually imposed in order to avoid 
 exponentially growing modes when perturbing around the value $R=R_1$. 

On the other hand, if there exists other value $R=R_c=const$, that we call 
$R_2$ (in order to avoid confusion with $R_1$) such that 
$f_{RR}(R)$ blows up at $R=R_2$, but $R_2\neq R_1$ then $R_2$ can be also a possible trivial solution of Eq.~(\ref{traceR}). However, in such 
a case one must be extremely cautious as in Eq.~(\ref{fieldeq3}) may appear products of the 
sort $\infty \times 0$ (or $\infty/\infty$). Even if such a product is finite, namely zero, still the interpretation of such solution 
would be problematic as $f_{RR}$ would be singular everywhere. At what extent such ``singularity'' is physical, is something 
that one should clarify. In Sec.~\ref{sec:exactsols}, we shall be dealing with an SSS solution where the Ricci scalar is not constant 
but $R\rightarrow R_2$ as $r\rightarrow \infty$. Hence, in that example $f_{R}\rightarrow \infty$, $f_{RR}\rightarrow \infty$ and 
$f_{RRR}\rightarrow \infty$ as $r\rightarrow \infty$. However, these pathologies, as peculiar as they may be, does not concern us 
too much in this article, since we will be mainly interested in situations where they are absent.

Now, in the particular and simplest scenario where $R=R_1$ is one of the trivial solutions of 
Eq.~(\ref{traceR}), we see that the field equation (\ref{Einsred}) corresponds to GR with 
the usual ``bare'' cosmological constant $\Lambda$ and the bare Newton's 
constant $G_0$ replaced by $\Lambda_{\rm eff}$ and $G_{\rm eff}$, respectively. Therefore, in that occurrence all the solutions that 
exist in GR exist in $f(R)$ gravity as well when taking into account the above replacements. In particular, the AFBH solutions with $R=0$ that exist in GR, 
like the Kerr--Newman family and its SSS limit, exist also in $f(R)$ gravity if 
$R_1=0$, i.e., if $f(0)\equiv 0$. On the other hand, 
the De Sitter or Anti-De Sitter BH solutions associated with the Kerr--Newman family with a cosmological 
constant~\cite{exactsolsL,Davis1989} exist also in $f(R)$ gravity if $R_1\neq 0$. 

It is note that BH solutions $R=R_1={\it const}$, which from the point of view 
of $f(R)$ gravity are {\it trivial}, have been systematically reported in the literature as something new or special 
(e.g. see Refs.~\cite{trivsols,Nzioki2010}). However, as we just showed, the existence of such solutions stems 
from the fact that $R_1$ exists in various $f(R)$ models, like the ones we consider in 
Section~\ref{sec:expmodels}. In turn, the existence of such a trivial solution is just a standard demand for $f(R)$ theories to produce a late 
acceleration expansion: the cosmological constant $\Lambda_{\rm eff}$ emerge while the universe evolves towards the solution 
$R\rightarrow R_1$ as the universe expands and matter dilutes. So nothing exceptional, astonishing or radically different from the already known BH solutions in GR are to be 
expected in this kind of trivial solutions in $f(R)$ gravity. Furthermore, and as a consequence of these remarks, the Bekenstein--Hawking 
entropy defined for such BH's, has $G_{\rm eff}$ instead of $G_0$ in the formula. That is, the modified entropy is 
$S= A/(4 G_{\rm eff})= Af_R(R_1)/(4 G_0)$~\cite{Briscese2008}, where $A$ is the area of the BH event horizon, instead of just $S= A/(4 G_0)$.  
\bigskip

Let us consider now the ``odd'' scenario where $R_1$ is such that $f_R(R_1)= 0$, and $R_1= 0$. Then
Eq.~(\ref{traceR}) admits $R\equiv 0$ as one possible trivial solution. Nonetheless, the possible solutions for the metric that satisfy 
Eq.~(\ref{fieldeq3}) {\it degenerate} and a whole spectrum of solutions can emerge, besides the usual 
AF vacuum solutions mentioned above. Such solutions are {\it all} the possible solutions of the Einstein equation 
$G_{ab}= \kappa T_{ab}$ compatible with a {\it null} Ricci scalar $R\equiv 0$. 
More specifically, if one considers an $f(R)$ model in vacuum such that $2f - Rf_{R}=0$ at $R=0$ 
(implying in turn $f(0)=0$) but assuming $f_{RR}(0)\neq 0$, then Eq.~(\ref{traceR}) is solved trivially. Moreover the field Eq.~(\ref{fieldeq3}) in vacuum reduce to $f_R(0) G_{ab}=0$ which is trivially 
satisfied for $f_R(0)=0$ even if $G_{ab}\neq 0$. Thus, in that instance the model admits 
unexpectedly all the possible solutions associated with the 
non-vacuum Einstein equation  $G_{ab}= \kappa T_{ab}$ compatible with $G^a_{\,\,a}=-R= -T=0$. 
Now, given such scenario one has to deal with a problematic interpretation of the 
new global quantities that appear in the BH solutions as integration constants since in reality we are dealing with 
an $f(R)$ model in vacuum. In non-vacuum GR the global quantities (other than $M$ and $J$) are ascribed to properties associated 
with the matter described by $T_{ab}$ (like the electric charge $Q$ or the nodes of the Yang--Mills field). Therefore the same quantities 
would appear in this degenerate scenario, but clearly they cannot have the same interpretation in vacuum. In section~\ref{sec:expmodels} within the framework of Model 1, 
we shall encounter one such example which is associated with the model $f(R)= k R^2$. In this model $R=0$ is a trivial 
solution and clearly $f_R(0)=0$, thus all possible solutions of Einstein equations with a traceless energy-momentum 
tensor are in principle allowed, for instance, AFSSS solutions. In fact, one such solution is exactly 
the analogue of the RN, where the roll of the charge is played by a new quantity that appears as an integration constant. This 
interesting situation was also analyzed recently in Ref.~\cite{Kehagias2015} and remarked before in Ref.~\cite{Nzioki2010}. We shall elaborate 
more about this in Sections~\ref{sec:exactsols} and~\ref{sec:expmodels}, notably, concerning the issue of {\it uniqueness} and the BT.
\bigskip

Having clarified the fact that $f(R)$ gravity naturally admits trivial BH solutions where $R={\it const}$ everywhere, 
notably the ones with $R=0$, and which correspond to the same solutions found in GR modulo a trivial redefinition of 
constants, the main goal of this paper is an effort to find AFSSS solutions where $R(r)$ is a non-trivial solution of the field 
equations, where $r$ is some radial coordinate. This means that if such solution exists, $R(r)$ should interpolate in a non-trivial manner 
between the event (Killing) horizon of the BH and the asymptotic region. If exist, this is what we might call 
a {\it hairy} solution. In order to find such a solution we enforce suitable regularity conditions at the horizon. 
These conditions provide a very specific form of derivatives of several variables at the horizon, notably, the first derivative of the Ricci scalar. 
These regularity conditions are extremely important as they prevent the presence of BH solutions that are pathological at the 
horizon. In particular, they prevent a singularity in the scalar-degree of freedom at the horizon, 
a singularity that would be otherwise considered as {\it physical} as opposed to a coordinate singularity. 
In Section~\ref{sec:regcond} and in Appendix~\ref{sec:regcond2} we provide such conditions in its full form and compare them with 
similar regularity conditions reported first in~\cite{Bergliaffa2011}, but amended in~\cite{Halilsoy2012}. 
Examples of such pathologies (singularities) are common in STT when using conformal methods to generate SSS exact solutions without 
enforcing the regularity condition of the scalar field at the BH horizon~\cite{singularsols}. Thus, the fact that the field equations 
may not be satisfied at the horizon in those examples cast serious doubts about its relevance as genuine counterexamples to the no-hair 
conjecture.


\section{Static and spherically symmetric vacuum solution in $f(R)$ gravity}
\label{sec:SSS}

The problem of describing a SSS space-time in vacuum  ($T_{ab}=0$) within the framework of 
$f(R)$ gravity reduces to solving the field equations~(\ref{fieldeq3}) and~(\ref{traceR}) for the following SSS metric:
\begin{equation}
\label{metric}
ds^2 =  - \bigg(1 - \frac{2M(r)}{r}  \bigg)e^{2\delta(r)}dt^{2} +  \bigg(1 - \frac{2M(r)}{r} \bigg)^{-1}dr^{2} + 
r^{2}\left(d\theta^2 + \sin^2\theta d\varphi^2 \right)\,\,\,,
\end{equation}
where the {\it mass function} $M(r)$ provides the ADM mass in the asymptotic region provided the spacetime is AF
\footnote{Certain spacetimes that are not AF posses a well defined ``ADM'' mass. In those cases, the Minkowski background is then replaced 
by a suitable background spacetime with respect to which the mass indicates deviations 
as one approaches the asymptotic region (the mass is then part of the ``monopole'' term in the $g_{tt}$ component). 
In those spacetimes the equivalent of the ADM mass corresponds typically to a suitable renormalization of 
$M(r\rightarrow \infty)$ (see footnote~\ref{f:ADS}).}. 
The function $\delta(r)$ indicates the extent to which the equality $G^{t}_{\,\,\,t}= G^{r}_{\,\,\,r}$ 
is satisfied or infringed by the components of the Einstein tensor or equivalently, by the corresponding components of the effective 
energy-momentum tensor given by the r.h.s of Eq.~(\ref{fieldeq3}) taking $T_{ab}=0$ there~\cite{gtt=g^rr}. 
So, if $\delta= const$, in particular zero, it means that the equality holds exactly 
everywhere in the spacetime. This situation includes some of the the best known SSS spacetimes. 
The field equations for the metric Eq.~(\ref{metric}) have been obtained in a rather convenient form 
in~\cite{Jaime2011} with a slightly different but equivalent parametrization. We present their final form based on those equations 
but without the matter terms: 
\begin{widetext}
\begin{eqnarray}
\label{TraceRsss}
R'' &=& \frac{1}{3f_{RR}} \bigg[ \frac{ ( 2f - Rf_{R} )r }{ r - 2M} - 3f_{RRR}R'^2 \bigg]  +  
\bigg[ \frac{  2( rM^{\prime}  - M ) }{ ( r - 2M )r }   - \delta' - \frac{2}{r} \bigg]R'  \,\,\,,\\
\label{Msss}
 M'  &=& \frac{M}{r} + \frac{1}{2(2f_{R} + rR'f_{RR})}\bigg\{ -\frac{4f_{R}M}{r} + \frac{r^2}{3}(Rf_{R} + f) \nonumber\\
& + &  \frac{rR'f_{RR}}{f_{R}} \bigg[ \frac{r^2}{3}(2Rf_{R} - f) - \frac{4 M}{r} f_{R} + 2rR'f_{RR}\Big(1 - \frac{ 2M}{r}\Big)\bigg] \bigg\} \,\,\,,\\
\label{deltasss}
\Big(1- \frac{2M}{r}\Big)\delta' &=&   \frac{1}{2(2f_{R} + rR'f_{RR})r}\bigg\{ \frac{2r^2}{3}(2f -  Rf_{R}) +  
\frac{rR'f_{RR}}{f_{R}}\Big[ \frac{r^2}{3}(2Rf_{R} - f ) - 2f_{R} \Big]   \nonumber\\
& - &    4(r -  2M)R'f_{RR}  + \frac{2(f_{R}   + rR'f_{RR} )(r -  2M) R'f_{RR}}{f_{R}} \bigg\}. 
\end{eqnarray}  
\end{widetext}

In order to find BH solutions, Eqs.~(\ref{TraceRsss})--(\ref{deltasss}) have to be solved from the BH horizon at $r=r_h$ to 
the asymptotic region which in the AF or Anti-De Sitter cases corresponds to spatial infinity. On the other hand, in a De Sitter 
background a cosmological horizon can be reached at some $r_h^c > r_h$. In this paper we shall focus only in AF solutions, therefore 
we deal only with the BH horizon.

Suitable boundary conditions at $r_h$ are to be imposed. Typically one assumes {\it regularity conditions} 
that enforce the verification of the field equations there (see Section~\ref{sec:regcond}). 
For instance, the value  $M(r_h)= r_h/2$ enforces the existence of the Killing horizon, where $r_h$ can have, in principle, any arbitrary 
non-negative value. The condition $\delta(r_h)$ is also arbitrary as it simply determines the value of $\delta(r)$ 
in the asymptotic region which can be redefined by a suitable change of the $t-coordinate$. 
In other words, the field equations are left invariant if one performs the transformation $\delta(r)\rightarrow \delta(r) + const$, a 
property that holds due to the existence of a static Killing vector field. 
So without loss of generality we can take $\delta(r_h)=0$. As concerns $R(r_h)$, hereafter $R_h$, this value is usually fixed so as to obtain  
the desired asymptotic value for $R$. Typically, but not necessarily (see below), a value $R(r_\infty)=0$ 
(where $r_\infty$ stands for $r\rightarrow \infty$) gives rise to AF spacetimes~\footnote{In Anti-De Sitter backgrounds one 
demands $R(r_\infty)= R_1$ with $R_1<0$. For De Sitter backgrounds enforcing the boundary conditions can be more involved 
since a cosmological horizon at $r_h^c$  ($r_h\ll r_h^c<\infty $) is present, and in this case the value $R_h$ is to  be fixed so as 
to recover the regularity conditions at $r_h^c$ as well.}. 

Due to the complexity of the field equations~(\ref{TraceRsss})--(\ref{deltasss}) one usually resorts to a numerical solution, 
in which case, the value $R_h$ is fixed by a {\it shooting method} (cf. Section~\ref{sec:numerical}). 
This value, which provides the adequate asymptotic behavior for $R(r)$, is also related with the 
the convergence of $M(r)$ to the {\it ADM mass}. Equation~(\ref{Msss}) can be written as in GR in the form
\begin{equation}
M'= 4\pi r^2 \rho_{eff} \,\,\,,
\end{equation}
where $\rho_{eff}$ can be readoff from Eq.~(\ref{Msss}). 
In this kind of coordinates the ADM mass $M_{ADM}$ for AF spacetimes is given by $M(r_\infty)$, also called the Komar mass~\cite{Wald1984}. 
In order for $M$ to converge to $M_{ADM}$ one requires $\rho_{eff}\sim 1/r^{2+\epsilon}$ asymptotically with $\epsilon >1$. 
That is, we require that the effective energy-density falls off faster than $1/r^3$. Otherwise $M(r)$ can diverge asymptotically as 
$M(r) \sim {\rm ln}(r)$ if $\epsilon=1$ or $M(r) \sim r^{1-\epsilon}$ if $\epsilon <1$
~\footnote{\label{f:ADS}In the case of Anti-De Sitter 
backgrounds the corresponding ``ADM'' mass is given by $M_{ADM}= M_{\rm ren}(r_\infty)$ where $M_{\rm ren}(r)=M(r) - \Lambda r^3/6$ ($\Lambda<0$). 
A rigorous definition for the {\it mass} in this kind of backgrounds has been given by several 
authors~\cite{adsmass,Magnon1985}. This mass is like the Komar mass~\cite{Magnon1985} but specialized to the 
SSS scenario. For instance, when the mass function is given by $M(r)=m + \Lambda r^3/6$ where $m$ is a constant 
(like in Kottler--Schwarzschild--De Sitter metrics), $m$ coincides with the mass found from the formal definitions~\cite{adsmass,Magnon1985}.
In the context of $f(R)$ gravity $\rho_{eff} -\Lambda_{\rm eff}/(\kappa)$ must behave as $1/r^{2+\epsilon}$ 
with $\epsilon >1$ for $M_{\rm ren}(r_\infty)$ to converge.}.
As a consequence, if the spacetime containing a BH possesses a non-trivial Ricci scalar $R(r)$, this must behave as $T_{eff}\sim -\rho_{eff}$, namely
$R\sim 1/r^{2+\epsilon}$ with $\epsilon >1$ in the AF scenario. Therefore, the shooting method is implemented such that $R(r)$ 
behaves asymptotically in the previous manner using $R_h$ as a control parameter. If no such behavior exists for all $R_h$, 
this implies that AF spacetimes with a {\it non-trivial} 
Ricci scalar do {\it not} exist. Thus, scalar-hair would be absent and the only possible AF solutions are $R=0$. 

There exist spacetimes with asymptotic behaviors different from AF or ADS/AADS with interesting properties. 
One of them correspond to a spacetime that is AF except for a deficit solid angle $0\leq \Delta<1$. This kind of spacetimes are typically generated by 
topological defects, like strings and global monopoles~\cite{Vilenkin1994}. The deficit angle is related with the ``symmetry breaking'' scale $\eta$ of the 
Mexican hat potential by $\Delta= 8\pi \eta^2$~\cite{Vilenkin1994}. 
The deficit angle produces typically a repulsion instead of an attraction of test bodies in the 
gravitational field generated by these defects. In this kind of spacetimes the mass function $M(r)\sim r\Delta/2 $ asymptotically. 
It is somehow remarkable that such spacetimes can have a well defined mass~\cite{Nucamendi1997}
which, in the case of spherical symmetry, can be computed, in practice, by a suitable renormalization of $M(r_\infty)$. This situation corresponds 
to $\epsilon=0$. It is then tantalizing to define the ``ADM'' mass in this case simply by $M_{ADM\Delta}= M_{\rm ren}^\Delta(r_\infty)$, where 
$M_{\rm ren}^\Delta(r)= M(r) - \Delta r/2$. In the next section we shall see that in fact such a mass is not exactly $M_{\rm ren}^\Delta(r_\infty)$ 
but rather proportional to it where the constant of proportionality depends on the deficit angle $\Delta$. 
Furthermore, $R\sim 2\Delta/r^{2}$ asymptotically. Hence, we have here an explicit example where the Ricci scalar 
is not trivial and it also vanishes asymptotically, and yet the spacetime is not AF. 
Below we present an explicit $f(R)$ model for which an exact SSS of this sort can be found, except that the deficit angle is not produced by a 
topological defect, but by the underlying modified-gravity model itself.
\bigskip

We finish this section by giving explicitly the kind of trivial solution
$R=R_1=const$ that we alluded in Section~\ref{sec:theory}. For SSS spacetimes we have
\begin{equation}
\label{SDS}
ds^2 =  - \Big(1 - \frac{2M_0}{r} - \frac{\Lambda_{\rm eff} r^2}{3} \Big)
dt^{2} +  \frac{dr^{2}}{\Big(1 - \frac{2M_0}{r} -\frac{\Lambda_{\rm eff} r^2}{3}\Big) }  + r^{2}d\Omega^{2} \,\,\,,
\end{equation}
with 
\begin{eqnarray}
\label{Mrtriv}
M(r) &=&  M_0 + \frac{\Lambda_{\rm eff} r^3}{6} \,\,\,,\\
\label{RfRSZ2}
R(r) &=& R_1= 4\Lambda_{\rm eff}  \,\,\,, \\
\label{delta0}
\delta (r) &\equiv & 0 \,\,\,, \\
\label{derpotcond}
2f(R_1) &=& R_1 f_R (R_1) \,\,\,.
\end{eqnarray}
This is the Kottler--Schwarzschild--De Sitter solution, where $M_0$ is a constant of integration of Eq.~(\ref{Msss}) which is identified with the ``ADM'' mass 
({\rm cf.} Ref.~\cite{adsmass,Magnon1985}). In the cases where $R_1\equiv 0$, the solution reduces to the 
Schwarzschild solution. It is a straightforward exercise to check that Eqs.~(\ref{Mrtriv})--(\ref{derpotcond}) 
solve exactly Eqs.~(\ref{TraceRsss})--(\ref{deltasss}), regardless of the $f(R)$ model, provided the model satisfies Eq.~(\ref{derpotcond}) 
and $f_R(R_1)\neq 0$, $f_{RR}(R_1)\neq 0$, $f_{RR}(R_1)<\infty$ and $f_{RRR}(R_1)<\infty$. As we show in the next section, when the $f(R)$ model is given explicitly $R_1$ writes in terms of its fundamental parameters [cf. Eq.~(\ref{LeffFRSZ}) ].


\subsection{Exact solutions}
\label{sec:exactsols}

Let us consider the model

\begin{equation}
\label{fRSZ}
f(R) =  2a\sqrt{R - \alpha}= 2a^2\sqrt{\left(\frac{R}{a^2}\right)- \left(\frac{\alpha}{a^2}\right)}\,\,\,,
\end{equation}
where $a>0$ is a parameter with units $[{\rm distance}]^{-1}$, and $\alpha$ is another parameter of the model which is related to an 
effective cosmological constant as we show below. This $f(R)$ model and a variant of it was considered in the past 
by several authors~\cite{Sebastiani2011,nonvacuum}. In the SSS scenario the metric
\begin{equation}
\label{gfRSZ}
ds^2 =  - \frac{1}{2} \Big(1 - \frac{\alpha r^2}{6} +  \frac{2Q}{r^2}   \Big)
dt^{2} +  \frac{dr^{2}}{\frac{1}{2} \Big(1 - \frac{\alpha r^2}{6} +  \frac{2Q}{r^2}   \Big) }  + r^{2}d\Omega^{2} \,\,\,,
\end{equation}
with the mass function, Ricci scalar, and $\delta(r)$ given, respectively, by
\begin{eqnarray}
\label{Mrexact}
M(r) &=&  \frac{r}{4}+ \frac{\alpha r^3}{24}- \frac{Q}{2r} \,\,\,,\\
\label{RfRSZ}
R(r) &=&  \alpha + \frac{1}{r^{2}} \,\,\,, \\
\label{deltafRSZ}
\delta (r) &\equiv & 0 \,\,\,,
\end{eqnarray}
solve Eqs.~(\ref{TraceRsss})--(\ref{deltasss}) exactly, as one can verify by straightforward substitutions. Here, $Q$ is an integration constant. 
Taking into account electromagnetic and Yang--Mills fields, this solution was extended in Ref.~\cite{nonvacuum}. When $\alpha=0$ this solution was part 
of a more general class of solutions associated with the model $f(R)= kR^n$~\cite{Clifton,Nzioki2010}. However, for $n=1/2$ the asymptotic behavior of those solutions was not analyzed by those authors as we do here.

The coordinates are defined such that,  $-\infty < t < \infty$, $0 \leq \theta \leq \pi$, $0 \leq \phi < 2\pi$ 
with  $r_h \leq r$ if $\alpha <0$ and $r_h \leq r \leq r_h^c$ if $\alpha >0$, where $r_h$ and $r_h^c$ corresponds to the location of the event and cosmological horizons, 
respectively, that we analyze below.

The metric (\ref{gfRSZ}) possesses a deficit angle $\Delta= 1/2$, $M_{ADM\Delta}\equiv 0$, ``charge'' $Q$ 
and a cosmological constant 
$\Lambda_{\infty}:= R(r_\infty)/4= \alpha/4$ (see Appendix~\ref{sec:AFdefang} for more details).
In the current case $M_{ADM\Delta}= (1-\Delta)^{-3/2}M_{\rm ren}^\Delta(r_\infty)$~\cite{Nucamendi1997} where 
$M_{\rm ren}^\Delta(r)= M(r)- \Lambda_{\infty} r^3/6 - r \Delta/2$. The divergent terms (linear and cubic in $r$) 
appear in this renormalization of mass since, as we remarked before, the spacetime has a deficit angle (associated with the linear term) and also a 
cosmological constant (associated with the cubic term). Using (\ref{Mrexact}) we conclude $M_{\rm ren}^\Delta(r)=- Q/(2r)$, thus, 
$M_{ADM\Delta}= M_{\rm ren}^\Delta(r_\infty) \equiv 0$. Examples of spacetimes with a deficit angle and with zero-mass BH 
are not new~\cite{Nucamendi2000}.

The metric with a deficit angle given by Eq.~(\ref{gfRSZ}) is a solution with a cosmological constant, $\Lambda_\infty = \alpha/4$. For instance, 
taking $\alpha>0$ we can introduce $\ell^{2}= 6/\alpha$ for convenience. The location of the black hole horizon depends on the value of $Q$. 
There are three possibilities: a) If $Q>0$, $Q:= q^{2}$, 
the event horizon of the black hole is located at $r_{h} = \frac{1}{2}\sqrt{2\ell^{2}  +  2\ell^{2}\sqrt{1  + 8q^{2}/\ell^{2}}}$; b) 
If $Q<0$ ($Q:= -q^{2}$) and $1  - 8 q^{2}/\ell^{2} > 0$, there are two horizons which are located at 
$r_{h}^c = \frac{1}{2}\sqrt{2\ell^{2}  +  2\ell^{2}\sqrt{1  - 8q^{2}/\ell^{2}}}$, and 
$r_{h} = \frac{1}{2}\sqrt{2\ell^{2}  -  2\ell^{2}\sqrt{1  - 8q^{2}/\ell^{2}}}$. In particular, for an {\it extreme} black hole with  
$q^2 = \ell^{2}/8$ the horizon is given by 
$r_{h} = |\ell|/\sqrt{2}$; c) If $Q= 0$, then $r_{h}^c = |\ell|$.

The AADS spacetime with a deficit angle is a solution with a negative cosmological constant $\Lambda_\infty = \alpha/4$ (with $\alpha<0$) 
and $\ell^{2}_{\rm AADS}= -6/\alpha$. Notice that for $Q\equiv 0$ an horizon 
does not exist like in the usual Anti-De Sitter (i.e. Anti-De Sitter spacetime without the deficit angle). 
However, for $Q= - q^{2} < 0$ an horizon exists and is located at $r_{h} = \frac{1}{2}\sqrt{-2\ell^{2}_{\rm AADS}  +  2\ell^2_{\rm AADS}
\sqrt{1  + 8q^{2}/\ell^2_{\rm AADS}}}$. 

When $\alpha = 0$, the spacetime turns to be AF except for a deficit angle. 
In this case only for $Q= - q^{2}<0$ there is an horizon at 
$r_{h} = \sqrt{2 q^{2}}$. 
Finally when $\alpha = 0=Q$, the spacetime is simply the Minkowski spacetime 
with a deficit angle. All other cases correspond to naked singularities. 

In absence of naked singularities a straightforward calculation of other scalars, like $R_{cd}R^{cd}$ and $R_{abcd}R^{abcd}$, show that the only {\it physical} singularity 
appear at $r=0$. Even if $\alpha = 0=Q$, such scalars are $R_{cd}R^{cd}= 1/2r^4$ and $R_{abcd}R^{abcd}= 1/r^4$  
while $R=1/r^2$, and thus, the singularity at $r=0$ is entirely due to the deficit angle. Since the coordinates used so far do not cover the entire manifold one can look for analytic extensions using Kruskal-like coordinates. 
These extensions and the construction of Penrose diagrams are out of the scope of the current paper and will be reported elsewhere.

Now, for $\alpha<0$, there is no cosmological horizon, and thus, one can analyze the solution as 
$r\rightarrow \infty$. It is then interesting to note that the asymptotic value of the Ricci scalar 
$R(r_\infty)= \alpha$ is not an algebraic solution of $2f(R) - Rf_{R}= 0$ which is $R_1= 4\alpha/3$ but rather a pole of 
$f_{RR}= -a [R-\alpha]^{-3/2}/2$. The trivial solution $R_1= 4\alpha/3$ provides a cosmological constant $\Lambda_{\rm eff}= R_1/4= \alpha/3$ 
which is different from the actual one $\Lambda_{\infty}=\alpha/4$. Now, even if $f_R= a [R-\alpha]^{-1/2}$, $f_{RR}= -a [R-\alpha]^{-3/2}/2$ and
$f_{RRR}= 3a [R-\alpha]^{-5/2}/4$ blow up at $R= \alpha$, the quantities that appear in Eq.~(\ref{TraceRsss}) 
behave well asymptotically (i.e. they are finite) as $R\rightarrow \alpha$:
\begin{eqnarray}
\frac{f_{RRR}(\nabla R)^2 }{f_{RR}} &=& \frac{f_{RRR} g^{rr} R'^2}{f_{RR}} = -\frac{g^{rr} R'^2}{R-\alpha} \sim \frac{\alpha}{2r^2}\,\,\,,\\
\frac{2f-Rf_R}{f_{RR}} &=& -2 (R-\alpha)(3R-4\alpha)\sim \frac{2\alpha}{r^2} \,\,\,
\end{eqnarray}

Moreover the quantities that appear in the r.h.s of Eq.~(\ref{fieldeq3}) behave also well asymptotically and give rise 
to the cosmological constant $\Lambda_{\infty}= \alpha/4$ as we show next. Take for instance the $r-r$ component of 
Eq.~(\ref{fieldeq3}) in vacuum. For our purposes it is more convenient to take the mixed components. Then
\begin{equation}
\label{Grr}
G^r_{\,\,\,\,r}= g^{rr}\left(\frac{f_{RR} \nabla_r\nabla_r R}{f_R} + \frac{f_{RRR} {R'}^2}{f_{R}}\right) -\frac{1}{6}\left(R+ \frac{f}{f_{R}}\right)
\end{equation}
Now let us analyze the asymptotic behavior of each of the terms at the r.h.s of Eq.~(\ref{Grr}):
\begin{eqnarray}
g_{rr} &\sim& -\frac{12}{\alpha r^2} \,\,\,,\\
g^{rr} &\sim& -\frac{\alpha r^2}{12} \,\,\,,\\
\frac{f_{RR} \nabla_r\nabla_r R}{f_R} &=& -\frac{\nabla_r\nabla_r R}{2(R-\alpha)}= -\frac{r^2}{2}\nabla_r\nabla_r R \nonumber \\
 &=&  -\frac{r^2}{2}\left( R'' - R'\Gamma^r_{\,\,\,\,rr}\right) =-\frac{r^2}{2}\left( \frac{6}{r^4} - R'\Gamma^r_{\,\,\,\,rr}\right)
\sim -\frac{r^2}{2} \left( \frac{6}{r^4} - \frac{\alpha r}{12} g_{rr} R'\right) \nonumber  \\
&=&  -\frac{r^2}{2} \left( \frac{6}{r^4} + \frac{\alpha g_{rr}}{6r^2}\right)\sim -\frac{r^2}{2} \left( \frac{6}{r^4} - \frac{2}{r^4}\right)
 = -\frac{2}{r^2} \,\,\,,\\
\frac{f_{RRR} {R'}^2}{f_{R}} &=& \frac{3 {R'}^2}{4(R-\alpha)^2} = \frac{3}{r^2} \,\,\,,\\
R+ \frac{f}{f_{R}} &=& R + 2 (R-\alpha)= \alpha + \frac{3}{r^2} \sim \alpha
\end{eqnarray}
Therefore, to leading order we obtain
\begin{equation}
\label{Grrasym}
G^r_{\,\,\,\,r} \sim g^{rr}\left(-\frac{2}{r^2} + \frac{3}{r^2}\right) -\frac{\alpha }{6} \sim -\frac{\alpha}{12}
-\frac{\alpha}{6}= -\frac{\alpha}{4} \,\,\,,
\end{equation}
so the r.h.s of Eq.~(\ref{fieldeq3}) is well behaved asymptotically, and it is just a constant which 
we can precisely identify with effective cosmological constant $\Lambda_{\infty}= \alpha/4$, with $\alpha<0$. 
Notice that this constant emerged not only from the 
last two terms at the r.h.s of Eq.~(\ref{Grr}) but also from the contribution of the first two, which one would naively think that they do not contribute 
as $R\rightarrow \alpha$ asymptotically. However, as mentioned in Sec.~\ref{sec:theory}, a closer look shows that one actually has in those two terms  
something like $\infty \times 0$ asymptotically. This is why it was necessary to perform the correct asymptotic analysis which leads 
then to contribution $-\alpha/12$ due to the first two terms of the r.h.s. of Eq.~(\ref{Grr}). Of course one can perform the same asymptotic analysis 
in the full set of equations (\ref{TraceRsss})--(\ref{deltasss}) to find that all of them behave well and consistently as 
$R\rightarrow \alpha$, since from both the l.h.s and the r.h.s one obtains exactly the same behavior. This is otherwise expected as we 
have explicitly the exact solution from which one can compute $M'$ and $R'$ and $R''$ to confirm that nothing diverges as $R\rightarrow \alpha$. 
The definition of this cosmological constant $\Lambda_{\infty}= \alpha/4$ is consistent with the canonical form 
that the metric coefficients $g_{tt}$ and $g_{rr}$ take in (\ref{gfRSZ}) in these coordinates. For instance, 
in terms of $\Lambda_{\infty}$ they read $g_{tt}=1/g_{rr}= - (1-\Delta - \frac{\Lambda_{\infty} r^2}{3} +  \frac{Q}{r^2})$ where $\Delta=1/2$.

Hence, we conclude that when $f_{RR}$ has a pole precisely at the Anti-De Sitter point, 
the cosmological constant $\Lambda_{\infty}$ does not arise simply from the last term of Eq.~(\ref{Grr}), 
like in the analysis performed in Sec.~\ref{sec:theory} where $\Lambda_{\rm eff}= R_1/4$. That analysis 
was valid provided that as $R\rightarrow R_1$ the following two necessary conditions were 
satisfied: $R_1= 2f(R_1)/f_R(R_1)$ and $f_{RR}(R_1)<\infty$, which as emphasized above, is not the actual case for this exact solution. 
Finally, we mention that for $\alpha\equiv 0$, which corresponds to a null cosmological constant, the quantity $(2f-Rf_R)/f_{RR}= -6R^2$
that appears in Eq.~(\ref{TraceRsss}) vanishes as the solution approaches 
the asymptotic value $R= 0$, 
even if $f_{RR}\rightarrow \infty$. Moreover, the term $f_{RRR} R'^2/f_{RR}= -3R'^2/(2R)$ also vanishes asymptotically since $R'\sim 1/r^3$ 
and $R\sim 1/r^2$. Therefore, a posteriori one can understand why Eq.~(\ref{TraceRsss}) is well behaved asymptotically.

It is important to stress that in the previous works~\cite{Clifton,Nzioki2010,Sebastiani2011}
the above physical and geometric interpretation of the metric (\ref{gfRSZ}) was completely absent 
and therefore, its meaning was rather unclear. In the nonvacuum case, some but not all of the aspects discussed above 
for this exact solution were elucidated~\cite{nonvacuum}. 

So far we have mainly discuss the case $\alpha<0$. Regarding $\alpha>0$, the Ricci scalar will tend to 
$\alpha$ asymptotically but will never reach this value since well before $r\rightarrow \infty$ the cosmological horizon 
is reached by the solution. Therefore, $R'(r_h^c)\neq 0$ and $R''(r_h^c)\neq 0$. That is, the possible 
solution $R\rightarrow \alpha$ with $R'=0=R''$ is never reached asymptotically due to the presence of the cosmological horizon. 
\bigskip

Now, as concerns the trivial solution $R=R_1= const$, the model (\ref{fRSZ}) admits the solution $R_1= 4\alpha/3$, which solves 
$R_1= 2f(R_1)/f_R(R_1)$. Notice that $f_{RR}(R_1)= -a [\alpha/3]^{-3/2}/2$. Therefore the SSS solution 
is given by the metric (\ref{SDS}) where 
\begin{equation}
\label{LeffFRSZ}
\Lambda_{\rm eff}= \frac{R_1}{4}= \frac{\alpha}{3} \,\,\,.
\end{equation}
In this case the cosmological constant is given by $\Lambda_{\rm eff}$, instead of $\Lambda_{\infty}= \alpha/4$.

We emphasize that the solution (\ref{SDS}) is {\it not} approached 
asymptotically by the solution~(\ref{gfRSZ}), since as we mentioned before, the latter has a deficit angle $\Delta=1/2$, while (\ref{SDS}) has  $\Delta=0$. 
In summary, the model (\ref{fRSZ}) admits the two exact solutions~(\ref{gfRSZ}) and ~(\ref{SDS}), one 
where the Ricci scalar is constant everywhere and one where it varies with the 
radial coordinate $r$. However, it is important to remark that both solutions are 
not two different solutions with the same boundary conditions, but two different solutions with 
different boundary conditions. Even if we put $M_0\equiv 0$, $Q\equiv 0$, one solution still has a deficit angle while the other does not. 
Moreover, both solutions have different cosmological constants $\Lambda_{\rm eff}$ and $\Lambda_\infty$, 
and so the spacetimes are not even the same asymptotically when $\alpha\neq 0$. On the other hand, 
one could try to make coincide the inner boundaries (i.e. the event horizons) of both solutions artificially 
as well as the value $R_h$ by fixing $Q$ and $M_0$, but still both spacetimes would have different global 
quantities (mass and charge) besides the respective deficit angles ($\Delta=0$ and $\Delta=1/2$) and 
the respective cosmological constants ($\Lambda_{\rm eff}$ and $\Lambda_\infty$). 
Namely, one solution would have zero ``ADM'' mass and non zero charge, while the other solution would have non-zero ``ADM'' mass and zero charge.

Two final remarks are in order. The first one concerns 
the deficit angle solution~(\ref{gfRSZ}). It is possible to show that in fact one can 
obtain the $f(R)$ model (\ref{fRSZ}) by using a kind of ``reconstruction method''. This consists by imposing a solution 
in the form of Eqs. (\ref{gfRSZ}) with (\ref{Mrexact})--(\ref{deltafRSZ}). That is with $g_{rr}= -1/g_{tt}$ and with $M$ and $R$ written in terms of 
finite powers of $r$. The condition $g_{rr}= -1/g_{tt}$ leads then to $\delta(r)= const$, i.e., $\delta'(r)\equiv 0$, which in turn provide a differential 
equation for $f(R)$ in terms of $r$, $R(r)$ and $M(r)$. However, given such a power expansion, one can invert and write $r=r(R)$, 
to obtain a differential equation for $f(R)$ and $R$ solely, which when solved provides (\ref{fRSZ}). This kind of ``tricks'', which are also common to find 
exact solutions in GR with exotic sources~\cite{Ayon-Beato1998}, have been applied in modified theories of gravity 
in the past~\cite{Sebastiani2011,nonvacuum}. 
Physically, one usually proceeds in the opposite way we just described. That is, one constructs or proposes an explicit $f(R)$ model
in order to fit some observations, for instance, in cosmology, and then one asks if such a model admits or not an exact solution in one or other 
scenario. At this regard, and as a second remark related with our comment at the end of the previous section, we mention that almost all physically viable 
$f(R)$ models admit the trivial solutions 
$R=R_1=const$ given by the Schwarzschild--De Sitter/Anti-De Sitter solution Eqs.~(\ref{SDS})--(\ref{delta0}), where the parameter $\alpha$ is replaced 
by a different constant according to the different $f(R)$ models. For instance, in Section~\ref{sec:expmodels} we will see more specific examples. 
These are the kind of trivial solutions that we alluded in Section~\ref{sec:theory} that several 
authors have systematically reported as something ``special'' in $f(R)$ gravity, while we see that they are nothing more than the usual 
solution found in GR with $\Lambda\rightarrow \Lambda_{\rm eff}= R_1/4$ and $G_0\rightarrow G_{\rm eff}= G_0/f_R(R_1)$. In particular the mass parameter 
$M_0$ can be redefined as $M_0= G_{\rm eff} {\cal M}$ where ${\cal M}$ can be taken as the fiducial mass associated with the spacetime. 
Moreover, the RN or the Kerr--Newman BH solutions with or without $\Lambda_{\rm eff}$ can also be found in $f(R)$ gravity when considering 
matter with a traceless energy-momentum tensor as these corresponds also to the same trivial solutions $R=R_1={\it const}$, including the 
AF ones when $R_1=0$.

We remark that there are more exact solutions for other choices of $n$ in the model $f(R)=k R^n$~\cite{Clifton,Nzioki2010}, not only 
for $n=1/2$, which corresponds to the model we have just analyzed taking $\alpha\equiv 0$. Nevertheless, most of those solutions are still trivial 
or have exotic asymptotics. In Section~\ref{sec:expmodels} (cf. Model 1) we discuss the case $n=2$, which was not covered in Ref.~\cite{Clifton,Nzioki2010}, as their equations become singular precisely for $n=2$.
\bigskip

Finding physically interesting exact BH solutions different from the trivial ones proves to be difficult when the 
$f(R)$ model is complicated, like the physically viable models that have passed many cosmological and Solar-System tests 
(e.g. Models 4--5 of Section~\ref{sec:expmodels}). In particular, if one is interested in genuine 
AF or ADS/AADS type of spacetimes. In such an instance, one has then to appeal to a numerical analysis. At this regard, we stress that 
we have used the two exact solutions presented in this section as a testbed for a FORTRAN code developed to solve numerically 
Eqs.~(\ref{TraceRsss})--(\ref{deltasss}) for more complicated and ``realistic'' $f(R)$ models, submitted to suitable boundary (regularity) conditions 
that represent the presence of a black hole (i.e. when an horizon is present). These {\it regularity conditions} are presented next.


\section{Regularity conditions}
\label{sec:regcond}

In order to obtain the regularity conditions at the horizon $r=r_H$, whether is the event (inner) horizon or the cosmological (outer) horizon, we 
expand the variables as follows:
\begin{equation}
\label{Fexp}
F(r) = F(r_H) + (r-r_H)F'(r_h) + \frac{1}{2}(r-r_H)F''(r_H) + \frac{1}{6}(r-r_H)^3 F'''(r_H) + {\cal O}(r-r_H)^4
\end{equation}
where $F(r)$ stands for $M(r)$, $R(r)$, $\delta (r)$. When replacing these expansions in Eq.~(\ref{TraceRsss}) 
and demanding that the derivatives of these variables are finite at the horizon one obtains after long but straightforward algebra 
the following regularity condition:
\begin{equation}
\label{Rprimereg}
R'|_{r=r_H} =  {\frac{ 2r\bigg(Rf_{R} -  2 f \bigg) f_{R} }
{ \bigg[r^2( 2Rf_{R}   -   f )   -  6f_{R} \bigg]f_{RR} }}\left|\rule{0mm}{0.6cm}\right._{r=r_H} \,\,\,,
\end{equation}
where as stressed, all the quantities in this equation are to be evaluated at the horizon $r=r_H$, which in principle is any positive value, 
whether $r_H$ is the event (inner) horizon (denoted by $r_h$) or the  (outermost) cosmological horizon 
($r_h^c$). In this paper we will only be interested in finding AF solutions, and thus, the spacetime will contain only the event (inner) horizon.

In turn, Eq.~(\ref{Msss}) provides:
\begin{equation}
\label{Mprimereg}
M'|_{r=r_H} =  {\frac{r^2\bigg(2Rf_{R} - f\bigg)}{12f_{R}}}\left|\rule{0mm}{0.6cm}\right._{r=r_H} \,\,\,.
\end{equation}

Hereafter the quantities evaluated at the horizon will be written with a subindex ``$H$''~\footnote{If one takes $r_h=0$, the regularity conditions 
correspond to a spacetime with a {\it regular origin.} In this case, $R'|_{r=0}=0$, $M|_{r=0}=0$, and $M'|_{r=0}=0$. 
The so called {\it solitons} are localized field configurations that are {\it globally regular}, 
including the origin. In GR there exist examples of spacetimes with matter that allow for 
{\it solitons} and SSSAFBH with hair, like in the 
Einstein--Yang--Mills system~\cite{Bartnik1988,Bizon1990} and in the Einstein-scalar-field system with an asymmetric scalar-field potential~\cite{Nucamendi2003}.}.
\bigskip

It is to be mentioned that regularity conditions of this kind were proposed first in~\cite{Bergliaffa2011}, and then rectified 
in~\cite{Halilsoy2012}. Those authors did not use exactly the same differential equations~(\ref{Msss}) and ~(\ref{deltasss}) 
that we used here, which were a consequence of Eqs.~(\ref{fieldeq3}) and~(\ref{traceR}). Instead, they departed from the much more involved 
field equation (\ref{fieldeq1}). Nevertheless our regularity conditions~(\ref{Rprimereg}) and~(\ref{Mprimereg}) are equivalent to those 
of~\cite{Halilsoy2012}.

In order to illustrate the consistency of these two regularity conditions, we take three $f(R)$ models and their exact solutions. 
First, $f(R)= R-2\Lambda$, which corresponds to 
GR plus a cosmological constant. The SSS solution is like in (\ref{SDS}) but with $\Lambda$ instead of $\Lambda_{\rm eff}.$ 
In this case Eq.~(\ref{Rprimereg}), when multiplied by $f_{RR}$, gives an identity $0\equiv 0$, since the numerator 
at the r.h.s $Rf_{R} -  2 f=R-4\Lambda\equiv 0$. On the other hand, since $R= 4\Lambda = const$ then 
$R'\equiv 0$. Therefore, $R' f_{RR} \equiv 0$. On the other hand, the r.h.s of Eq.~(\ref{Mprimereg}) yields 
$M'_H= \Lambda r^2_H/2$, which corresponds precisely to the De Sitter/Anti-De Sitter value obtained from Eq.~(\ref{Mrexact}).

The second case corresponds to the trivial constant solution discussed in previous sections: 
$R(r)=R_1= const$, $R'(r)\equiv 0$, where $R_1$ given by $(Rf_{R} -  2 f)_{R_1}=0$ 
and such that $f_{RR}(R_1)\neq 0$, $f_{RR}(R_1)< \infty$ and $f_{RRR}(R_1)< \infty$. Then Eq.~(\ref{Rprimereg}) gives $R'_H\equiv 0$, which is 
clearly compatible with the trivial solution, whereas the r.h.s of Eq.~(\ref{Mprimereg}) yields 
$M'_H= R_1 r^2_H/8$, where $R_1= 4\Lambda_{\rm eff}$.

A less trivial test to our regularity conditions is provided by the exact solution given by Eqs.~(\ref{Mrexact})--(\ref{deltafRSZ}). 
For that solution it is easy to verify that both sides of Eq.~(\ref{Rprimereg}) give $R'_H= -2/r_H^3$ while Eq.~(\ref{Mprimereg}) yields $M'_H= 1/4 + \alpha r^2/8 + Q/(4r_H^2)$ 
in both sides.
\bigskip

Now, Eqs.~(\ref{TraceRsss})--(\ref{deltasss}) do not provide the regularity conditions for $R''$ and $\delta'$ at the horizon. 
We require to differentiate Eqs.~(\ref{TraceRsss}) and (\ref{deltasss}) with respect to $r$ one more time. 
When doing so and when replacing the expansions in the 
form Eq.~(\ref{Fexp}) it is possible to obtain the values $R''_H$ and $\delta'_H$. This is a lengthy but otherwise straightforward 
calculation. In the Appendix~\ref{sec:regcond2} we provide the explicit expressions. 


\section{No hair theorems and Scalar-Tensor approach to $f(R)$ theories.}
\label{sec:STT}
As remarked in the Introduction, there are very well known NHT's for the Einstein-scalar field system, hereafter Einstein-$\phi$ system
 (i.e. Einstein-Hilbert gravity minimally coupled to a {\it real} scalar field 
$\phi$)  when the potential associated with a scalar field $\phi$ verifies the 
non-negativity condition $\mathscr{U}(\phi)\geq 0$. On the other hand, it is also well known that provided $f_{RR}>0$ one can 
map $f(R)$ into a Scalar-Tensor Theory (STT) by defining the scalar $\chi=f_R$. This is the so called Jordan-frame 
representation of the $f(R)$ theory. Under this framework and in the presence of matter, the scalar field $\chi$ turns out to be coupled {\it non minimally} 
to the curvature but minimally to matter. 
A further conformal transformation allows one to write the theory in the so called Einstein frame, where a new scalar field $\phi$ with a new potential 
$\mathscr{U}(\phi)$, is now coupled minimally to the curvature but non-minimally 
to the matter sector. Nonetheless, in the absence of matter, and still under the Einstein frame, the theory look exactly as an Einstein-$\phi$ system, 
and thus, one can turn to the NHT's and see whether they apply or not for the $f(R)$ model at hand. Since we are interested precisely in finding SSS black holes in 
vacuum, we can thus exploit this equivalence to see if hair is absent or if it can exist. Thus, given a specific $f(R)$ model, we carry out 
the following protocol: 1) We check if the model allows a ``trivial'' solution $R=0$ (leading to Schwarzschild solution). 
If the model does, then it means that the model is a priori able of allowing AF hairy solutions 
(i.e. solutions where $R$ would be a non-trivial function of $r$ that interpolates from the 
event horizon, with value $R_h$, to spatial infinity where $R=0$). Many $f(R)$ models used as geometric dark-energy admit also the trivial 
solution $R=0$. Thus, in this paper we focus only in such kind of models and analyze if geometric hair can also exist. In a future work, 
we shall analyze the existence of hair in ADS or AADS black-hole solutions; 2) We write the model as a STT in the Einstein frame. If the potential 
$\mathscr{U}(\phi)$ is not negative, then the NHT's apply and we conclude that AFSSS hairy solutions {\it cannot} exist in such model. In fact, if 
$\mathscr{U}(\phi)$ turns to be strictly positive, the Schwarzschild solution does not even exist, only the De Sitter type of solutions exist at best; 
3) If the potential has negative branches then we proceed 
to analyze if hairy solutions exist by solving numerically the field equations of Sec.~\ref{sec:SSS} under the regularity conditions provided in Secs.
~\ref{sec:regcond} and the Appendix~\ref{sec:regcond2}. 

In order to fix the ideas, let us now briefly review the different scalar-field transformations required to analyze properly an $f(R)$ model under 
the STT approach.

Let us consider for simplicity the following gravitational action without matter 
\begin{equation}
\label{Qframe}
I_{\rm grav} =\!\! \int \!\! d^4 x \sqrt{-g}\:\frac{1}{2\kappa}\Big[f_Q(Q)(R-Q) + f(Q)\Big],
\end{equation}
where at this point $Q$ is an auxiliary scalar-field that depends on $R$ and $f_Q= df/dQ$~\footnote{Do not confuse the field $Q$ with the parameter appearing in 
the metric (\ref{gfRSZ}).}. Now, if $f_{QQ}(Q)\neq0$ (in particular $f_{QQ}(Q)>0$) the variation of this action with respect to 
the metric give rise to field equations equivalent to the original theory (\ref{f(R)}) in vacuum provided $Q\equiv R$. Thus the action (\ref{Qframe}) 
is equivalent to (\ref{f(R)}) if we enforce $Q=R$ (see Appendix~\ref{sec: BDidentification} for a further discussion).

Furthermore, one can introduce the scalar field $\chi: = f_Q(Q)$ and write a dynamically equivalent action as
\begin{equation}
\label{JordanF}
I_{\rm JF} =
\!\! \int \!\! d^4 x  \: \sqrt{-g}  \Big( \frac{1}{2\kappa} R\:\chi - \chi^{2}V(\chi) \Big), 
\end{equation}
where $V(\chi)$ is defined as follows:
\begin{equation}
\label{JFpot}
V(\chi) = \frac{1}{2\kappa\chi^{2}}\Big[ Q(\chi)\chi - f(Q(\chi)) \Big].
\end{equation}
By {\it dynamically equivalent}, we mean that we obtain field equations that are completely equivalent to the original action 
(see Appendix~\ref{sec: BDidentification}). To do so, 
we can treat the scalar-field $\chi$ as metric-independent, and $R$ as independent of $\chi$, 
and thus take the action as a functional of both 
$g_{ab}$ and $\chi$. Thus, variation of this action with respect to the metric leads to field equations equivalent to Eq.~(\ref{fieldeq1}), 
while variation with respect to $\chi$ simply leads to $R=Q$. We see then that the action (\ref{JordanF}) is equivalent to the action of a Brans--Dicke 
like theory, with a Brans--Dicke parameter $\omega_{\rm BD} = 0$ (implying that the kinetic term associated with the gradients 
of $\chi$ is absent) and with a potential $U(\chi)= 2\kappa \chi^2V(\chi)$ (see Appendix~\ref{sec: BDidentification} for the details). 
Clearly a STT of this sort (i.e. one with $\omega_{\rm BD} = 0$) but without a potential would be incompatible with the bound 
$\omega_{\rm BD} \gtrsim 4\times 10^4$~\cite{Bertotti2003}, which is required for the theory to pass the Solar System tests, 
and thus, would be automatically ruled out. However, the presence of this potential makes possible for certain $f(R)$ models to pass those tests even if 
$\omega_{\rm BD} = 0$. This depends if the potential allows for the emergence of the {\it chameleon} mechanism (cf. Ref.~\cite{Hu2007,Lombriser2015}), 
but not all the potentials have this property.
 
While we will consider some $f(R)$ models that seem to pass such tests thanks to the chameleon mechanism, our main purpose in this paper is 
to analyze the issue about the existence or absence of hairy BH solutions in such models rather to test their observational viability. Thus, there are models that we use only for that purpose and which may fail the observational tests.

It is said that, the actions (\ref{Qframe}) and (\ref{JordanF}) are written in the so called Jordan frame (JF), where by {\it frame} is to be 
understood as a set of physical variables. As we remarked, in reality we have only introduced a new scalar-field variable without changing the metric in any way, 
thus this formulation of the theory is completely equivalent to the original one (at least in the sectors where $f_{RR}>0$), and thus, the field equations are also 
equivalent. 

As mentioned before, the energy-momentum tensor of matter is conserved in the JF (see the Appendix~\ref{sec:consEMT}), and therefore it is with respect to this frame 
that point-test particles follow geodesics.

Furthermore one can define a new scalar field and a conformal metric as follows:
\begin{eqnarray}
\label{phichi}
\phi &=& \sqrt{\frac{3}{2\kappa}} {\rm ln}\chi \,\,\,,\\
\label{confT}
\tilde{g}_{ab} &=& \chi \: g_{ab} = e^{ \sqrt{\frac{2\kappa}{3}}\phi}\: g_{ab} \,\,\,.
\end{eqnarray}
In terms of these new variables the gravitational action (\ref{JordanF}) takes the form~\cite{Maeda2003}\footnote{In the action (\ref{EinsteinF}) we have omitted 
in the integral the term $\sqrt{\frac{3}{2\kappa}}{\tilde \nabla}^a{\tilde \nabla}_a \phi$, which can be converted into a surface term, 
that we assume to vanish.}
\begin{equation}
\label{EinsteinF}
I_{\rm EF} =
\!\! \int \!\! d^4 x  \: \sqrt{-\tilde{g}}  \Big[ \frac{1}{2\kappa}\tilde{R} - \frac{1}{2}\tilde{g}^{ab}({\tilde \nabla_a} \phi)({\tilde \nabla_b}\phi) 
- \mathscr{U}(\phi) \Big],
\end{equation}
where all the quantities with a {\it tilde} are defined with respect to the metric $\tilde{g}_{ab}$. 
This corresponds precisely to the Einstein-Hilbert action coupled minimally to a scalar field $\phi$. Notice that, unlike the JF action, a kinetic 
term appears due to the conformal transformation. The action (\ref{EinsteinF}) is written in what is known as the {\it Einstein frame} (EF), where the potential 
$\mathscr{U}(\phi)$ is defined as $\mathscr{U}(\phi) \equiv V(\chi[\phi])$, which will be given explicitly when the model $f(R)$ be provided 
(see Section~\ref{sec:expmodels}). If we include the matter action then the scalar field $\phi$ will be coupled non minimally to the matter fields. 
In this paper we are only interested in the vacuum case~\footnote{See Ref.~\cite{Damour1992} for a thorough discussion of STT in the EF 
with the presence of matter.}, 
so the field equations obtained from the action (\ref{EinsteinF}) are simply those of the Einstein-$\phi$ system:
\begin{eqnarray}
{\tilde G}_{ab} &=& \kappa T_{ab}^\phi \,\,\,,\\
T_{ab}^\phi &=& ({\tilde \nabla_a}\phi) ({\tilde \nabla_b}\phi) - {\tilde g}_{ab}\left[ \frac{1}{2}{\tilde g}^{cd}({\tilde \nabla_c}\phi) ({\tilde \nabla_d}\phi) + 
\mathscr{U}(\phi)\right]\,\,\,,\\
\label{KGEF}
\Box^{\!\!\!\tilde{}} \,\,\phi &=& \frac{d\mathscr{U}}{d\phi} \,\,\,.
\end{eqnarray}

As we emphasized previously, based on this equivalence and in view of our interest in finding SSS and AF non-trivial black holes in $f(R)$ gravity, 
we have to take into account the NHT's~\cite{Sudarsky1995,Bekenstein1995} which are valid when 
$\mathscr{U}(\phi)\geq 0$~\footnote{The theorems actually account for multiple scalar fields.}. The theorems 
roughly establish that {\it whenever the condition $\mathscr{U}(\phi)\geq 0$ holds, given an AFSSS spacetime containing a black-hole 
(with a regular horizon) within the Einstein-$\phi$ system, the only possible solution is the hairless Schwarzschild solution.} Here by {\it hairless} we mean 
that the scalar field $\phi(r)=\phi_0$, i.e., the scalar field is constant everywhere in the domain of outer communication of the BH and it is such  
that $\mathscr{U}(\phi_0)\equiv 0$, in order to prevent the presence of a cosmological constant which would spoil the AF 
condition.

The NHT's can be avoided if the potential has negative branches, notably at the horizon~\cite{Nucamendi2003,Anabalon2012}. So in our case, given an 
$f(R)$ model, we have only to check if the corresponding potential satisfies or not 
the condition $\mathscr{U}(\phi)\geq 0$. In the affirmative case, we conclude that SSS and AF hairy black holes are absent. Nevertheless, when 
this condition fails, one usually need to resort to a numerical treatment in order to analyze if a black hole can support scalar hair or not.

Before concluding this section, a final remark is in order. There is an important relationship between the critical points of the potential $\mathscr{U}(\phi)$, 
notably the extrema, and those of the ``potential'' $\mathscr{V}(R)$ defined such that~\cite{Jaime2011}(see also Appendix~\ref{sec: BDidentification})
\begin{equation}
\label{derpotVf}
\mathscr{V}_R(R)= \frac{d\mathscr{V}}{dR}= \frac{1}{3}(2f- Rf_R)\,\,\,,
\end{equation}
In Section~\ref{sec:theory} we denoted the extrema of $\mathscr{V}(R)$ by $R_1$. It is straightforward to verify 
\begin{equation}
\frac{d\mathscr{U}}{d\phi} = \left(\frac{d\chi}{d\phi}\right)\left(\frac{d V}{d\chi}\right) = 
\frac{1}{\kappa f_R^2} \left(\frac{3}{2\kappa}\right)^{3/2}\Big(\frac{2f-Rf_R}{3}\Big) = 
\left(\frac{3}{2\kappa}\right)^{3/2} \frac{1}{f_R^2}\frac{d\mathscr{V}}{dR}
\end{equation}
Therefore, provided $f_R(R_1) \neq 0$ and $f_R(R_1)<\infty$ (i.e. the conditions for a well defined map to the Einstein frame), 
we see that the extrema of $\mathscr{U}(\phi)$ correspond precisely to $R_1$.  
However, care must be taken when $f_R(R_1)=0$ or $f_R(R_1)=\infty$, as it may happen in some models that we will encounter in the next section. 
When this happens, the conformal transformation (\ref{confT}) becomes singular or ill defined as $\chi=f_R$.

Finally, for $f(R)$ models where $f_{RR}$ is not strictly positive, notably, where $f_{RR}$ can vanish at some $R=R_w$, called {\it weak singularity} 
(cf. Model 5 in Section~\ref{sec:expmodels}) it will be useful to introduce the ``potential'' $\mathscr{W}(R)$ defined via
\begin{equation}
\label{derpotVfmod}
\mathscr{W}_R (R)= \frac{d\mathscr{W}}{dR}= \frac{2f- Rf_R}{3 f_{RR}}= \frac{\mathscr{V}_R(R)}{f_{RR}}\,\,\,.
\end{equation}
The finite or divergent behavior of $\mathscr{W}_R (R_w)$ also provides insight about $\mathscr{V}_R (R_w)$. For instance, if $\mathscr{W}_R (R_w)$ is 
finite, it means that $\mathscr{V}_R (R_w)$ vanishes like $f_{RR}(R_w)$. Furthermore, $\mathscr{W}_R(R)$ can supply further information about the possible trivial 
solutions $R=const$. As we discussed in Section~\ref{sec:theory} and also in Section~\ref{sec:exactsols}, if at $R=R_2$ where 
$f_{RR}(R_2)=\infty$ and $\mathscr{W}_R (R_2)$ vanishes, $R=R_2$ can be one trivial solution of Eq.~(\ref{traceR}) in vacuum or more generally, 
when the matter has a traceless energy-momentum tensor, a solution that can be different from $R=R_1$ if $\mathscr{V}_R(R_2)\neq 0$.


\subsection{$f(R)$ models and the NHT's}
\label{sec:expmodels}

In this section we focus on some $f(R)$ models that also admit the trivial solution $R=0$, and check whether or not they 
satisfy the condition $\mathscr{U}(\phi)\geq 0$ for which the NHT's apply. The results are summarized in Table~\ref{tab:models} at the 
end of this section. In order to obtain the EF potential $\mathscr{U}(\phi)$ from $V(\chi)$ let us recall the relationships
\begin{eqnarray}
\label{chiphi}
\chi &=& e^{ \sqrt{\frac{2\kappa}{3}}\phi}\,\,\,,\\
\label{EFpot}
\mathscr{U}(\phi) &:=& V(\chi[\phi]) \,\,\,,
\end{eqnarray}
the first one is obtained from Eq.~(\ref{phichi}), while the second one is a definition.

{\bf Model 1:} 
$f(R) = \lambda_n \Big(\frac{R}{R_n}\Big)^{n}$, where $n$, $\lambda_n$ and $R_n$ are positive parameters of the theory. 
$R_n$ fixes the scale for each $n$ and $\lambda_n$ can be chosen to be proportional to $R_n$. 
This model has been thoroughly analyzed in the past in several scenarios (see \cite{Jaime2013} and references 
therein). For this model,
\begin{equation}
\label{mod1chiR}
\chi= f_R= \frac{n\lambda_n}{R_n}\Big(\frac{R}{R_n}\Big)^{n-1}\,\,\,.
\end{equation}
If we focus in the domain $R\in [0,\infty)$, then for $n>1$, $\chi \in [0,\infty)$. In fact the value $\chi=0$ corresponds to a degenerate situation 
$G_{\rm eff}\rightarrow \infty$ that we discuss below. For $0<n<1$, $\chi\rightarrow \infty$ as $R\rightarrow 0$ and 
vice versa. So, $\chi \in (0,\infty)$ for $0<n<1$.

By inverting the relationship (\ref{mod1chiR}) and using Eq.~(\ref{JFpot}) followed by the use of Eqs.~(\ref{chiphi}) and (\ref{EFpot}), 
the two potentials read
\begin{equation}
V(\chi) = \frac{(n-1)}{2 \kappa n} \Big( \frac{R_n^n}{n \lambda_n} \Big )^{
\frac{1}{n-1} } \chi^{ \frac{2-n}{n-1} } \hskip 1cm (n\neq1) \,\,\,,
\end{equation}
defined for $\chi>0$, and
\begin{equation}
\mathscr{U}(\phi) = \frac{(n-1)}{2 \kappa n} \Big( \frac{R_n^n}{n\lambda_n} \Big
)^{ \frac{1}{n-1} } e^{\big(\frac{2-n}{n-1}\big) \sqrt{\frac{2\kappa}{3}}\phi} 
\hskip 1cm (n\neq1) \,\,\,,
\end{equation}
defined for $-\infty<\phi< +\infty$. The condition  $\mathscr{U}(\phi) >0$ holds if  $n>1$, while for $n<1$, the potential is  $\mathscr{U}(\phi) <0$, but 
in both cases the potential does not have minima, at least not for a finite $\phi$. Moreover, we do not consider 
$n<0$ because $f_R<0$ which can give rise to an effective negative gravitational constant. For $n=1$, corresponding to GR, 
$V(\chi) \equiv 0 \equiv \mathscr{U}(\phi)$ as one can see directly from Eq.~(\ref{JFpot}). For this model the NHT's a priori apply since 
$\mathscr{U}(\phi) >0$. The fact that the potential is strictly positive and have no minima imply that the solution $R=0$ cannot even exist in 
the EF. The point is that for $0<n<1$ the solution $R=0$ corresponds to $\chi\rightarrow \infty$ ($\phi\rightarrow \infty$), while for $n>1$ 
the same solution corresponds to $\chi\equiv 0 \equiv R$ ($\phi\rightarrow -\infty$). We see then that in both cases the mapping to the STT in the EF is ill 
defined precisely at $\chi=0$ where $\tilde g_{ab}=0$, while $\tilde g^{ab}\rightarrow 0$ as $\chi\rightarrow \infty$
~\footnote{This particular model illustrates the care that one has to take when transforming the original variables to the STT counterpart. 
In the next section we shall encounter situations where $\mathscr{U}(\phi)$ is not even well defined as it turns to be multivalued.}. This problem 
at $\chi=0$ exacerbates for $n=2$ that we discuss below. We conclude that for this $f(R)$ model AFSSS simply cannot exist 
under the EF. In the original formulation, the theory also degenerates at $R=0$ for $n>1$ since $f_R(0)=0$.  AFSSS solutions exist but they are no unique as we are 
about to see.

As we remarked briefly at the end of Section~\ref{sec:theory}, for this class of $f(R)$ models a quite degenerate situation may occur. 
To fix the ideas, let us focus on the case $n=2$ in the original formulation, since in the STT approach the maps breakdown at $R=0$ as we just 
mentioned, given that $\chi=f_R= const \times R$. So in this case the field Eqs.~(\ref{fieldeq3}) and~(\ref{traceR}) in vacuum reduce to
\begin{eqnarray}
\label{field1R2}
 G_{ab} &=& \frac{1}{R}\Bigl{[} \nabla_a \nabla_b R -\frac{g_{ab}}{4}R^2 \Bigl{]} \,\,\,,\\
\label{field2R2}
\Box R &=& 0 \,\,\,.
\end{eqnarray}
where we used $f_{RRR}\equiv 0$, $(Rf_R + f)/f_R= 3R/2$ and $(2f-Rf_R)/f_{RR}=0$ in Eqs.~(\ref{field1R2}) 
and~(\ref{field2R2})\footnote{For this $f(R)$ model $(2f-Rf_R)/f_{RR}= \frac{(n-2)  R^2}{n(n-1)}$ and $(Rf_R + f)/f_R= \frac{(n+1)R}{n}$. 
Thus we can expect degenerate solutions (in the sense described in the main text) for $n>2$ as well.}.

Therefore we see that $R=0$ is a trivial solution of Eq.~(\ref{field2R2}). On the other hand, for such trivial solution 
Eq.~(\ref{field1R2}) is satisfied for any $G_{ab}\neq 0$ compatible with $G^{a}_{\,\,a}= -R=0$. For instance, this can be satisfied 
for any solution of the metric satisfying the Einstein equation $G_{ab}=\kappa T_{ab}$, with $T^{a}_{\,\,a}=0$. This {\it degeneracy} is 
somehow remarkable as shows that solutions of the field equations in $f(R)$ gravity may not be unique, as illustrated by this simple model. 
In the AFSSS scenario, one BH solution is clearly the Schwarzschild solution, but other solutions are possible~\cite{Nzioki2010,Kehagias2015}, 
which would be interesting to know to what kind of matter content they correspond in pure GR. In this example, 
the AFSSSBH solutions are unique as concerns the solution $R=0$, since at the horizon $R=0=R'$, but they are highly non unique as concerns the metric. 
That is, to the trivial solution $R=0$ of Eq.~(\ref{TraceRsss}), one can associate any solution for the metric satisfying  
$G_{ab}=\kappa T_{ab}$, with $T^{a}_{\,\,a}=0$, like the Schwarzschild solution, the RN solution, a solution within the Einstein--Yang--Mills system, etc.

As emphasized in~\cite{Nzioki2010,Kehagias2015}, this $f(R)$ model provides a specific example showing that a generalization of the
 Birkhoff's theorem similar to the one elucidated in the Introduction, simply cannot exist in general. It is enlightening to stress that this 
degenerate situation in vacuum is in a way similar, 
but opposite, to the ``degeneracy'' that appears in GR with matter sources: given $f(R)= R$, Eq.~(\ref{fieldeq1}) or 
Eq.~(\ref{fieldeq3}) reduce to the Einstein field equation, whereas ~(\ref{traceR}) reduce to $f_{RR} \Box R\equiv 0$, with $f_{RR}\equiv 0$. 
This means that this equation is satisfied identically regardless of the value $\Box R$, which in general, is not zero because $R=-\kappa T$. 
In other words, in GR the metric is constrained to satisfy the Einstein's field equation, but $R$ is not constrained to satisfy any differential 
equation like Eq.~(\ref{fieldeq1}).

Finally we mention that when $R= const=R_1$, the model with $n=2$ also admits the trivial solution 
Eqs.~(\ref{SDS})--(\ref{derpotcond})~\footnote{For $n=2$ the potential $\mathscr{U}(\phi)$ is constant in the EF, thus the trivial solution 
$\phi={\it const}$ gives rise to the Schwarzschild--De Sitter solution just like in the original variables.}. Incidentally, for this model the 
algebraic condition (\ref{derpotcond}) is satisfied for any $R_1$. That is, $R_1$ emerges as an integration constant independent of the parameters 
of the model. Therefore the value $\Lambda_{\rm eff}=R_1/4$ depends on the assigned 
value for $R_1$. In particular, taking $R_1=0$ we just recover the usual Schwarzschild solution as mentioned above. 

For any other $n > 1$ the model admits the trivial solution $R=0$, but the solution for the metric is not unique either as the degeneracy emerge as 
well in a similar way to the case $n=2$.
\bigskip

{\bf Model 2:} $f(R) = R + c_2 R_{I}(R/R_{I})^{2}$, where $c_2$ is a positive dimensionless constant and $R_I$ is a positive parameter that fixes the {\it scale}. This model was proposed by Starobinsky 
as an alternative to explain the early inflationary period of the universe~\cite{Starobinsky1980}. 
 For this model,
\begin{equation}
\chi=  f_R= 1 + 2c_2 R/R_{I}\,\,\,.
\end{equation}
In principle the model is defined for $-\infty < R<\infty$. However, if we impose $\chi=f_R>0$, then we require $R> -R_{I}/(2c_2)$. In fact if we allow $\chi \leq 0$ the 
transformation to STT is not well defined, and the model degenerates at $\chi=0$ in the original variables as $G_{\rm eff}\rightarrow \infty$. If we focus on 
solutions with $R \geq 0$ then $\chi \geq 1$. Furthermore, $f_{R}$ and $f_{RR}$ are both finite at $R=0$ for any $c_2 \in \mathbb{R}$.

Proceeding like in the previous model, this one has associated the following potentials
\begin{eqnarray}
V(\chi) &=& \frac{R_{I}}{8 c_2\kappa}\Big( \frac{\chi -1}{\chi} \Big)^{2} \,\,\,,\,\,\,\\
\label{inflacion}
\mathscr{U}(\phi) &=& \frac{R_{I}}{8 c_2\kappa}\Big( 1 - e^{ -\sqrt{\frac{2\kappa}{3}}\phi} \Big)^{2} \,\,\,.
\end{eqnarray}
The potential $\mathscr{U}(\phi)$ is defined for $-\infty <\phi <\infty$. The region $\phi \geq 0$ corresponds to $\chi \geq 1$, while $-\infty<\phi<0$ 
corresponds to $0<\chi <1$.

\begin{figure}[t]
\includegraphics[width=8.5cm,height=5cm]{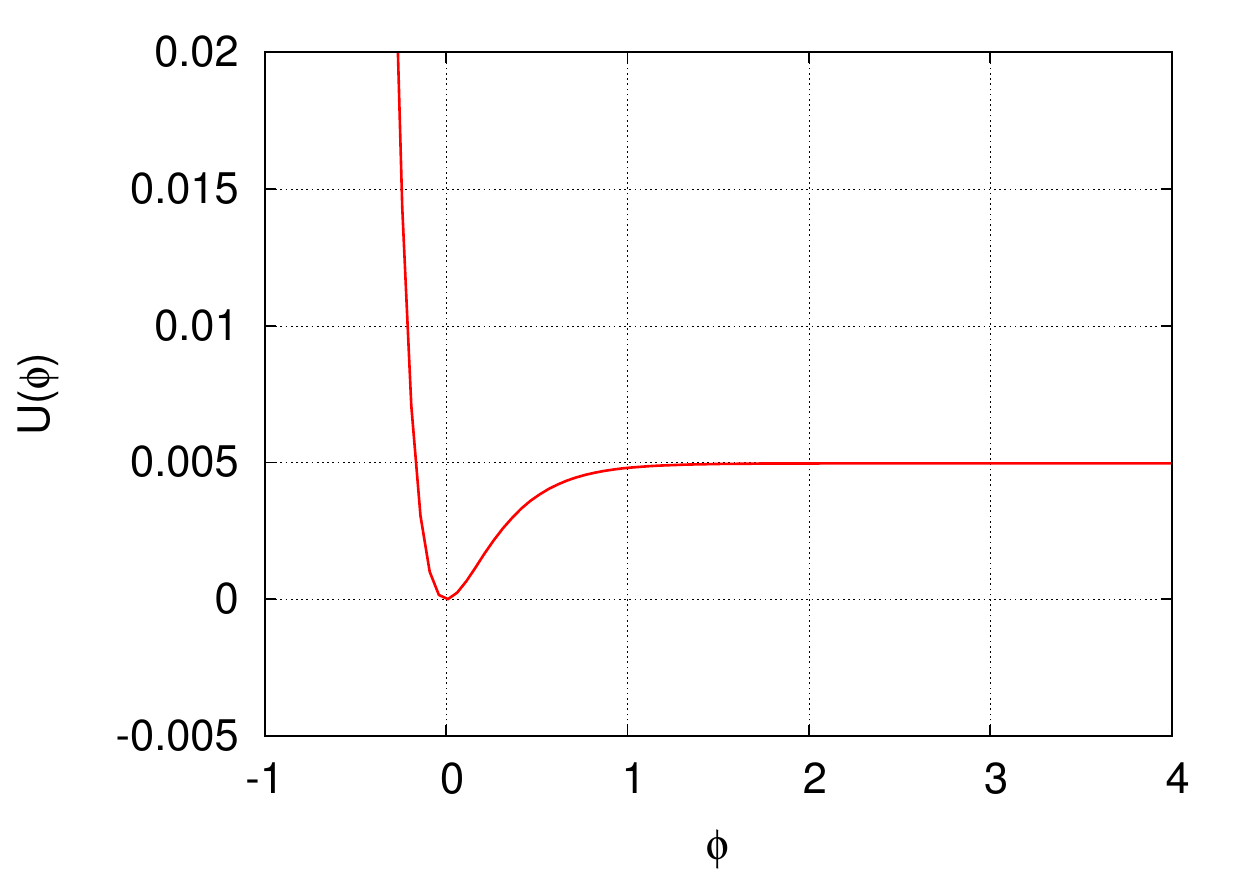}
\caption{(color online). Potential $\mathscr{U}(\phi)$ (in units of 
$R_{I}/G_0$; $\phi$ is given in units of $G_0^{-1/2}$) associated with the Model 2 with $c_2=1$. The potential verifies $\mathscr{U}(\phi)\geq 0$ for any $c_2>0$ 
and therefore, the NHT's apply. The minimum at $\phi=0$ is associated with the Schwarzschild solution where $R(r)\equiv 0$.}
\label{fig:Staroinf}
\end{figure}

Clearly $\mathscr{U}(\phi)\geq 0$, with a global minimum located at $\phi=0$ where the potential vanishes (see Figure~\ref{fig:Staroinf}). 
The model admits the solution $R=0$ corresponding to $\chi=1$ and $\phi=0$. The trivial solution $R=0$ is the only root of 
$\mathscr{V}_R(R)= R/3$. For this model the NHT's apply, and therefore AFSSSBH hairy solutions 
are absent. 

Although the model is defined for $c_2>0$ in order to be compatible with inflation, in the context of BH's
one can in principle consider $c_2<0$ as a way to evade the NHT's because then $\mathscr{U}(\phi)\leq 0$
(in that instance the condition $\chi >0$, implies $R< R_{I}/(2|c_2|)$ 
and the global minimum becomes a global maximum). 
It turns out, however, that potentials that are negative around a maximum but vanishes there (in this case $ \mathscr{U}(0)=0$) sometimes admit non-trivial 
solutions that seem hairy and AF\footnote{In more general 
scenarios such potentials lead to instabilities because perturbations around the 
maximum leads to growing modes. Thus the trivial solution $\phi=0$ would be unstable 
and never settle into a stationary configuration.}. We shall discuss a numerical example of this kind later, but suffice is to say that the BH solutions that one finds 
may vanish asymptotically but can have an {\it oscillatory} behavior that make them not genuinely AF. In order to illustrate this, suppose that asymptotically 
the metric component $g_{rr}$ behaves like $g_{rr}\sim 1 + 2 C_1r^\sigma \sin(C_2 r)/r$ (where $\sigma$ and $C_{1,2}$ are some constants, 
and $0\leq \sigma <1$),  then the mass function $M(r)\sim C_1r^\sigma \sin(C_2 r)$ oscillates (it may even diverge if $\sigma \neq 0$), and thus, 
it does not really converge to a finite value in the limit $r\rightarrow \infty$, a value that 
one would identify with the ADM mass. Yet  $g_{rr}\rightarrow 1$ as $r\rightarrow \infty$. Thus, for this kind of solutions the spacetime is not 
authentically AF. These arguments can be justified using the following heuristic 
analysis. Let us consider Eq.~(\ref{KGEF}) and neglect the non-flat spacetime contributions from the metric. Moreover, expanding $\mathscr{U}(\phi)$ around its maximum 
(which is equivalent to expand $-\mathscr{U}(\phi)$ around its minimum) 
gives $\mathscr{U}(\phi)= -m^2\phi^2/2$. With these simplifying assumptions 
it is easy to see that the SSS solution of Eq.~(\ref{KGEF}) is 
\begin{equation}
\phi= \phi_0 \frac{\sin(x + x_0)}{x}\,\,\,.
\end{equation}
where $x= mr$, and $x_0$, $\phi_0$ are constants. Now, at leading order when $r\rightarrow \infty$, we can take the conformal factor $\chi=1$ 
in Eq.~(\ref{confT}) and both metrics (the Jordan and the Einstein frame metrics) 
coincide asymptotically. Thus the energy-density contribution 
$\tilde{\rho}_\phi= -T_{t}^{\,\,t\,,\,\phi}$ is given by
\begin{equation}
\tilde{\rho}_\phi= \frac{1}{2}\phi'^2 + \mathscr{U}(\phi)= 
\frac{\phi_0^2\cos[2(x+x_0)]}{r^2} \,\,\,,
\end{equation}
which is not positive definite. As a consequence, the mass functions is not positive 
definite either. In fact, the mass function behaves asymptotically as $M(r)\sim
\phi_0^2\sin[2(x+x_0)]/m$, which oscillates with $r$ and does not converges to a definite value (the Komar mass). As we mentioned 
above, this heuristic analysis confirms the behavior of the full numerical solution 
for potentials of this kind. One such sort of solutions are shown for the Model 4 
in Section~\ref{sec:numerical} below.
\bigskip

{\bf Model 3:} $f(R) = R - \alpha_1 R_{\ast}\ln\Big( 1 + \frac{R}{R_{\ast}}\Big)$, where $\alpha_1$ is a dimensionless constant, and $R_{\ast}$ is a positive 
parameter that fixes the scale. 
This model was proposed  by Miranda {\it et al.}~\cite{Miranda2009} as a possible solution for the accelerated expansion of the universe while being free of 
singularities during the cosmic evolution. This model is also viable for constructing relativistic extended objects. Notwithstanding, it has problems at the level of the Solar System and grow of structure~\cite{Mirandadebate}. Although it may not be a realistic model 
in all the scenarios, it is worth consider it in this context due to its simplicity. The model is defined for 
$R>- R_{\ast}$ and the condition $f_R>0$ restricts $R$ in the range $R>R_{\ast}(\alpha_1-1)$. In particular 
for $R_{\ast}$ and $\alpha_1$ positive, as we assumed for this model, 
whenever $f_R>0$ is satisfied the condition $R>- R_{\ast}$ is also satisfied. Furthermore $f_{RR}= \alpha_1 R_*^{-1}(R/R_*+1)^{-2}$ is finite at $R=0$, as well 
as $f$ and $f_R$. For this model 
\begin{equation}
\chi=f_R=  1 - \frac{\alpha_1 R_{\ast}}{R_{\ast} + R} \,\,\,.
\end{equation}
The domain $R_{\ast}(\alpha_1-1) < R < \infty$ corresponds to $ 0<\chi <1$, and to $-\infty < \phi < 0$. The potentials are 
\begin{eqnarray}
V(\chi) &=& \frac{R_{\ast}}{2\kappa \chi^2}\bigg[  \alpha_1\ln\bigg( \frac{\alpha_1}{ 1 - \chi}\bigg) + 1-\chi - \alpha_1 \bigg], \\
\mathscr{U}(\phi) &=& \frac{R_{\ast}}{2\kappa}e^{
-2\sqrt{\frac{2\kappa}{3}}\phi}\bigg[  \alpha_1\ln\bigg( \frac{\alpha_1}{ 1 -
e^{\sqrt{\frac{2\kappa}{3}}\phi}}\bigg) - e^{\sqrt{\frac{2\kappa}{3}}\phi}
+ 1 - \alpha_1 \bigg].
\end{eqnarray}
The potential $\mathscr{U}(\phi)$ is thus defined for $-\infty < \phi < 0$.

\begin{figure}[t]
\includegraphics[width=5.9cm,height=5cm]{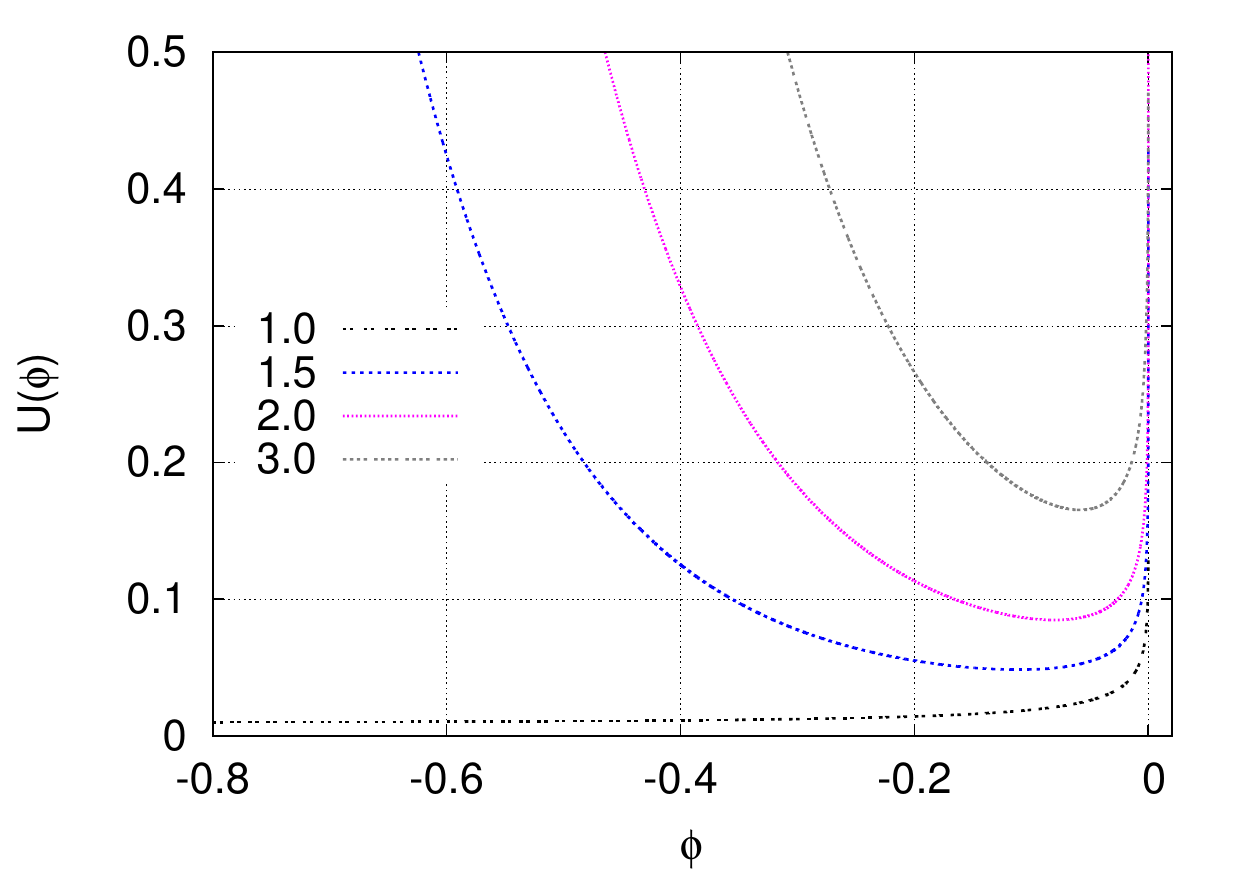}
\includegraphics[width=5.9cm,height=5cm]{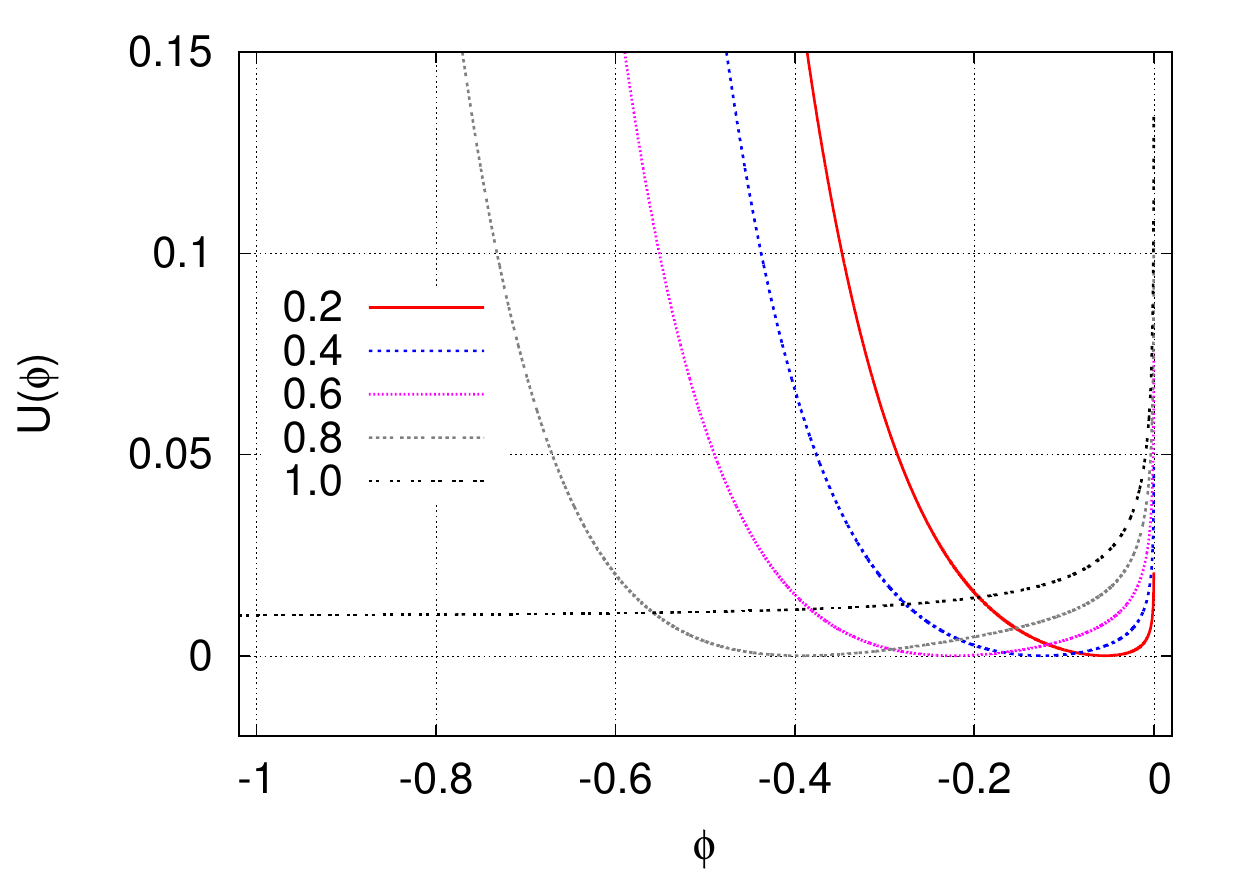}
\includegraphics[width=5.9cm,height=5cm]{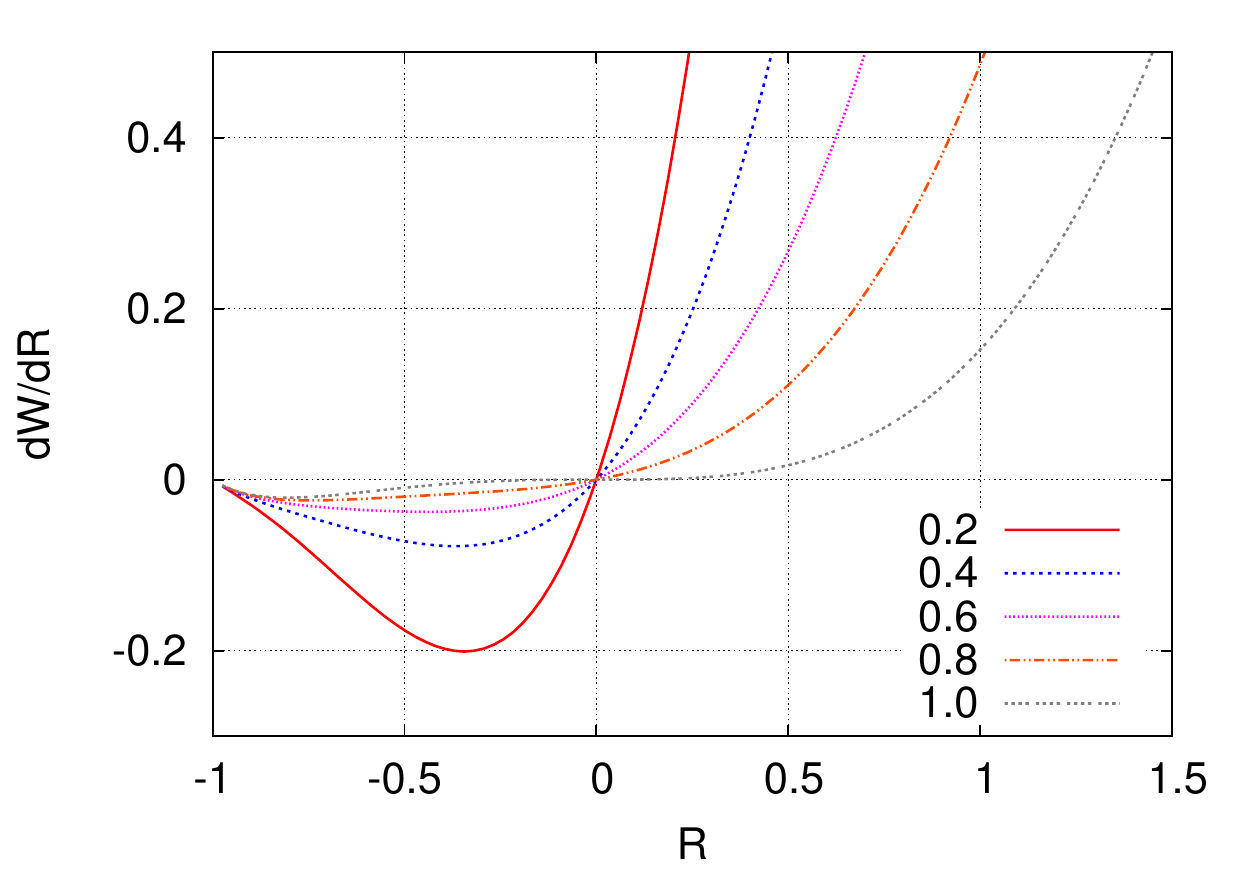}
\caption{(color online). Left panel: Potential $\mathscr{U}(\phi)$ (in units of $R_*/G_0$; $\phi$ is given in units of 
$G_0^{-1/2}$) associated with the Model 3 for different values of $\alpha_1$ in the range $1 \leq \alpha_1$. For such values $\mathscr{U}(\phi)$ is strictly positive. 
In particular for $\alpha_1=1$ the potential vanishes when $\phi\rightarrow -\infty$. For $\alpha_1>1$ the potential has a global minimum at 
$\phi_{\rm 1,min}$ which leads to a De Sitter type of solutions $\phi(r)= \phi_{\rm 2,min}$ in the EF and $R(r)= const$ in the JF. Middle panel: 
Potential $\mathscr{U}(\phi)$ for five values of $\alpha_1$ in the interval $0 < \alpha_1 \leq 1$. For $0 < \alpha_1 < 1$, the potential satisfies 
$\mathscr{U}(\phi) \geq 0$ and it vanishes at the global minimum $\phi_{\rm 2,min}$ leading to the Schwarzschild solution where $\phi(r)= \phi_{\rm 2,min}$ 
in the EF and $R(r)\equiv 0$ in the JF (see the right panel). 
Right panel: the function $d\mathscr{W}/dR$ is depicted. The zeros of this function at  $R=0$, which are mapped to $\phi_{\rm 2,min}$ in the 
EF, lead to the Schwarzschild solution. The zeros at $R= -1$ (in units of $R_*$) in principle lead to 
a Schwarzschild--Anti-De Sitter solution, but $f\rightarrow \infty$ and $f_R\rightarrow -\infty$, $f_{RR}\rightarrow \infty$, $f_{RRR}\rightarrow -\infty$ there.}
\label{fig:MirandaU}
\end{figure}

The solution $R=0$ corresponds to $\chi=1-\alpha_1$ in this model. However, for $\alpha_1 \geq 1$ one is led to $\chi \leq 0$, which 
by construction is not allowed in the EF frame. Therefore this solution cannot be present in that frame for those values of $\alpha_1$. 
In fact, for such values of $\alpha_1$ the potential 
$\mathscr{U}(\phi)$ has a minimum at some $\phi_{\rm 1,min}$, but $\mathscr{U}(\phi_{\rm 1,min}) > 0$ (see Figure~\ref{fig:MirandaU})~\footnote{The minimum at 
$\phi_{\rm 1,min}$ corresponds to one of the roots $R_1\neq 0$ of $\mathscr{V}_R(R)$, which leads to a Schwarzschild--De Sitter type of solution.}. According to our protocol, 
a strictly positive potential cannot allow for the Schwarzschild solution. Therefore, such solution cannot be recovered from the EF approach. As a consequence, 
the NHT's also rule out the existence of AFSSS hairy BH solutions. In the original variables the trivial solution can be recovered since $R=0$ is a 
root of $\mathscr{V}_R(R)$, but then the effective gravitational constant $G_{\rm eff}$ becomes negative, as the condition $f_R > 0$ 
fails at $R=0$.

Now, for $\alpha_1=1$, 
$\mathscr{U}(\phi)\rightarrow 0$ if $\phi\rightarrow -\infty$, which is the minimum, a situation similar to the Model 1.  
Finally, for $0 < \alpha_1 <1$, the potential satisfies $\mathscr{U}(\phi) \geq 0$ (see the middle panel of Figure~\ref{fig:MirandaU}).
In this case the NHT's also apply, and the Schwarzschild solution in the 
EF is associated with a local minimum $\phi= \phi_{\rm 2,min}$ at which $\mathscr{U}(\phi_{\rm 2,min})=0$. The trivial solution  $\phi(r)= \phi_{\rm 2,min}$ 
is associated with the trivial solution $R=0$. 

The overall conclusion for this model is that hair is in general forbidden, i.e., non-trivial AFSSSBH solutions $R(r)$ attempting to interpolate between $R_h$ 
and $R=0$ cannot exist if $f_R>0$. More specifically: 
1) For $\alpha_1 \geq 1$ the EF representation precludes the presence of 
hair in AFSSSBH according to the NHT's since the solution $R=0$ does not even exist as the potential 
$\mathscr{U}(\phi)$ never vanishes (the potential is only defined for $f_R>0$). If in the original JF variables one permits 
the possibility of having negative values for $f_R$ (which in turn implies negative values for $G_{\rm eff}$), then the trivial solution $R=0$ may exist. 
But then when looking for a hairy solution that interpolates from $R_h$ to $R=0$, 
the solution may cross the value $f_R=0$ if $f_R|_{R_h}>0$, a value which is associated with the {\it singularity} 
$G_{\rm eff}=G_0/f_R\rightarrow \infty$. Such hairy solution would be rather pathological 
if it exists at all. On the other hand, if we impose the condition $f_R<0$, in order to avoid that singularity, 
then $R$ turns to be restricted in the range $- R_{\ast}< R < R_{\ast} (\alpha_1-1)$. In particular one would require $\alpha_1>1$ to allow for 
the solution $R=0$ to exist. This range for $R$ restricts severely the region in which one can look for 
an optimal shooting $R_h$ (see Section~\ref{sec:numerical}), unless $1 \ll \alpha_1$. All in all, the numerical exploration shows that hairy solutions seem to be 
absent in all these scenarios; 2) For $0 < \alpha_1 <1$ the  NHT's apply straightforwardly, and therefore hairy solutions cannot exist either. Unlike the 
previous subclass ($\alpha_1 \geq 1$), the solution $R(r)=0$ corresponding to $\phi(r)= \phi_{\rm 2,min}$ 
[$\mathscr{U}(\phi_{\rm 2,min})=0$] can also be recovered in the EF since for this range of values of $\alpha_1$ the condition $f_R(0)>0$ is fulfilled.
\bigskip

We turn now our attention to three $f(R)$ models that have been analyzed recently in cosmology and which are some of the most successful ones 
as concerns the cosmological and the Solar-System tests. However, as we show below, for two of these models (i.e. Models 5 and 6) 
the potential $\mathscr{U}(\phi)$ is not even well defined as it turns to be multivalued. Therefore, the conclusion about the absence or existence of hair 
in these two models has to be obtained numerically using the original formulation (Sections~\ref{sec:theory} and \ref{sec:SSS}) as opposed to the EFSTT approach.
\bigskip

\begin{figure}[t]
\includegraphics[width=5.9cm,height=5cm]{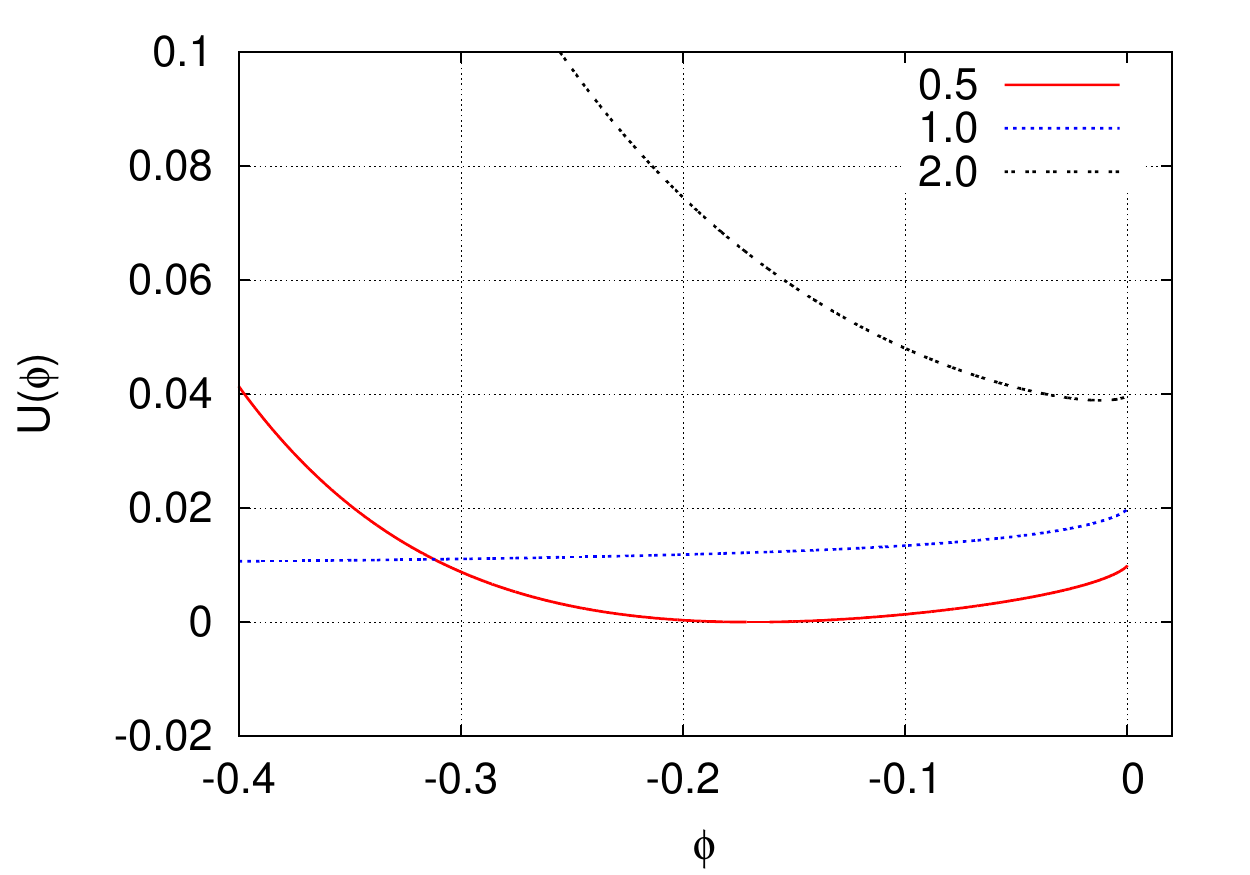}
\includegraphics[width=5.9cm,height=5cm]{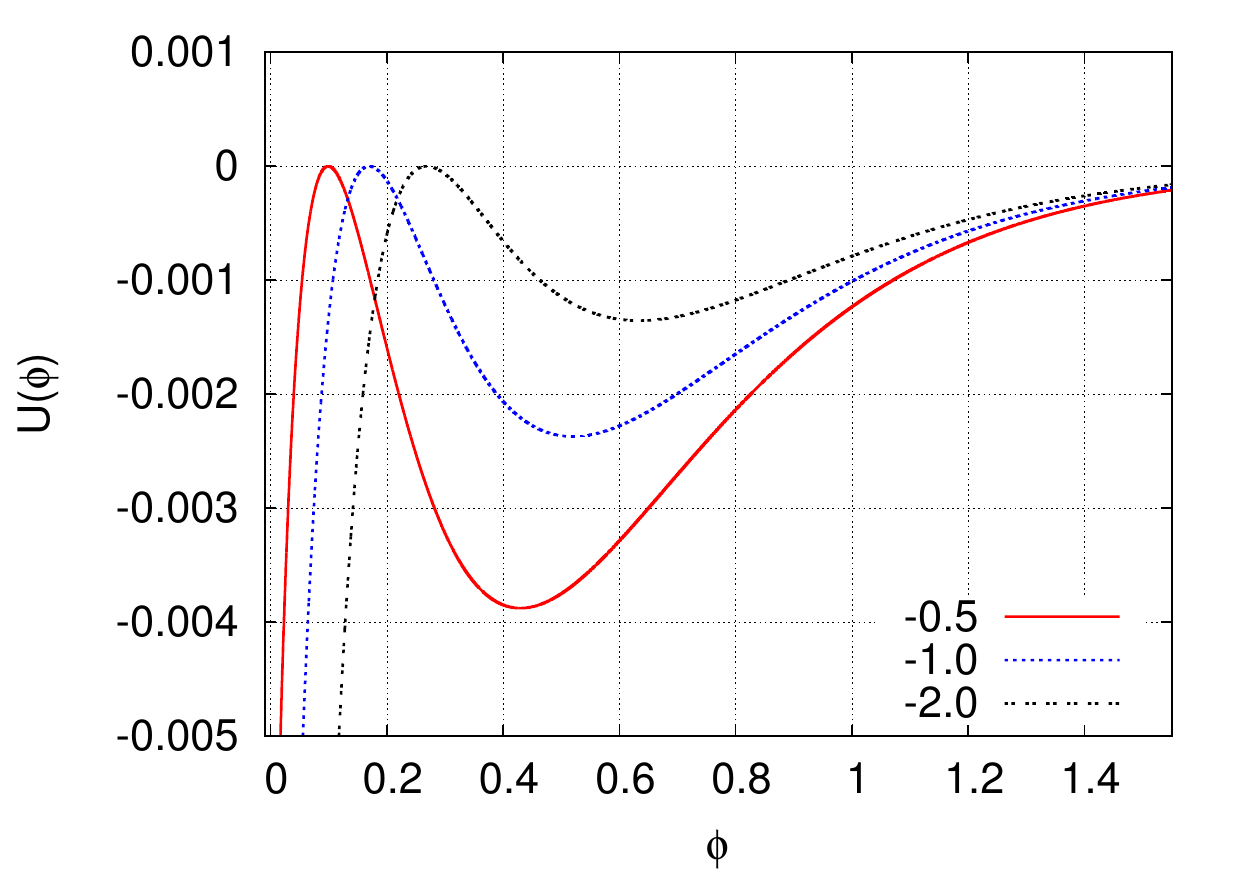}
\includegraphics[width=5.9cm,height=5cm]{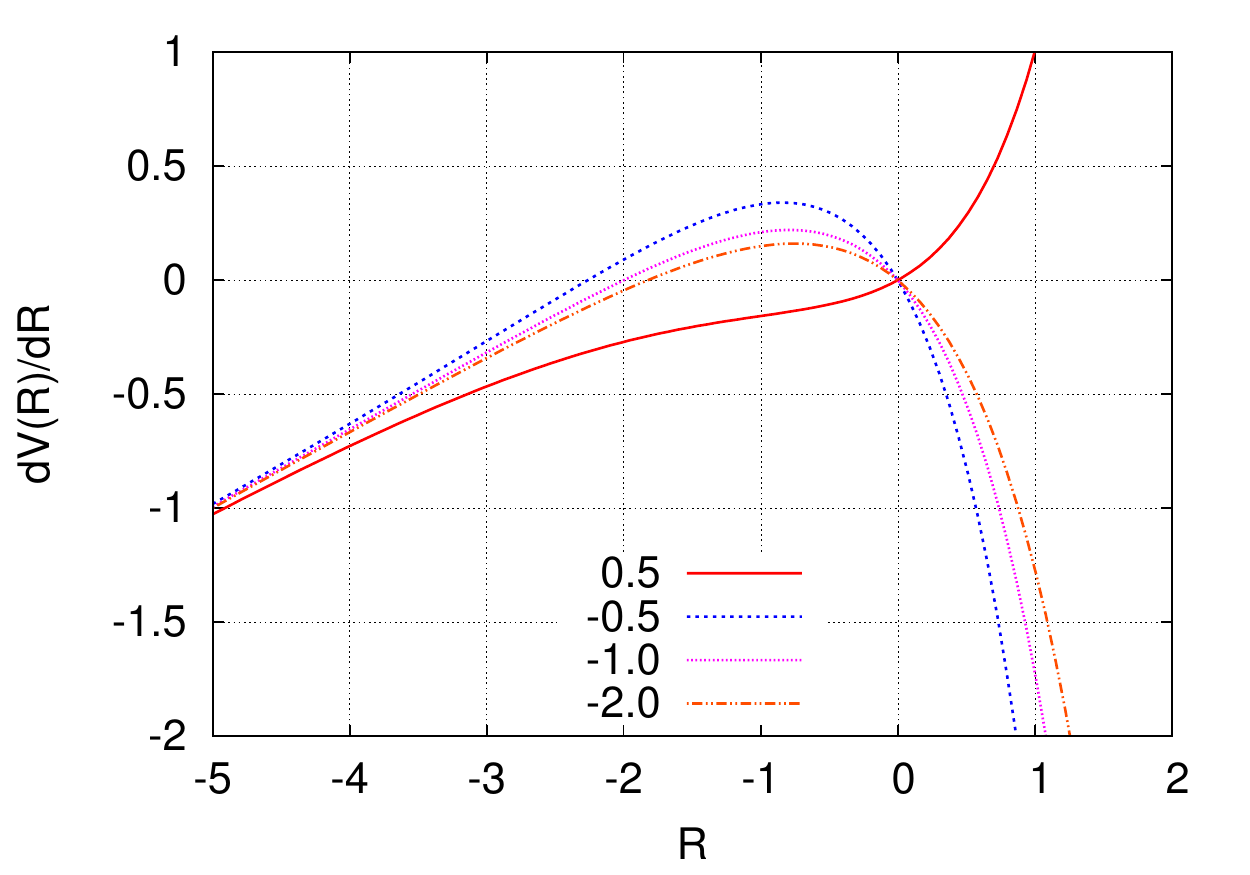}
\caption{(color online). Left-panel: Potential $\mathscr{U}(\phi)$ (in units of $R_e/G_0$; $\phi$ is given in units of 
$G_0^{-1/2}$) associated with the Model 4 depicted for three values of $\lambda_e>0$. 
The potential satisfies $\mathscr{U}(\phi) \geq 0$ for $0 < \lambda_e$ and therefore the NHT's
apply. In particular for $1< \lambda_e$ the potential is strictly positive and has a 
global minimum which leads to a De Sitter type of solution $R= R_1= const$. 
For $0<\lambda_e <1$ the potential can vanish at the global minimum 
which is associated with a solution $R= 0$. The panel displays one such example for $\lambda_e=0.5$ (cf. right panel). 
For $\lambda_e=1$ the potential 
vanishes when $\phi\rightarrow -\infty$. Middle-panel: Potential $\mathscr{U}(\phi)$ 
depicted for three values of $\lambda_e<0$. We appreciate that 
$\mathscr{U}(\phi) \leq 0$ and thus the NHT's can be evaded. The potential vanishes at the global 
maximum, and therefore the solution $R= 0$ exists (see the right panel). In principle the potential suggests that a hairy BH solution 
$\phi(r)$ might exist interpolating between some $\phi_h$ at the horizon and $\phi_{\rm max}$ at infinity where $\mathscr{U}(\phi_{\rm max})=0$. 
However, the numerical analysis shows (cf. Figure~\ref{fig:fRexpnohair}) that such solutions are not genuinely asymptotically flat as the 
mass function at infinity is not well defined (i.e. it never converges to a fixed value). The local minimum  $\mathscr{U}(\phi_{\rm min})<0$ 
that appears for each $\lambda_e<0 $ is associated with the Anti-De Sitter solutions (see right panel). Right panel: 
the function $d\mathscr{V}(R)/dR$ for four values of $\lambda_e$ shows the zeros at which the trivial solutions 
$R=R_1=const$ exist, including the solution $R= 0$ (corresponding to the 
places where $\mathscr{U}(\phi)$ vanishes as depicted in the left and middle panels) and the Anti-De Sitter solutions (one zero $R_1<0$ for each $\lambda_e<0$ 
is appreciated) corresponding to $\phi_{\rm min}$ where the local minimum 
$\mathscr{U}(\phi_{\rm min})<0$ as shown in the middle panel.}
\label{fig:expU}
\end{figure}

{\bf Model 4:} $f(R) = R - R_{e} \lambda_e\big( 1 - e^{-\frac{R}{R_e}}\big)$ where 
$\lambda_e$ is a dimensionless constant that usually is taken to be positive for a successful phenomenology and $R_e$ is a positive parameter that fixes the scale. 
For instance, $\lambda_e>0$ ensures $f_{RR}>0$ and $f_{RR}$ never vanishes at a finite $R$ regardless of the sign of $\lambda_e$. 
This exponential model has been analyzed in the past by several authors~\cite{Jaime2012e,exponential}. For this model the scalar field $\chi$ is
\begin{equation}
\label{chiexp}
\chi=f_R=  1 -  \lambda_e e^{-\frac{R}{R_e}}  \,\,\,.
\end{equation}
The condition $f_R>0$ holds provided $R > R_e {\rm ln}\lambda_e$, in which case $0<\chi <1$, where $\chi\rightarrow 1$ as $R\rightarrow \infty$. 
In particular, for $0<\lambda_e \leq 1$ the condition $f_R>0$ is satisfied if $R> -R_e |{\rm ln}\lambda_e|$. 
The potential $\mathscr{V}(R)$ has only a global minimum at $R=0$, i.e., $\mathscr{V}_R(0)=0$~\cite{Jaime2012e}, and thus, it leads 
to the Schwarzschild solution (cf. the right panel of Figure~\ref{fig:expU}). 
For $\lambda_e>1$ the potential $\mathscr{V}(R)$ has a local maximum at $R=0$, and the 
potential develops in addition a local minimum at some $R<0$ and a global minimum at some $R>0$~\cite{Jaime2012e}. These extrema correspond to trivial solutions $R=R_1$ associated 
with the Schwarzschild--Anti-De Sitter and Schwarzschild--De Sitter solutions respectively. 
Notice, however, that $f_R(0)<0$ since the condition $R > R_e {\rm ln}\lambda_e$ fails at 
$R=0$. Therefore, in this case the 
Schwarzschild solution has $G_{\rm eff} <0$. Since the mapping $\chi \rightarrow \phi$ is defined only for $\chi>0$, we conclude that 
for $\lambda_e>1$ the condition $R > R_e {\rm ln}\lambda_e$ implies $R>0$ for the EFSTT approach to be well defined and thus, like in the previous model, 
one cannot recover the solution $R=0$. As a consequence, we require the original formulation of the theory to analyze if hairy solution 
can exist for these values of $\lambda_e$. But, again, a hairy solution, if exists, can encounter the singularity at $f_R=0$ before approaching the asymptotic 
value $f_R(0)<0$, notably if $f_R|_{R_h}>0$.

The potentials are 
\begin{eqnarray}
V(\chi) &= & \frac{R_e}{2\kappa \chi^{2} }\Big[ (\chi
-1 )\ln(\frac{\lambda_e}{1-\chi}) + \lambda_e + \chi - 1 \Big]\,\,\,,\\
\mathscr{U}(\phi) &=&  = \frac{ R_{e}}{2\kappa}e^{-2\sqrt{\frac{2\kappa}{3}}\phi}\bigg[ \Big( e^{
\sqrt{\frac{2\kappa}{3}}\phi} -1 \Big)\ln\Big( \frac{\lambda_e}{1-e^{
\sqrt{\frac{2\kappa}{3}}\phi}} \Big) + \lambda_e + e^{\sqrt{\frac{2\kappa}{3}}\phi} - 1 \bigg] \,\,\,.
\end{eqnarray}
which are valid in the domain $0<\chi <1$ and $-\infty <\phi <0$, respectively. The potential $\mathscr{U}(\phi) \geq 0$ is depicted in Figure~\ref{fig:expU} 
(left panel) for various values of $\lambda_e >0$. For this model the NHT's apply. As emphasized above, for $\lambda_e \geq 1$ the 
potential $\mathscr{U}(\phi)$ is strictly positive, in particular its minimum, and therefore the corresponding solution $R=0$ cannot be recovered from the 
EFSTT approach, but only the De Sitter type of solution $R=R_1=const$ which is associated with the minimum of $\mathscr{U}(\phi)$.
On the other hand, for  $0<\lambda_e < 1$ the potential vanishes at its minimum 
which leads to the solution $R=0$ that allows one to recover the Schwarzschild solution.

If for a moment we dismiss the condition $f_{RR}>0$ and consider $\lambda_e<0$ 
we can evade the NHT's because now $\mathscr{U}(\phi)$, 
defined for $0<\phi <+\infty$,
is never positive (see the middle panel of Figure~\ref{fig:expU}). The new domain is 
a consequence of Eq.~(\ref{chiexp}) which yields 
$\chi= 1 + |\lambda_e| e^{-\frac{R}{R_e}}$. Thus for $-\infty<R < +\infty$ 
the scalar field $\chi$ is defined in the domain $1 < \chi < +\infty$, which in turns 
leads to  $0<\phi <+\infty$. The potentials depicted in Figure~\ref{fig:expU} (middle panel) 
suggest that hairy solutions might exist. For instance, in the JF variables, 
such solution $R(r)$ would interpolate between the horizon $R_h$ and its 
value $R=0$ at spatial infinity. It turns out that indeed such solution can be found numerically 
(see the next section) but the spacetime is not authentically asymptotically flat as discussed above within the framework of Model 2. The 
the mass $M(r_\infty)$ never converges to a well defined value as the mass function behaves asymptotically $M(r)\sim r^\sigma g(r)$, 
where  $0<\sigma \leq 1$ and $g(r)$ is an oscillating but presumably a bounded function (cf. Figure~\ref{fig:fRexpnohair}). 
\bigskip
\bigskip

\begin{figure}[t]
\includegraphics[width=5.9cm,height=5cm]{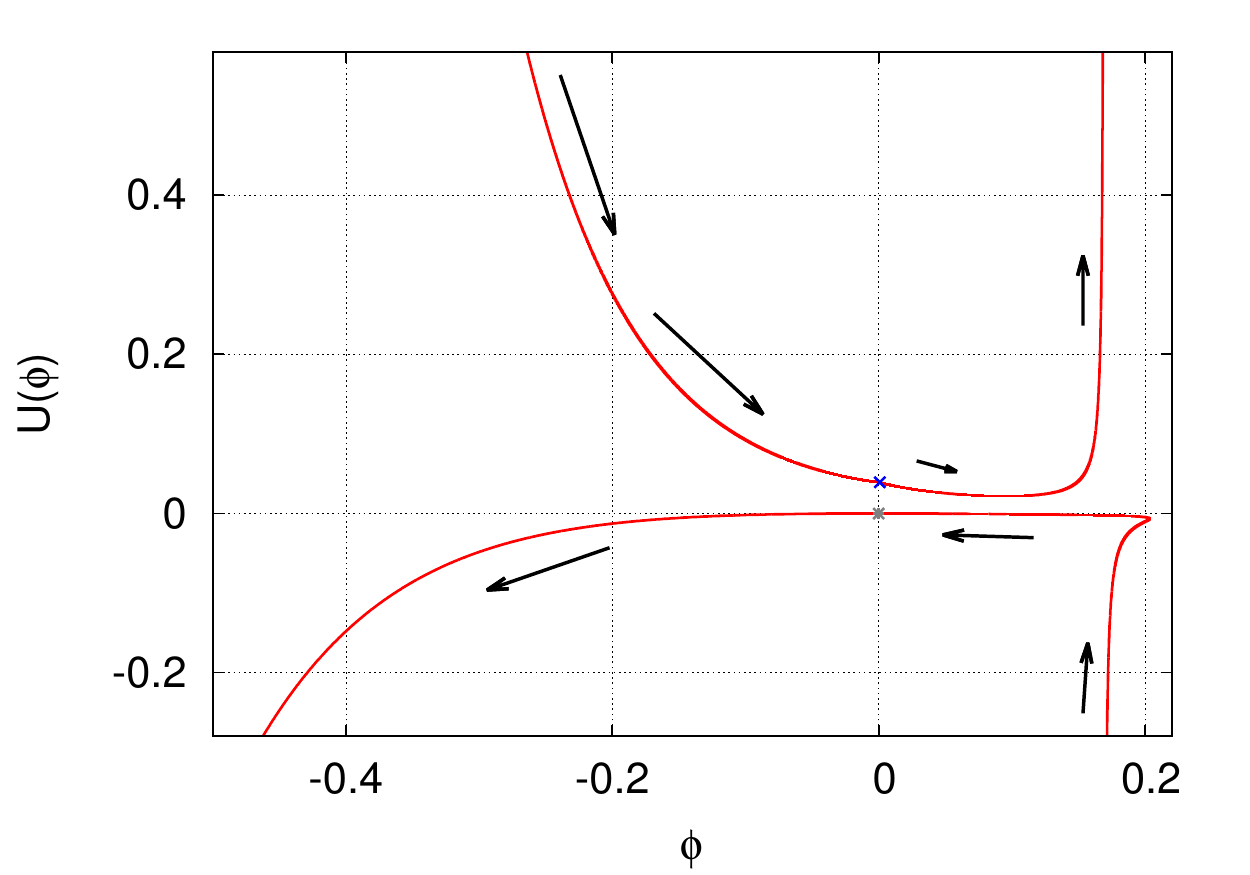}
\includegraphics[width=5.9cm,height=5cm]{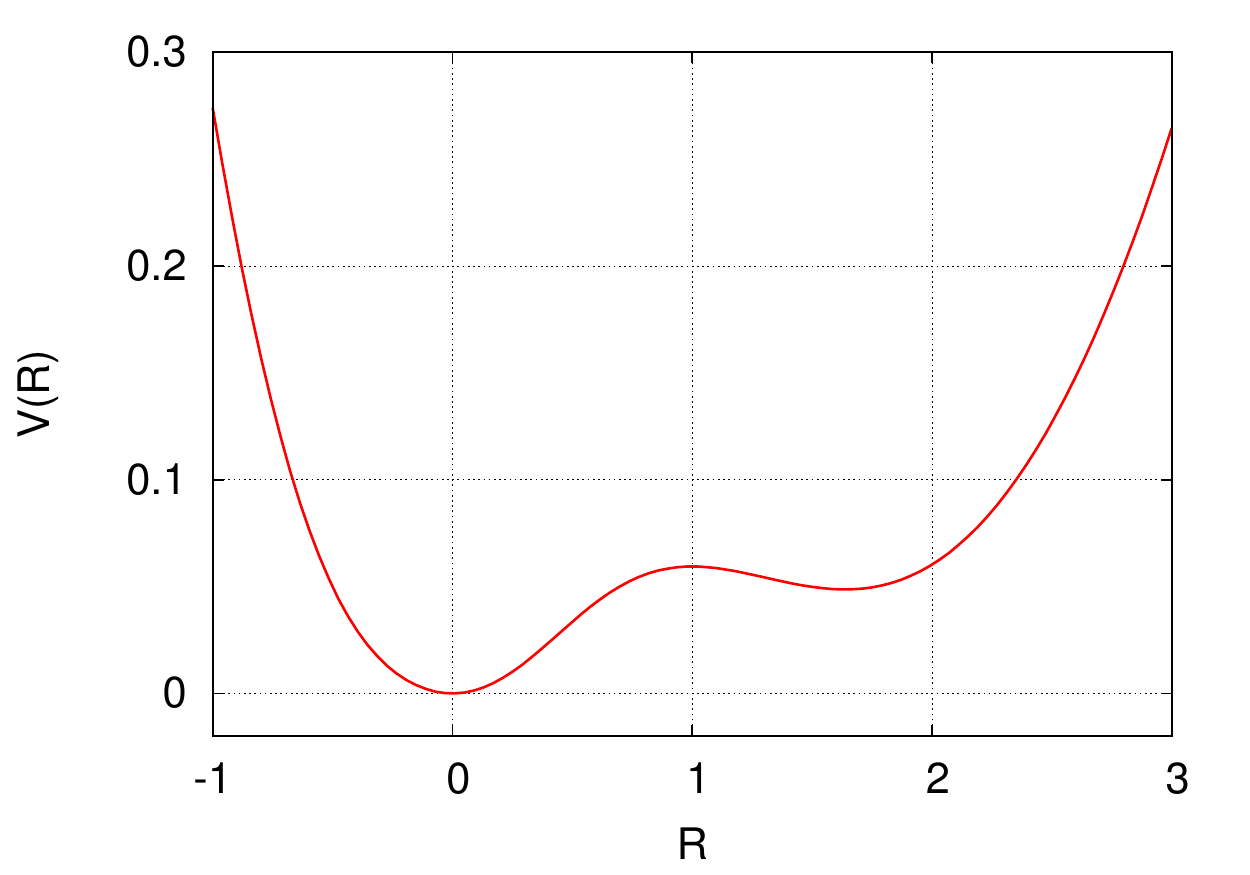}
\includegraphics[width=5.9cm,height=5cm]{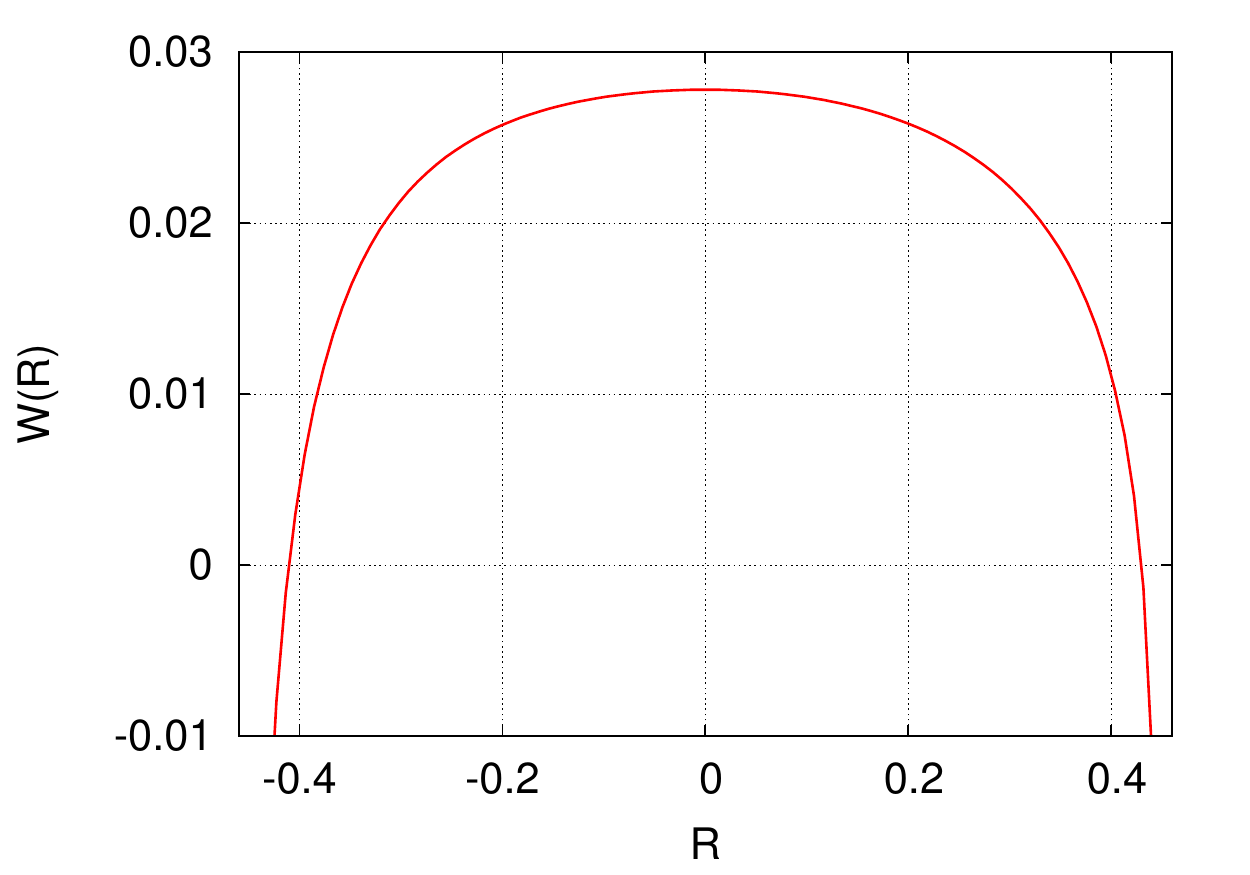}
\caption{(color online). Left panel: Potential $\mathscr{U}(\phi)$  (in units of $R_S/G_0$; $\phi$ is given in units of $G_0^{-1/2}$) 
 associated with the Starobinsky Model 5 with $\lambda_{S}=1$ and $q=2$. The potential $\mathscr{U}(\phi)$ is multivalued and has negative branches, therefore, 
the NHT's cannot be applied. The arrows indicate the trajectory of the parametric plot for increasing values of $R$ ($R$ is in units of $R_{S}$). 
One of the marks (in gray) indicate the value $R=0$ corresponding to $\phi=0$ at which $\mathscr{U}=0$. The second mark (in blue) indicates 
the starting point of the parametric plot at $R= -10$. 
Incidentally, for $R\rightarrow \infty$  the field $\phi\rightarrow 0$ [cf. Eq.~(\ref{f_RStaro}] and 
the potential returns to its starting point. Middle panel: Potential $\mathscr{V}(R)$ showing explicitly the extrema (maximum or minimum)
where the trivial solutions $R(r)=const$ exist. The global minimum at $R=0$ leads to the Schwarzschild solution with $R(r)\equiv 0$. 
The other extrema are associated with the Schwarzschild--De Sitter solutions $R(r)=const$. For this particular model any non-trivial solution $R(r)$ 
interpolating between $R_h$ (the value at the BH horizon) and $R=0$ (the asymptotic value) is confined within the 
range  $-1/\sqrt{5}< R < 1/\sqrt{5}$ which corresponds to $|R|\lesssim 0.447$, i.e., values of $R$ close to the global minimum at $R=0$. 
Outside this range, the {\it weak singularities} can be reached by $R$ (see the right panel). 
Right panel: the function $\mathscr{W}(R)$ is depicted showing the places where $\mathscr{W}_R(R)= d\mathscr{W}/dR$ diverges. These places 
called {\it weak singularities} are located at $R_\pm =\pm 1/\sqrt{5}$, where $f_{RR}=0$ (these values are denoted generically by $R_2$ in the main text). 
At such values Eq.~(\ref{TraceRsss}) blows up.}
\label{fig:StaroU}
\end{figure}

{\bf Model 5:} $f(R)= R+\lambda_S R_{S}\left[ \left( 1+\frac{R^2}{R^2_{S}}\right)^{-q}-1\right]$ where $\lambda_S$ is a dimensionless constant, $q$ 
a dimensionless parameter and $R_{S}$ is a positive parameter that fixes the scale.
This model was proposed by Starobinsky~\cite{Starobinsky2007} as a mechanism for generating the late accelerating expansion while satisfying several 
{\it local} observational tests. We analyzed this and the Model 3 in the past in the cosmological setting~\cite{Jaime2012} 
and for constructing star-like objects~\cite{Jaime2011} using the approach of Section~\ref{sec:theory}. In this paper 
we take $\lambda_s=1$ and explore several values of $q$ (see Section~\ref{sec:numerical}). 
For this model the conditions $f_R>0$ (for a positive $G_{\rm eff}$) and $f_{RR}>0$ do not hold in general. 
In fact, $f_{RR}$ vanishes at $R=R_2^\pm=\pm R_{S}/\sqrt{2q+1}$. Since $f_{RR}$ appears in the 
denominator of Eq.~(\ref{TraceRsss}), the vanishing of $f_{RR}$ was termed by Starobinsky a {\it weak singularity}. 
One can appreciate these features from Figure~\ref{fig:StaroU} (right panel) where the ``potential'' 
$\mathscr{W}(R)$ is depicted. We see that $|\mathscr{W}_R(R)|=\infty$ at $R_2^\pm$ where $f_{RR}$ vanishes. 
Thus, the {\it weak singularities} at $R_2^\pm$ cannot be ``cured'' by the term $2f-Rf_R$ in $\mathscr{V}_R(R)$ because such term does not vanish there, and 
which otherwise could have lead to a finite $\mathscr{W}_R$. Therefore any solution $R(r)$ intending to interpolate between $R_h$ and $R=0$ such that $R_h>R_2^+$ or $R_h<R_2^-$  will irremediably 
encounter the {\it weak singularities} at $R_2^\pm$ where we expect a singular behavior in Eq.~(\ref{TraceRsss}). As a consequence, our search for 
a numerical BH solutions with non-trivial $R$ 
was limited mostly in the range $|R_h|<R_2^+$ (see Section~\ref{sec:numerical})\footnote{In the cosmological scenario one usually aims at a De Sitter ``point'' $R_1\neq 0$ 
(as opposed to the Minkowski ``point'' $R_1=0$) in order to recover an effective cosmological constant asymptotically (in {\it time}), 
and thus, to mimic the dark energy. In that scenario the actual numerical solution $R(t)$ is always positive and larger than $R_2^+$, 
thus, the solution never crosses the {\it weak} singularity~\cite{Jaime2012}. Something similar takes place for the Model 6.}.

For this model the potential $\mathscr{V}(R)$ has several extrema (see the middle panel of Figure~\ref{fig:StaroU}), in particular, a global minimum 
at $R=0$ with $f_R(0)=1$, $f_{RR}(0)= -2\lambda_S q/R_{S}$, which allows one to recover the Schwarzschild solution. Notice that the global minimum 
at $R=0$ corresponds to the global maximum of $\mathscr{W}(R)$. This is because  $d^2\mathscr{W}/dR^2|_{R=0}= f_{RR}^{-1} d^2\mathscr{V}/dR^2|_{R=0}$ and 
$d^2\mathscr{V}/dR^2|_{R=0}$ is positive, whereas $f_{RR}^{-1}(0)<0$, and so $d^2\mathscr{W}/dR^2|_{R=0}$ is negative. On the other hand 
$\mathscr{V}_R(R)$ at $R_2^\pm$ is well behaved there. The other extrema, a local maximum and minimum, lead to two 
Schwarzschild--De Sitter solutions with positive $R=const.$

Now, the inversion $R=R(\chi)$ required to recover the potential $V(\chi)$ and then the potential $\mathscr{U}(\phi)$ demands 
$f_{RR}>0$ or $f_{RR}<0$. That is, the inversion is possible when 
$\chi=f_R$ is a monotonic function of $R$, which is not the case for this model. In principle one could perform the inversion picewise in very specific domains of the model 
but not in all the domain where the model is defined. In view of this drawback the potential $\mathscr{U}(\phi)$ is not well defined. In fact it is 
multivalued as we are about to see. Its expression cannot be given in closed form but only in parametric representation through the equations
\begin{eqnarray}
\label{f_RStaro}
\chi(R)&=& f_R =  1 - \frac{2\lambda_s q (R/R_{S})}{\Big[1+ (R/R_{S})^2\Big]^{1+q}} \,\,\,,\\
\phi (R) &=& \sqrt{\frac{3}{2\kappa}} {\rm  ln}\chi (R) \,\,\,,\\
\mathscr{U}(\phi(R)) &:=& V(\chi[\phi(R)]) \,\,\,.
\end{eqnarray}
The form of the potential $\mathscr{U}(\phi)$ is shown in Figure~\ref{fig:StaroU} (left panel). Given that $\mathscr{U}(\phi)$ is not single valued 
it is a priori unclear how to establish a method to solve the differential 
equations in the EFSTT approach and decide unambiguously which value of $\mathscr{U}(\phi)$ to assign for a given $\phi$. 
Hence we conclude that one cannot obtain any rigorous result from this frame using this potential, let alone 
trying to implement the NHT's. But even if we tried to do so, the lower branch of the potential does not satisfy the condition 
$\mathscr{U}(\phi) \geq 0$ required by the theorem to prevent the existence of hair. 
In view of this, any strong conclusion about the existence or absence of hair must 
be obtained from the original formulation of the theory that was presented in Section~\ref{sec:SSS}. 
Furthermore, due to the complexity of the model itself and of the differential equations, a numerical analysis is in order. In the next section 
we provide the numerical results that show evidence about the absence of hairy AFSSS black holes in this model. 
\bigskip

\begin{figure}[t]
\includegraphics[width=5.9cm,height=5cm]{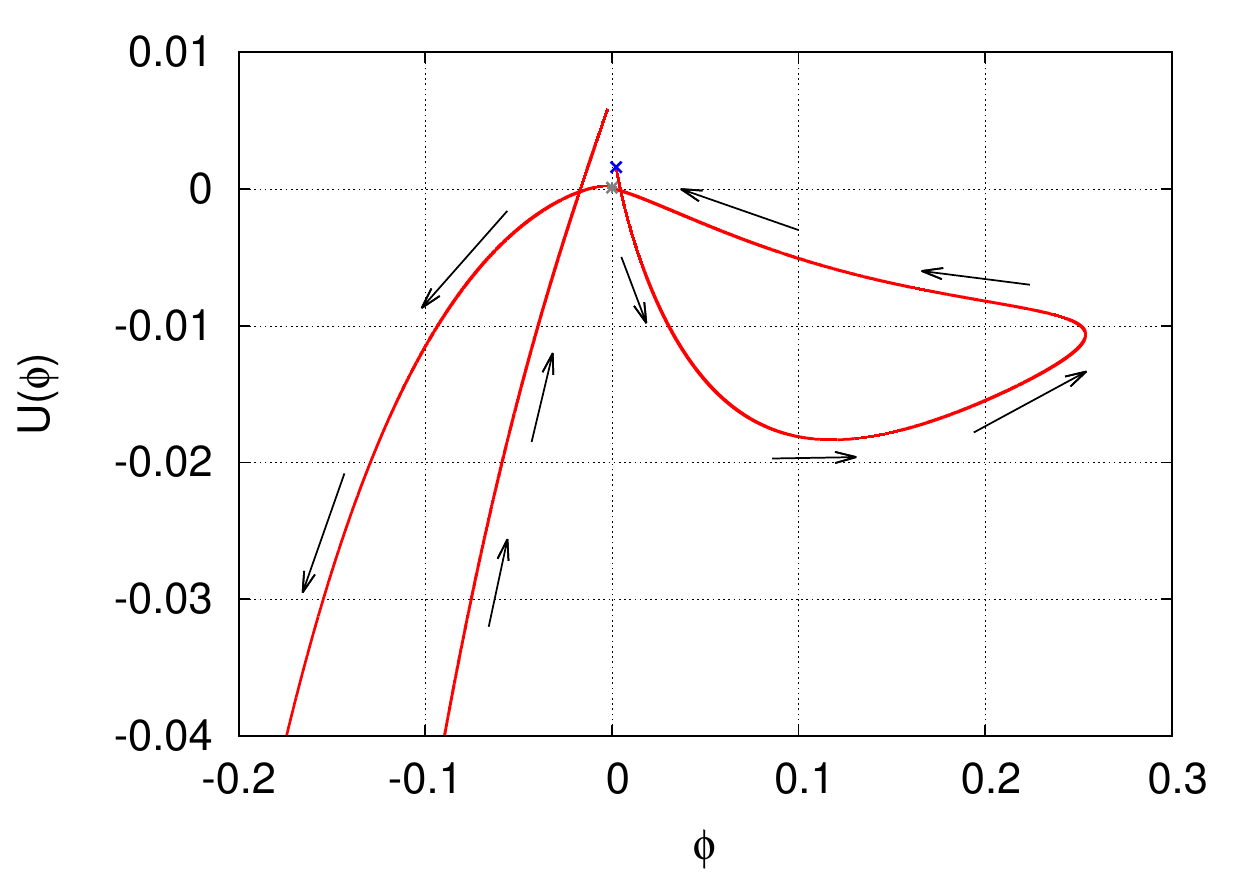}
\includegraphics[width=5.9cm,height=5cm]{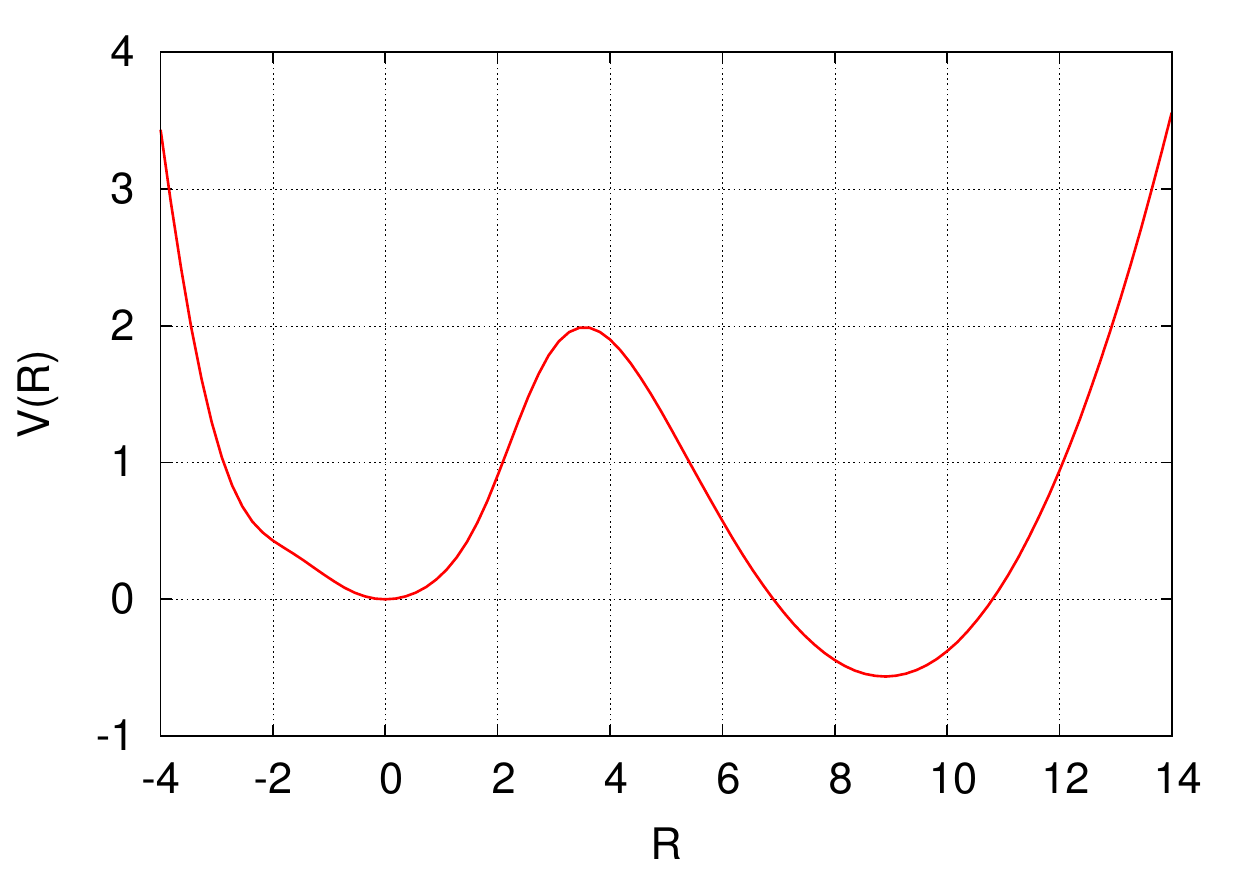}
\includegraphics[width=5.9cm,height=5cm]{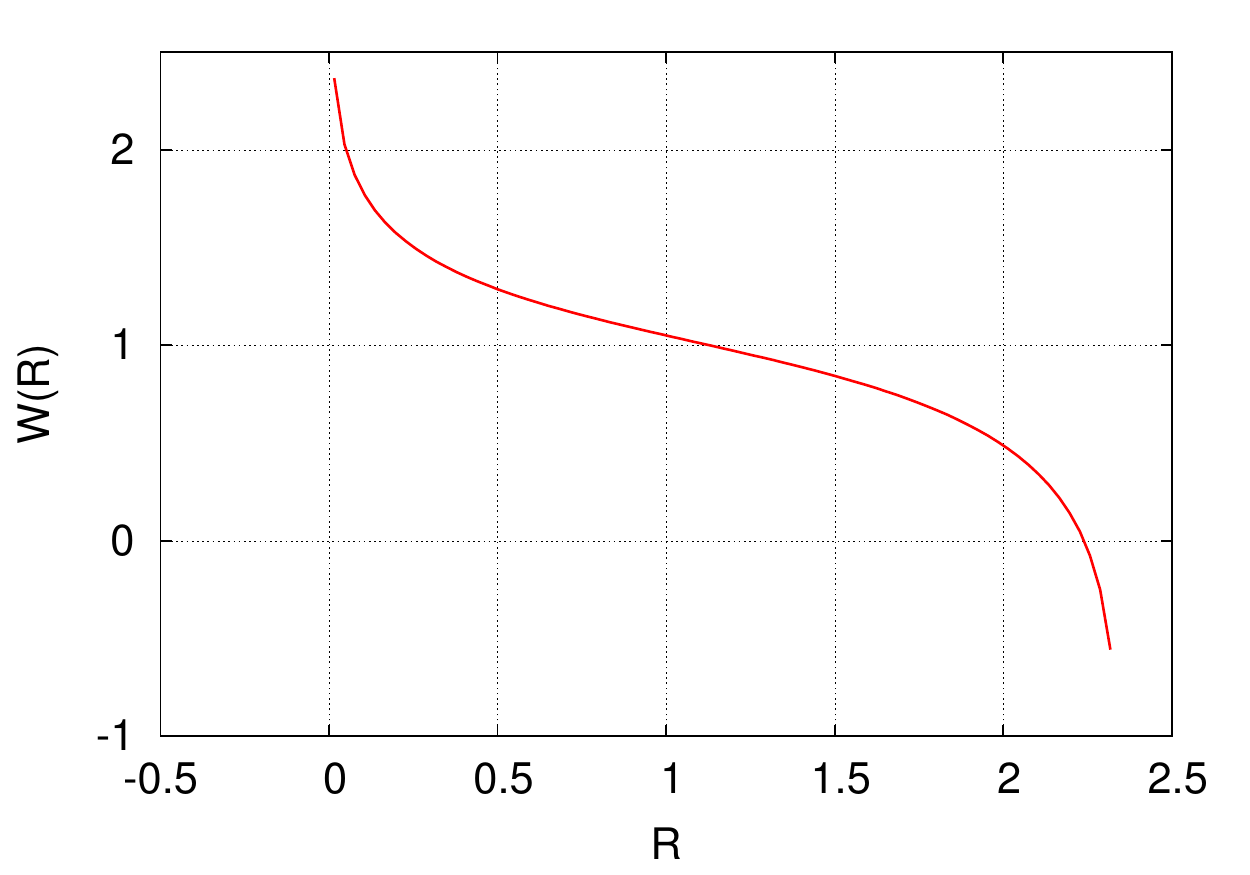}
\caption{(color online). Left panel: Potential $\mathscr{U}(\phi)$ (in units of $R_{\rm HS}/G_0$; $\phi$ is given in units of 
$G_0^{-1/2}$) associated with the  Hu--Sawicky Model 6 for $n=4$ with the values $c_1$ and $c_2$ as in the main text. The potential $\mathscr{U}(\phi)$ is multivalued and has negative branches and the NHT's cannot be applied. The arrows and marks have the same meaning as in Figure~\ref{fig:StaroU}. Middle panel: Potential $\mathscr{V}(R)$ showing 
explicitly the extrema where the trivial solutions $R(r)=const$ may exist. At 
the local minimum $R=0$ occurs a {\it weak} singularity where $f_{RR}=0$ 
(see the right panel). Right panel: Potential $\mathscr{W}(R)$. Two weak singularities where $\mathscr{W}_R(R)=\infty$ 
are located at $R=0$ and $R\approx 2.34$. In particular, the singularity at $R=0$ precludes the 
search for numerical AF hairy solutions with $R(r)\rightarrow 0$ as $r\rightarrow \infty$ since the ``singularity'' is encountered 
at finite $r$.}
\label{fig:HSU}
\end{figure}

{\bf Model 6:} $f(R)= R- R_{\rm HS}\frac{c_{1}\left(\frac{R}{R_{\rm HS}}\right)^n}{c_{2}\left(\frac{R}{R_{\rm HS}}\right)^n+1}$, 
where $c_1$ and $c_2$ are two dimensionless constants, and like in previous models, $R_{\rm HS}$ fixes the scale. 
This model was proposed by Hu and Sawicky~\cite{Hu2007}, and it is perhaps one of the most thoroughly studied $f(R)$ models 
(see Ref.~\cite{Jaime2012} for a review). 
In the cosmological context, $c_1$ and $c_2$ were fixed as to obtain adequate cosmological observables, like the actual dark and matter 
content in the universe. For instance taking $n=4$, their values are 
$c_1\approx 1.25 \times 10^{-3} $, and $c_2\approx 6.56 \times 10^{-5}$~\cite{Jaime2012}. 
Notice that the Model 5 with $q=1$ and this model with $n=2$ are essentially the same. Like in the previous model, the conditions $f_R>0$ and 
$f_{RR}>0$ are not met in general, therefore, the potential 
$\mathscr{U}(\phi)$ is multivalued and has negative branches as well. 
It can be plotted using a parametric representation as in the Model 5:
\begin{eqnarray}
\chi(R)&=& f_R =  1 - \frac{n c_1 (R/R_{\rm HS})^{n-1}}{\Big[1+ c_2 (R/R_{\rm HS})^n\Big]^{2}} \,\,\,,\\
\phi (R) &=& \sqrt{\frac{3}{2\kappa}} {\rm  ln}\chi (R) \,\,\,,\\
\mathscr{U}(\phi(R)) &:=& V(\chi[\phi(R)]) \,\,\,.
\end{eqnarray}
Figure~\ref{fig:HSU} depicts the potential $\mathscr{U}(\phi)$ (left panel) where 
one can appreciate the pathological features. In fact, in this model 
a {\it weak} singularity $f_{RR}=0$ is located precisely at $R=0$, i.e., the value that $R(r)$ 
should reach asymptotically in the AF scenario, and it is also the value corresponding to the (hairless) Schwarzschild solution that we should 
be able to recover. Nevertheless, and unlike Model 5, this singularity in Eq.~(\ref{traceR}) or in 
Eq.~(\ref{TraceRsss}) disappears for some values of $n$ because $\mathscr{W}_R$ is finite or vanishes at $R=0$ in this model. 
Namely, $\mathscr{W}_R$ vanishes at $R=0$ for $0 < n \leq 2$. Thus, in this range of $n$, the model admits the 
trivial solution $R=0$. We do not consider the case $n=0$ as this model reduces to $f(R)= R + const$, which 
amounts to GR plus a cosmological constant. 
For $n=3$ we find $\mathscr{W}_R(0)= - R_{\rm HS}^2/(18c_1)$, 
which does not even vanish. Therefore, this  means that $R=0$ does not solve Eq.~(\ref{TraceRsss}), not trivially nor asymptotically. 
For $n>2$ and $n\neq 3$, there is indeed a {\it weak} singularity at $R=0$ where $|\mathscr{W}_R|= \infty$ (cf. the right panel of Figure~\ref{fig:HSU}). 
For $n<1$ the quantity $f_R(0)$ blows up, thus we consider only $n \geq 1$, notably, for the numerical analysis of 
Section~\ref{sec:numerical}.

In this model the potential $\mathscr{V}(R)$ has an minimum at $R=0$ for any $n>0$, which allows for the Schwarzschild solution 
whenever $f_{RR}\neq 0$ at $R=0$ (i.e. for $|n|\leq 2$). However, when  $f_{RR}= 0$ at $R=0$ and for which $|\mathscr{W}_R(0)|= \infty$ 
the Schwarzschild solution may not even exist. In those situations, non-trivial solutions where $R(r)$ vanishes asymptotically 
will encounter such singularity (see Section~\ref{sec:numerical}).

Like in the Model 5, any analysis using the ill-defined potential $\mathscr{U}(\phi)$ for the Hu--Sawicky model is not robust. We then turn to a numerical analysis using the original formulation of the theory. This is presented in the next section.
 
\bigskip

\begin{table*}
\centering
\begin{tabular}{|c| c| c| c|}
\multicolumn{1}{c}{} & \multicolumn{1}{c}{} & \multicolumn{1}{c}{} & \multicolumn{1}{c}{} \\  
\hline 
\pbox{6cm}{Model} & Potential in EF & Properties &  \multicolumn{1}{l|}{No-hair} \\
& & & \multicolumn{1}{l|}{theorems} \\ 
\hline
& & & \\
$f(R) = \lambda_n \Big(\frac{R}{R_n}\Big)^{n}$ & $\mathscr{U}(\phi) = \frac{(n-1)}{2 \kappa n} \Big( \frac{R_n}{n\lambda_n} \Big
)^{ \frac{1}{n-1} } e^{\big(\frac{2-n}{n-1}\big) \sqrt{\frac{2\kappa}{3}}\phi}$ & $\mathscr{U}(\phi)> 0 $ \,,\,$R_n>0$\,,\,$\lambda_n>0$\,,\,$n>0$ & \checkmark\\
\hline
& & & \\
$f(R)=R + c_2 \left(\frac{R}{R_I}\right)^{2} $ & \multicolumn{1}{l|}{
 $\mathscr{U}(\phi) = \frac{R_{I}}{8 c_2\kappa}\Big( 1 - e^{ -\sqrt{\frac{2\kappa}{3}}\phi} \Big)^{2}$} & 
$\mathscr{U}(\phi)\geq 0 $ \,,\,$c_2>0$ & \checkmark \\
& \multicolumn{1}{l|}{}& $\mathscr{U}(\phi)\leq 0 $ \,,\,$c_2<0$ & \xmark \\
\hline
& & & \\
$f(R)= R - \alpha_1 R_{\ast}\ln\Big( 1 + \frac{R}{R_{\ast}}\Big)$ & \multicolumn{1}{l|}{
$\mathscr{U}(\phi) = \frac{R_{\ast}}{2\kappa}e^{
-2\sqrt{\frac{2\kappa}{3}}\phi}\bigg[  \alpha_1\ln\bigg( \frac{\alpha_1}{ 1 -
e^{\sqrt{\frac{2\kappa}{3}}\phi}}\bigg)$} & $\mathscr{U}(\phi)> 0 $ \,,\,$1 \leq \alpha_1$\,,\, $R_*>0$  & \checkmark \\
& \multicolumn{1}{l|}{ 
\hspace{3cm}$- e^{\sqrt{\frac{2\kappa}{3}}\phi}
+ 1 - \alpha_1 \bigg]$} & $\mathscr{U}(\phi)\geq 0 $ \,,\, $0 < \alpha_1 <1$ \,,\, $R_*>0$  & \checkmark \\
\hline
& & & \\
$f(R)= R - R_{e} \lambda_e\big( 1 - e^{-\frac{R}{R_e}}\big) $ & \multicolumn{1}{l|}{
$\mathscr{U}(\phi) = \frac{ R_{e}}{2\kappa}e^{-2\sqrt{\frac{2\kappa}{3}}\phi}\bigg[ \Big( e^{
\sqrt{\frac{2\kappa}{3}}\phi} -1 \Big)$} & $\mathscr{U}(\phi)> 0$ \,,\,$ 1\leq \lambda_e $\,,\, $R_e>0$  & \checkmark  \\
& \multicolumn{1}{l|}{
$\times\ln\Big( \frac{\lambda_e}{1-e^{
\sqrt{\frac{2\kappa}{3}}\phi}} \Big) + \lambda_e + e^{\sqrt{\frac{2\kappa}{3}}\phi} - 1 \bigg] $}  & 
$\mathscr{U}(\phi)\geq 0$ \,,\,$ 0< \lambda_e< 1 $\,,\, $R_e>0$    & \checkmark  \\
& \multicolumn{1}{l|}{} & $\mathscr{U}(\phi)\leq 0$ \,,\,$\lambda_e<0 $\,,\, $R_e>0$  & \xmark  \\
\hline
& & & \\
$f(R)= R+\lambda_S R_{S}\left[ \left( 1+\frac{R^2}{R^2_{S}}\right)^{-q}-1\right]$  & (ill defined: multivalued) & -- &  -- \\
\hline
& & & \\
$f(R)= R- R_{\rm HS}\frac{c_{1}\left(\frac{R}{R_{\rm HS}}\right)^n}{c_{2}\left(\frac{R}{R_{\rm HS}}\right)^n+1}$  & (ill defined: multivalued) & -- &  -- \\
\hline 
\end{tabular}
\caption{$f(R)$ models and their corresponding scalar-field potentials in the 
Einstein frame. The last column indicates if the no-hair theorems apply (\checkmark) or not (\xmark). The potentials that are strictly positive 
definite $\mathscr{U}(\phi)> 0 $, do not even admit Schwarzschild BH's with $R(r)=0$. Models 5 and 6 lead to potentials $\mathscr{U}(\phi)$ that are 
generically multivalued and which have no definite sign. In those models the EFSTT approach is not well defined and the 
applicability of the no-hair theorems is in jeopardy, thus, we perform a numerical analysis in the original variables 
in order to find evidence about the existence or absence of geometric hair in AFSSS black holes.}
\label{tab:models}
\end{table*}


\subsection{Numerical analysis and the quest for hairy solutions}
\label{sec:numerical}

As we discussed in the previous section, in some circumstances it is possible to formulate the original $f(R)$ model  
as a STT in the EF where the scalar-field turns to be coupled minimally to the EF metric but it is subject to a potential $\mathscr{U}(\phi)$. If this 
potential verifies the condition $\mathscr{U}(\phi) \geq 0 $, then the NHT's apply and, at least in the region where $f_R>0$ and $f_{RR}>0$, 
we can assert that hair (where $\phi(r)$ or equivalently $R(r)$ are not trivial solutions) is absent, in which case, the only possible AF solutions are at best 
$\phi(r)= const$ and $R(r)=0$. This conclusion follows for the Models 1-4 in the sectors where their parameters allow for the $R=0$ solution and led to 
$\mathscr{U}(\phi) \geq 0 $. On the other hand, we mentioned that Models 2 and 4 can have potentials $\mathscr{U}(\phi)$ with negative 
branches if we allow for the parameters $c_2$ and $\lambda_e$ to be negative. Negative values of such parameters are not usually considered 
in cosmology, but for the sake of finding hairy solutions, we can contemplate them. Because the NHT's do not apply when $\mathscr{U}(\phi)$ is 
negative, notably at the horizon, the problem of hair reopens when this happens. At this regard, several strategies are available 
to solve it: 1) Show an explicitly exact AFSSS black hole solution with hair; 2) Prove 
analytically the absence of it (i.e. extend the NHT's); 3) Show numerical evidence about one or the other.

Given that the differential equations presented in Section~\ref{sec:SSS} are very involved, strategies 1 or 2 might lead to a 
dead end, thus we opted for option three. In particular, this strategy seems even the most adequate as concerns the Models 5 and 6, where the potential 
$\mathscr{U}(\phi)$ is not even well defined.

We proceed to solve numerically Eqs.~(\ref{TraceRsss})--(\ref{deltasss}) subject to the regularity conditions at the horizon provided in 
Sec.~\ref{sec:regcond} and in the Appendix~\ref{sec:regcond2}. The only free conditions are the value $r_h$ and $R_h$. The methodology is 
roughly as follows. One starts by fixing the size of the black hole $r_h$, 
and then looks for $R_h$ so that $R\rightarrow 0$ as $r\rightarrow +\infty$. This ``boundary-value'' problem is solved using a 
{\it shooting method}~\cite{Press1990} within a Runge--Kutta algorithm. 
We have implemented a similar methodology for constructing star-like objects in $f(R)$ gravity in the past~\cite{Jaime2011}. Numerical solutions with non-trivial hair with {\it asymmetric} 
(non positive definite) potentials have been found previously within the Einstein-$\phi$ system using similar techniques~\cite{Nucamendi2003}. As we will see in the next section, for certain $f(R)$ models is not even necessary to perform a shooting as the dynamics of $R$ naturally 
drives $R\rightarrow 0$ asymptotically for a given $R_h$.

Now, as we mentioned previously, for the AF solutions to exist, it is not sufficient that $R\rightarrow 0$ as $r\rightarrow +\infty$. 
In Section~\ref{sec:exactsols} we analyzed one exact solution where this happens precisely, and yet, the solution is not AF but has a deficit 
angle. In that case the mass function $M(r)$ diverges at least linearly with $r$. It is then crucial to ensure that the mass function 
converges to a constant value (that we assume to be the Komar or, equivalently, the ADM mass) in order to claim for a genuinely AF solution. 

\begin{figure}[t]
\includegraphics[width=5.9cm,height=5cm]{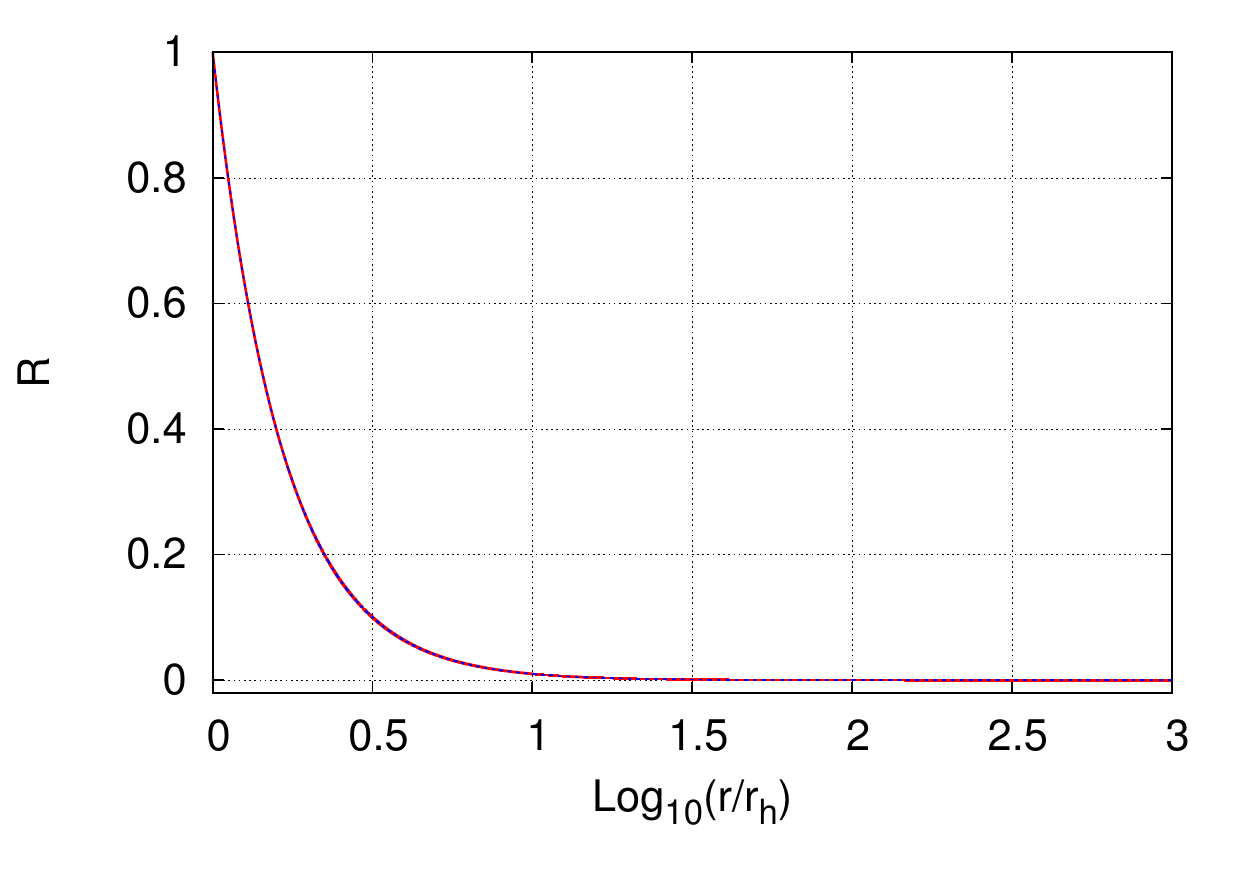}
\includegraphics[width=5.9cm,height=5cm]{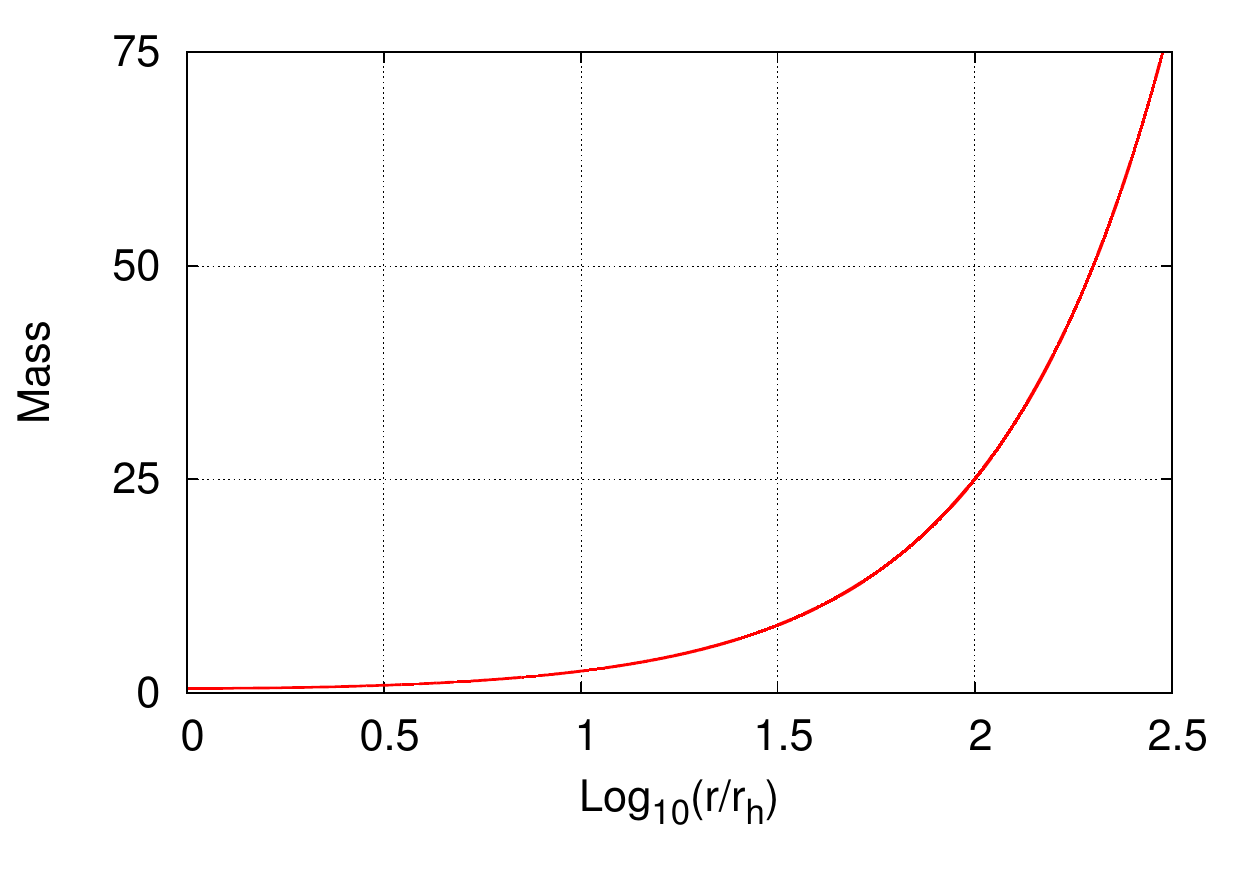}
\includegraphics[width=5.9cm,height=5cm]{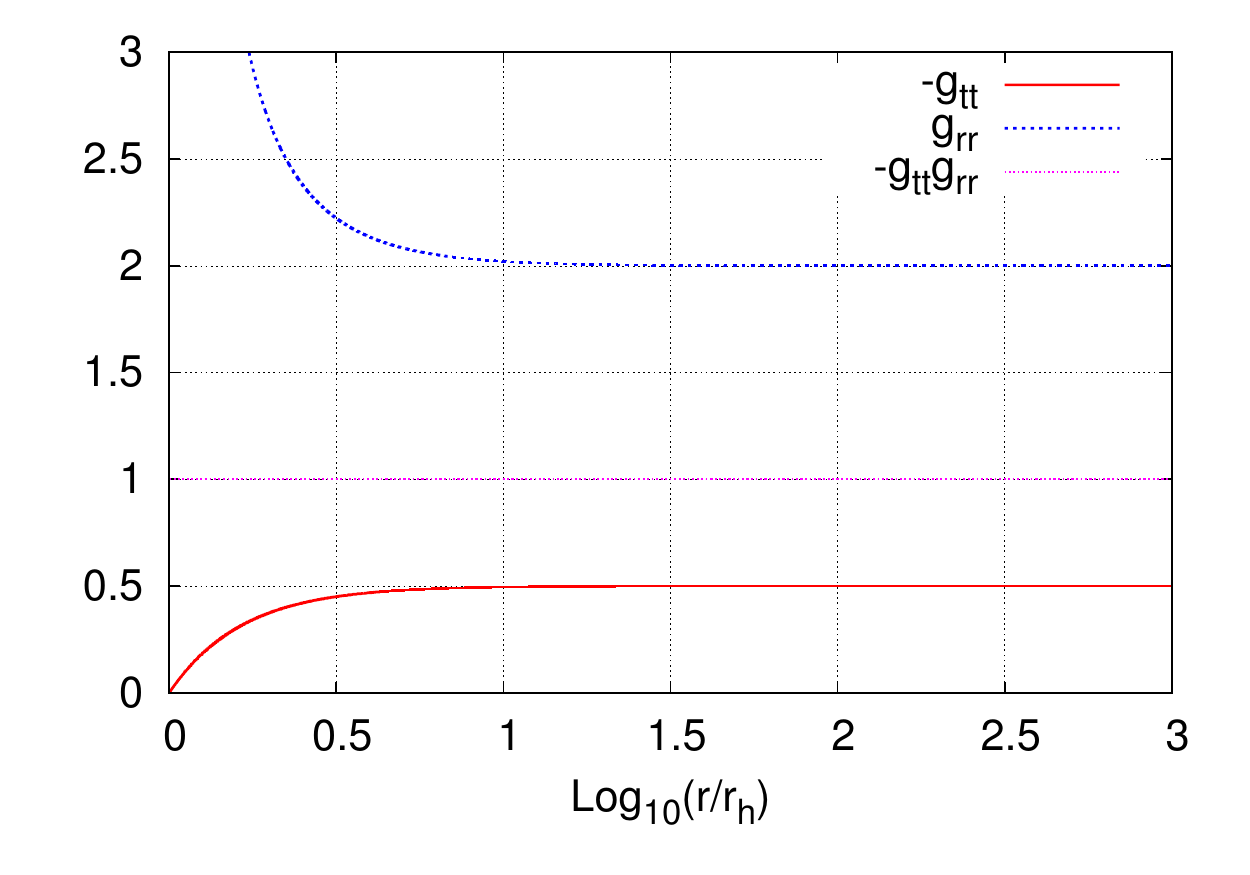}
\caption{(color online). Left-panel: Ricci scalar $R(r)$ (the 
exact and numerical solutions are superposed) computed using the model Eq.~(\ref{fRSZ}) for 
$\alpha=0$ (i.e. null cosmological constant). 
The Ricci scalar is not trivial and vanishes asymptotically, however, the spacetime is not exactly asymptotically flat but has a deficit 
angle [see Eqs.~(\ref{gfRSZ})--(\ref{deltafRSZ}) ]. At the horizon $r_h$ the Ricci scalar satisfies the regularity conditions. 
Middle-panel: the mass function $M(r)$ is not constant but grows linearly with the coordinate $r$ due to the deficit angle. 
Right-panel: metric components $-g_{tt}$, $g_{rr}$ and their product $-g_{tt}\times g_{rr}=e^{2\delta(r)}$. In the middle and right panels the exact and numerical solutions 
are superposed as well (cf. Figure~\ref{fig:relerr}).}
\label{fig:fRSZ}
\end{figure}

\begin{figure}[t]
\includegraphics[width=6.5cm,height=5cm]{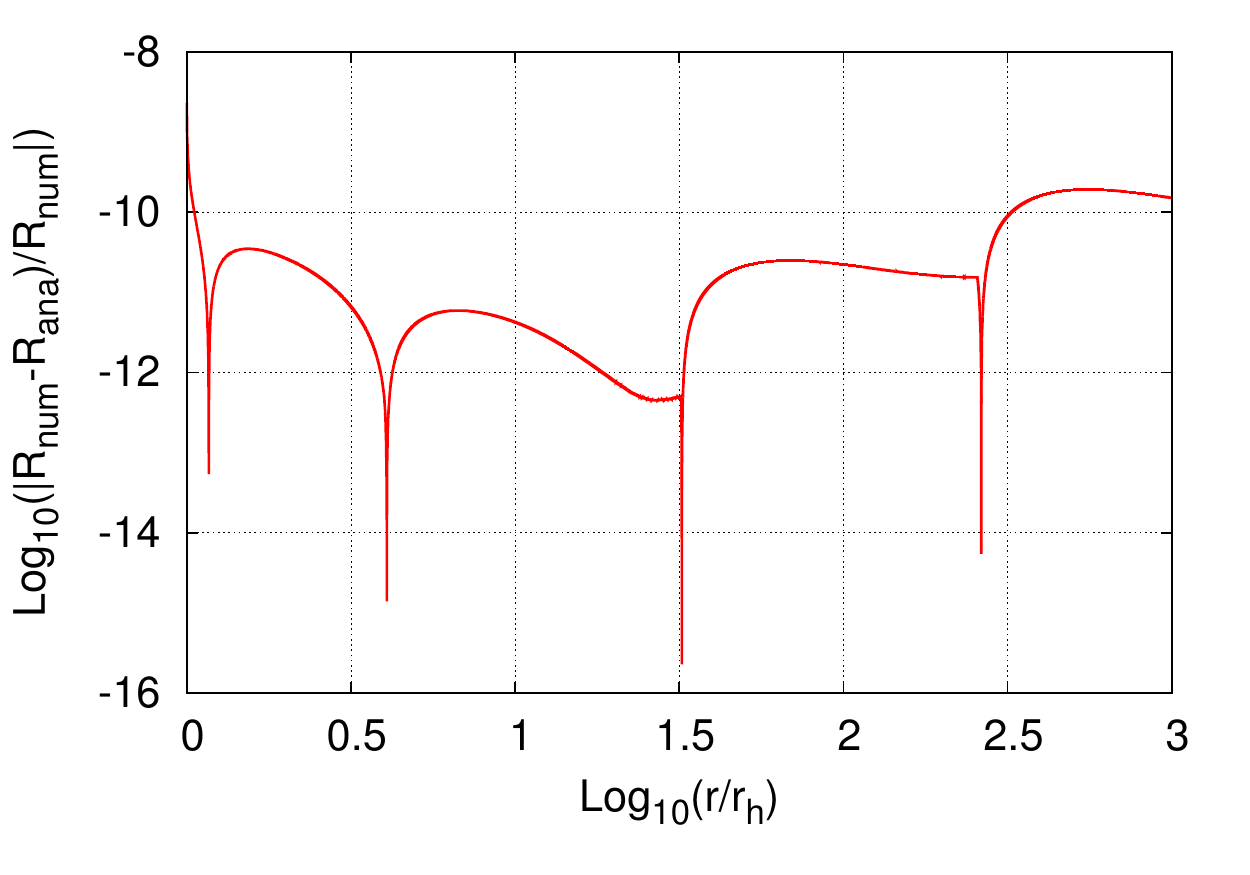}
\caption{(color online). Relative error between the numerical and the exact solution for $R(r)$ as depicted in Figure~\ref{fig:fRSZ} (left panel). 
Similar relative errors (not depicted) are found for the mass function $M(r)$ and the metric components. }
\label{fig:relerr}
\end{figure}

As a matter of fact, we used that exact solution as a testbed for our code. That is, we took the model Eq.~(\ref{fRSZ}) as {\it input} and recovered numerically the 
exact solution provided by Eqs.~(\ref{gfRSZ})--(\ref{deltafRSZ}), 
notably for $\alpha =0$. Notice that in this case $R(r)$ is not trivial. Figure~(\ref{fig:fRSZ}) depicts the analytic and the 
numerical solutions superposed, showing an excellent agreement between the two. Typical numerical errors are depicted in Figure~\ref{fig:relerr}. 

We also checked that the trivial solutions $R=R_1=const$ that exist in several 
of the Models 1-5 were recovered numerically when starting with $R_h=R_1$
and which lead to the {\it hairless} Kottler--Schwarzschild--De Sitter solutions, 
including the plain AF Schwarzschild solution when $R_1\equiv 0$.

Additionally we devised other {\it internal} tests to verify the consistency of our code. These tests are similar to those implemented in 
our analysis of star-like objects~\cite{Jaime2011}, and are independent of the fact that exact solutions are available or not.

Let us turn our attention to the specific models that deserved a detailed numerical exploration.

\bigskip

\bigskip

\begin{figure}[t]
\includegraphics[width=5.9cm,height=5cm]{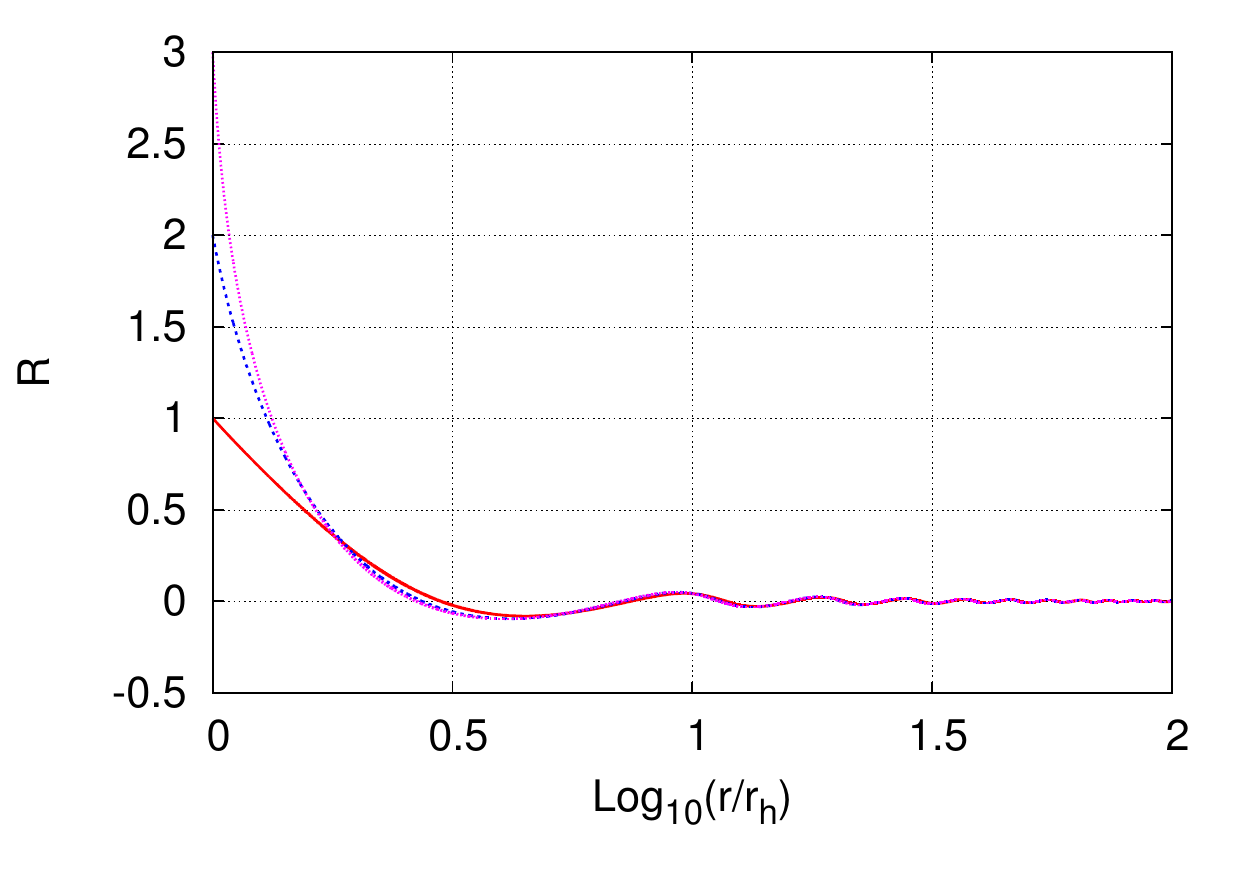}
\includegraphics[width=5.9cm,height=5cm]{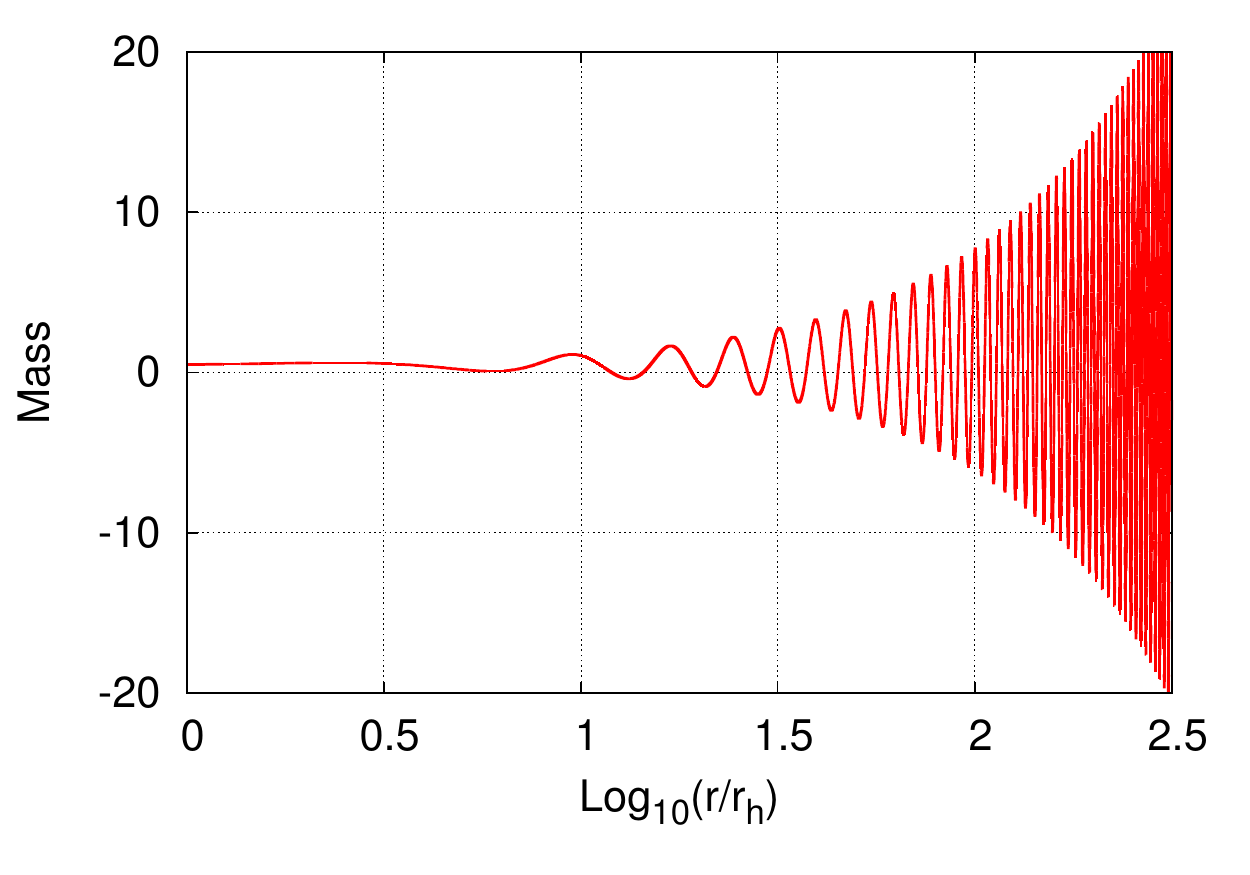}
\includegraphics[width=5.9cm,height=5cm]{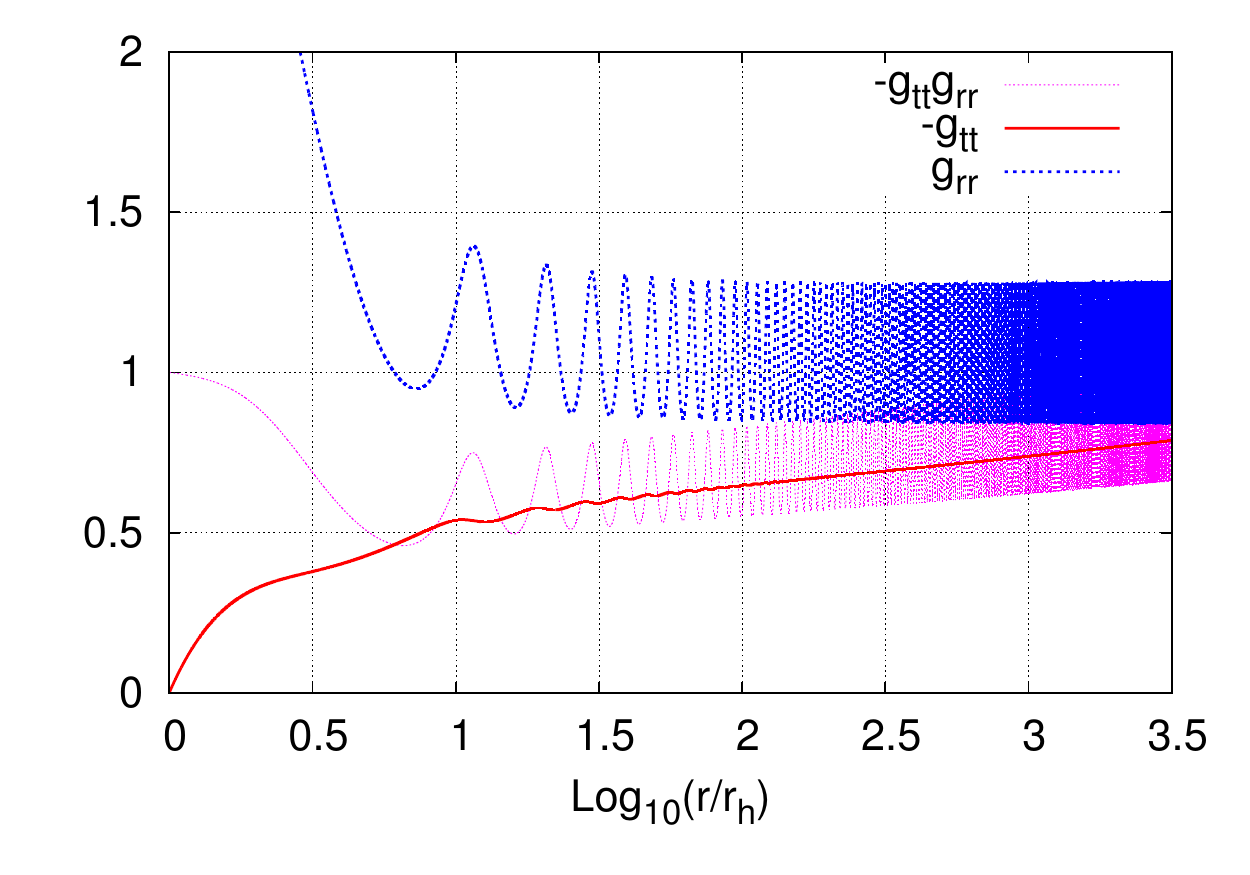}
\caption{(color online). Numerical solutions for the  Model 4 taking $\lambda_e=-3$. 
Left-panel: Ricci scalar for three different values of $R_h$. The solutions vanish asymptotically. Middle panel: the mass function $M(r)$ corresponds to 
the solution of $R$ with $R_h=1$ shown in the left panel. Similar plots for $M(r)$ are found for the other two solutions of $R$. 
The non-converging behavior of $M(r)$ as $r\rightarrow +\infty$ indicates that the spacetime is not AF. Right panel: metric components for the 
solution depicted in the middle panel. The metric components $-g_{tt}$ and $g_{rr}$ and their product $-g_{tt}\times g_{rr}=e^{2\delta(r)}$ 
are bounded but oscillate as $r\rightarrow +\infty$ corroborating that the resulting spacetime is not AF.}
\label{fig:fRexpnohair}
\end{figure}

{\bf Model 4:} We consider the Model 4 with $\lambda_e<0$. In this sector of $\lambda_e$ the potential $\mathscr{U}(\phi)$ is not positive definite and the 
model may admit hairy solutions because the NHT's do not apply. Notwithstanding, the only solutions with a non-trivial Ricci scalar $R(r)$ 
that we find numerically are not exactly AF. The Ricci scalar vanishes asymptotically in an oscillating fashion as $r\rightarrow +\infty$, but  
the mass function $M(r)$ does not converge but oscillates as well and grows unboundedly as $r^{\sigma}$ with $0<\sigma\leq 1$ 
(see Figure~\ref{fig:fRexpnohair}). This behavior is similar to the one provided by the heuristic analysis within the Model 2, except that here 
we take into account the full system of equations. Despite such behavior the metric components, which 
depend on $M(r)$, remain bounded in the asymptotic region. This can be partially understood by looking to $g_{rr}= 1-2M(r)/r$, and realize that 
for $r_h\ll r$ the non-oscillating part of this metric component 
behaves like $r^{\sigma-1}$. So if $\sigma\lesssim 1$, that part of $g_{rr}$ may converge to $1$ very slowly, 
so slowly that one cannot even notice it by looking to the numerical outcome.

This behavior seems to be generic for any $\lambda_e<0$ and any $R_h$. The conclusion is that we do not find any genuinely AFSSS black hole solution in this model.

\bigskip

Finally, let us focus on the Models 5 and 6 that led to pathological potentials in the EFSTT description.
\bigskip

\begin{figure}[t]
\includegraphics[width=5.9cm,height=5cm]{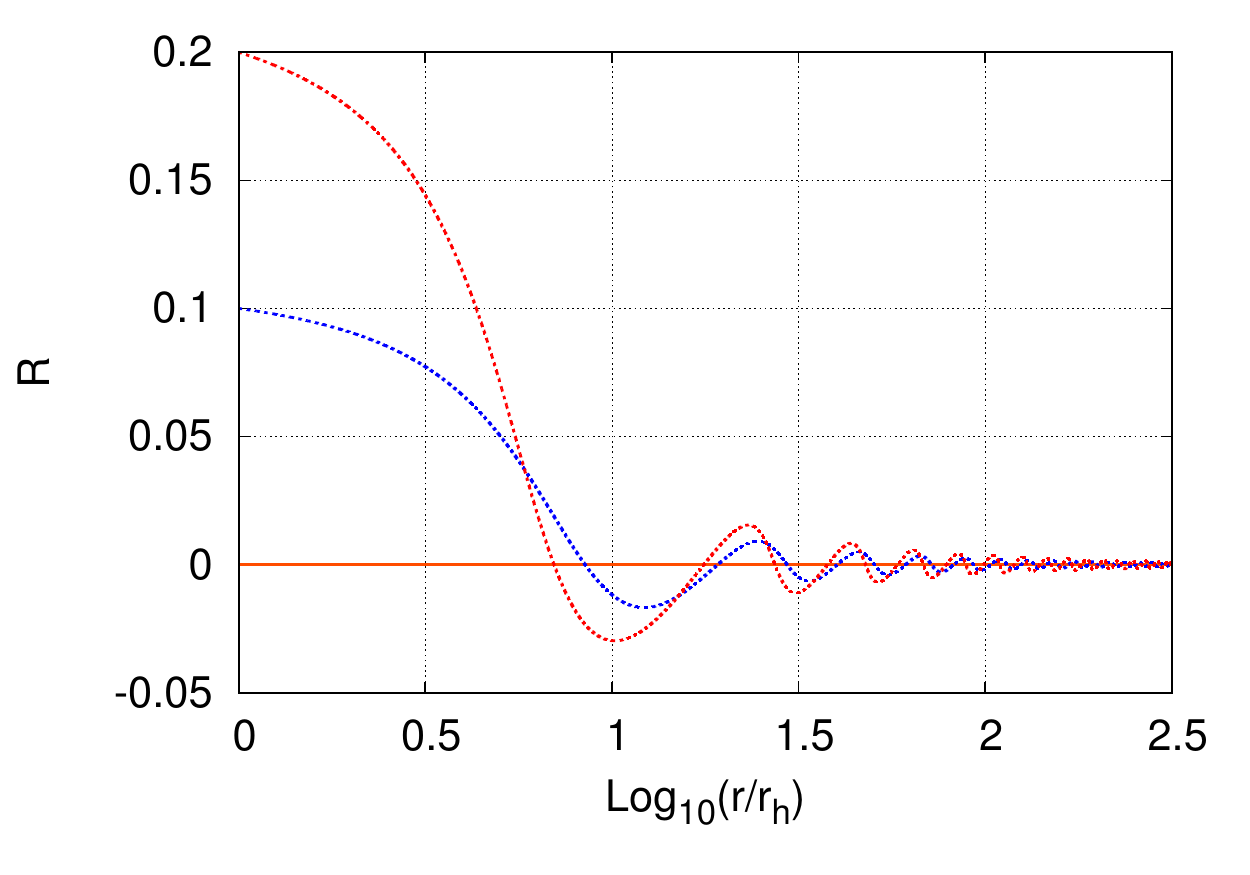}
\includegraphics[width=5.9cm,height=5cm]{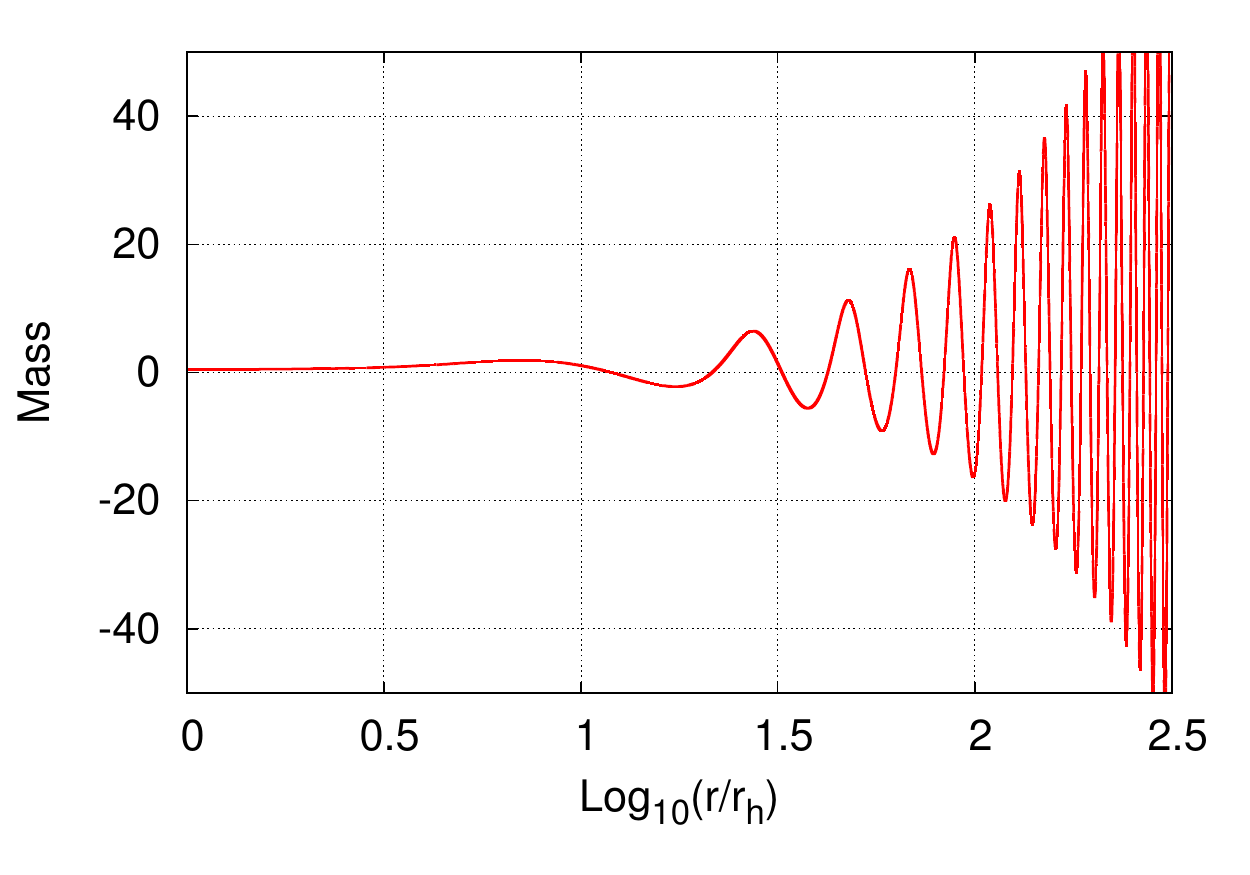}
\includegraphics[width=5.9cm,height=5cm]{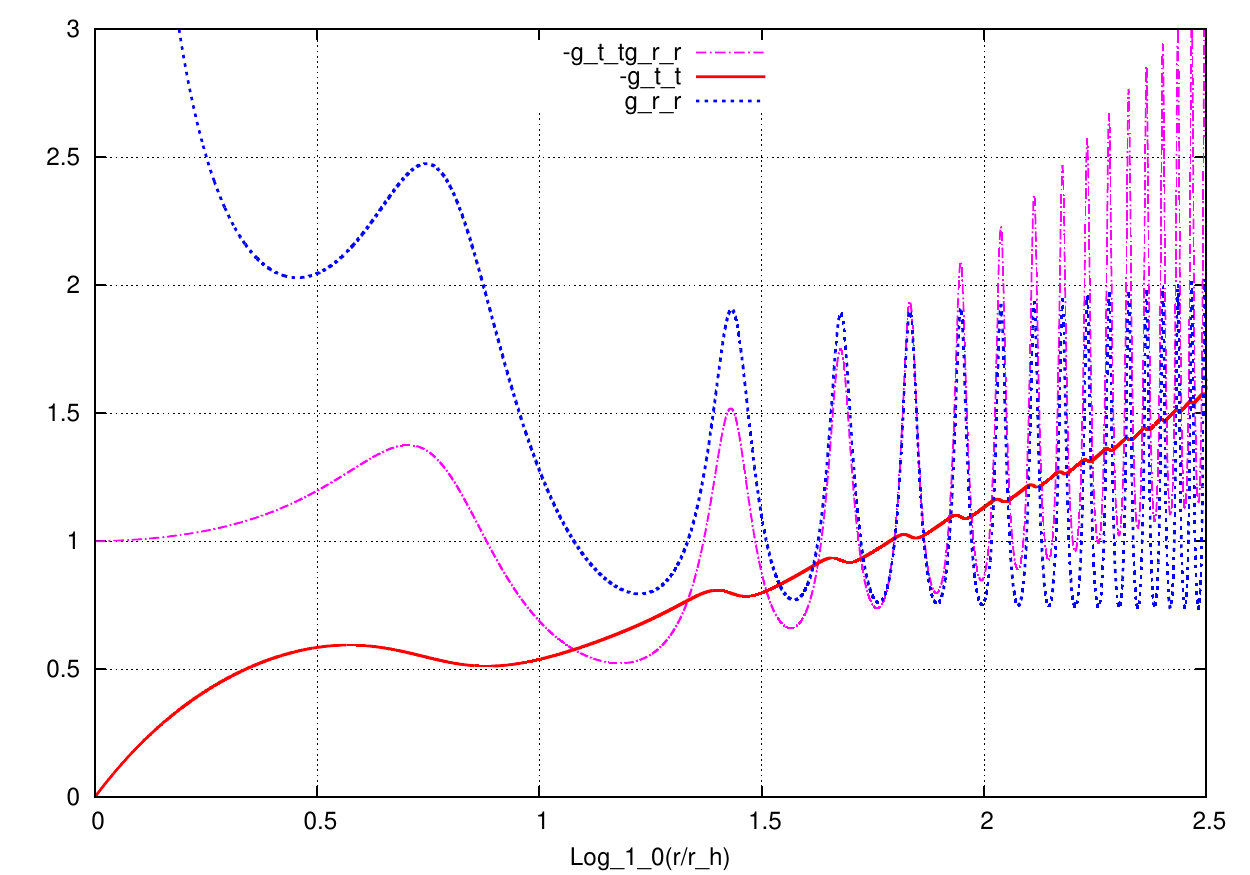}
\caption{(color online). Examples of numerical solutions for the Starobinsky Model 5 with $q=2$ and $\lambda_s=1$. Left panel: 
Ricci scalar for three values of $R_h$ (including $R_h=0$ leading to the trivial solution $R(r)\equiv 0$). 
$R(r)\rightarrow 0$ asymptotically, however, for the non-trivial solutions (dotted lines) the 
mass function $M(r)$ does not converge (see the middle panel). Middle panel: mass function $M(r)$ associated with the 
solution of $R$ with $R_h= 0.1$ as shown in the left panel. Similar plots for $M(r)$ are found, but not depicted, for the other non-trivial 
solution of $R$, while $M(r)= const= r_h/2$ when $R(r)\equiv 0$, corresponding to the Schwarzschild solution. 
Right panel: metric components associated with the solution of middle panel. 
The metric components $-g_{tt}$ and $g_{rr}$ and their product $-g_{tt}\times g_{rr}=e^{2\delta(r)}$ 
oscillate as $r\rightarrow +\infty$ corroborating that the resulting spacetime is not AF. 
Here $R$ is given in units of $R_{\rm ST}$, and $M$ and $r$ in units of $1/\sqrt{R_{\rm ST}}$ ($G_0=c=1$).}
\label{fig:Starono-hairoscillating}
\end{figure}

\begin{figure}[h]
\includegraphics[width=5.9cm,height=5cm]{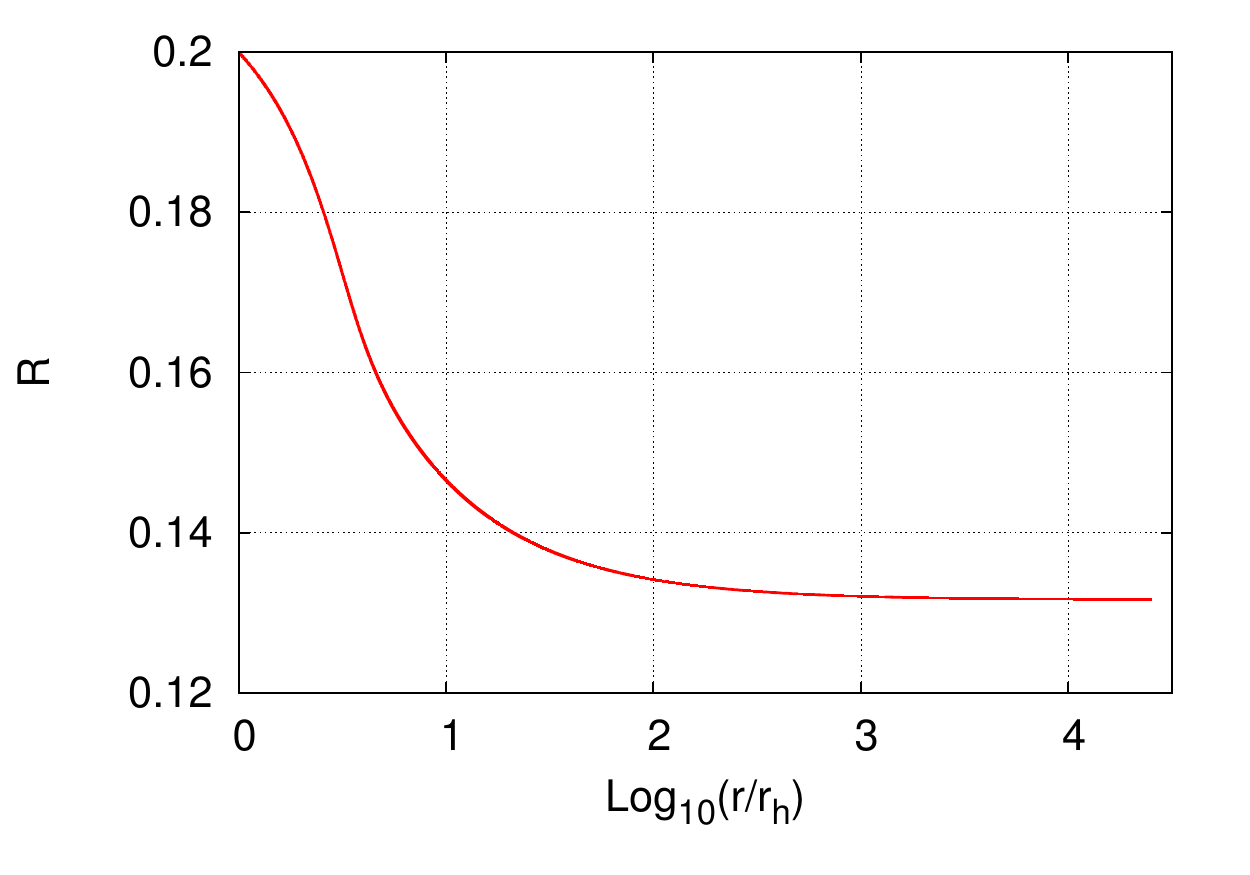}
\includegraphics[width=5.9cm,height=5cm]{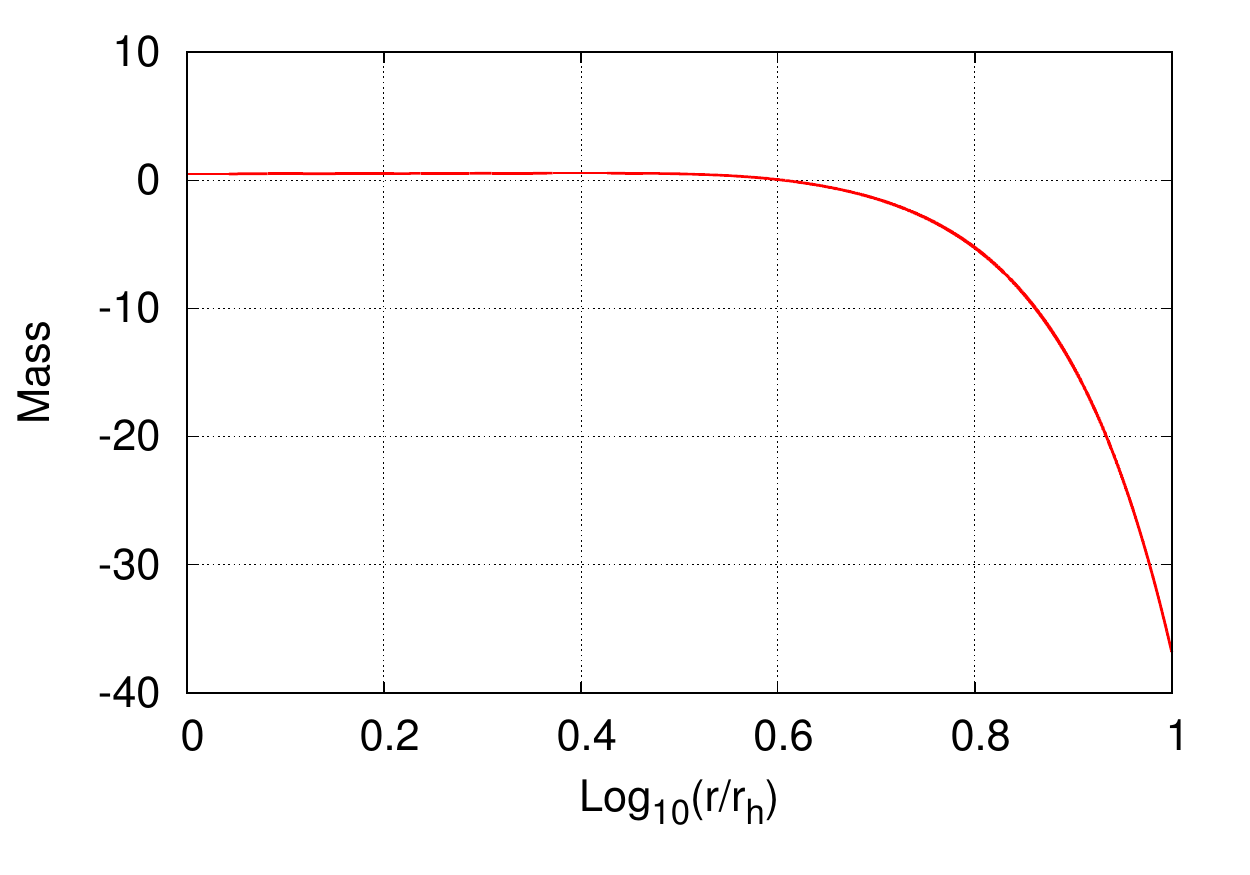}
\includegraphics[width=5.9cm,height=5cm]{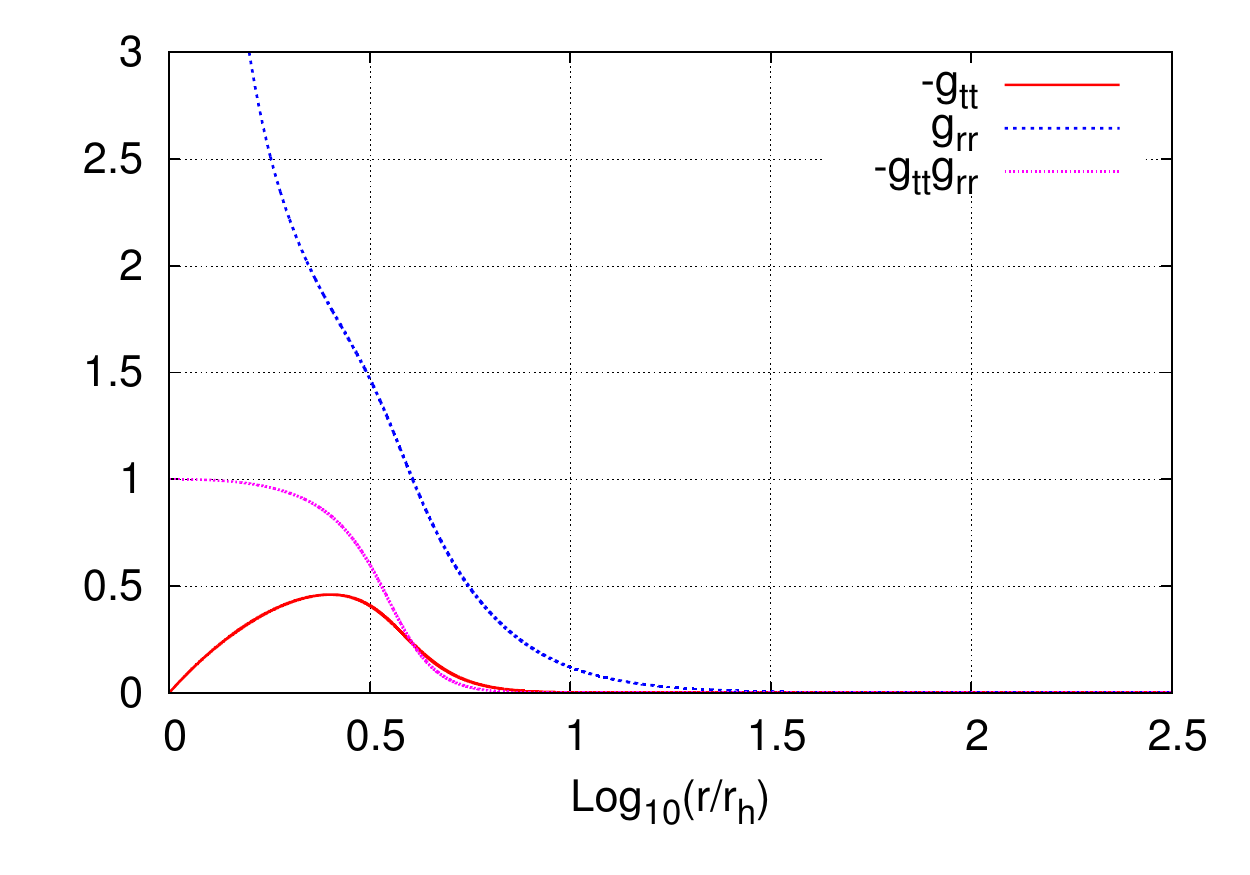}
\caption{(color online). Example of a numerical solution for the Starobinsky Model 5 with $q=4$ and $\lambda_s=1$. 
Left panel: the numerical solution shows that the spacetime is not AF as $R(r)$ does not vanish asymptotically. 
It has rather a spurious De Sitter behavior when $R(r)\rightarrow const >0$ asymptotically. 
Middle panel: the mass function $M(r)$ does not converge asymptotically but decreases to a very large {\it negative} values. 
This is even opposite to the growing behavior 
$M(r) \sim r^3 >0$ that should be expected if the spacetime were genuinely asymptotically De Sitter. Right panel: 
the metric components $-g_{tt}$, $g_{rr}$ and their product $-g_{tt}\times g_{rr}=e^{2\delta(r)}$ vanish asymptotically. 
This behavior confirms that the spacetime is not even genuinely asymptotically De Sitter where a cosmological horizon is 
expected at $r_h^c> r_h$ where $g_{tt}(r_h^c)=0$ and $g_{rr}(r_h^c)= +\infty$.}
\label{fig:Starono-hair}
\end{figure}

{\bf Model 5:} For the Starobinsky model we limit our search for a shooting value $R_h$ first in the region $0<R< R_s/\sqrt{2q+1}$ and 
then in  $-R_s/\sqrt{2q+1}< R_h< 0$ in order to avoid crossing the {\it weak} singularities at 
$\pm R_s/\sqrt{2q+1}$ when $R(r)$ tries to reach the asymptotic value $R=0$. We never found a successful shooting parameter leading to an authentic asymptotically flat solution. Two examples of this kind of solutions are depicted in Figures~\ref{fig:Starono-hairoscillating} 
and \ref{fig:Starono-hair}. Figure~\ref{fig:Starono-hairoscillating} shows that for $q=2$ 
the solutions are similar to the exponential Model 4 with $\lambda_e=-3$ depicted in Figure~\ref{fig:fRexpnohair}. Thus, the asymptotic behavior does not correspond to an AF spacetime.  
 
For $q=4$, we find situations where the Ricci scalar decreases monotonically to a constant value without oscillating as one can see in 
the left panel of Figure~\ref{fig:Starono-hair}. However, this constant is not related 
with the trivial solution $R=R_1=const$ which is the solution of the algebraic equation $2f(R_1)-R_1 f_R(R_1)=0$. In fact, what happens is that 
$M(r)\rightarrow -\infty$ as $r\rightarrow \infty$, as we can see from the middle panel of Figure~\ref{fig:Starono-hair}, and also 
$M(r)/r\rightarrow -\infty$, therefore, by looking at Eq.~(\ref{TraceRsss}), we appreciate that the combination $[2f(R)-Rf_R]/(1-2M/r)$ 
goes to zero even if $2f(R)-Rf_R\neq 0$. In this way, we see that $R'\rightarrow 0$ and $R''\rightarrow 0$, while 
$R\rightarrow const$ asymptotically, which solves Eq.~(\ref{TraceRsss}). This behavior of $M(r)$
explains why the metric components vanish asymptotically (see the right panel of Figure~\ref{fig:Starono-hair}). Before vanishing we 
see that $g_{rr}=1$ at ${\rm log}_{10}(r/r_h)\approx 0.6$ precisely where $M(r)=0$. At this value of $r$, the 
component $g_{tt}= -e^{-2\delta(r)}$. Thus, we conclude that the AF behavior is not recovered either for this and other values $q>0$ and different $R_h$.

In summary, we find strong numerical evidence that for the Starobinsky Model 5 AFSSSBH with geometric hair do no exist. 
This conclusion is obtained by changing the parameters in several combinations as well as the values $R_h$.
\bigskip

{\bf Model 6:} The numerical analysis of the Hu--Sawicky Model 6 requires more care because $f_{RR}$ 
vanishes at $R=0$ for several values of $n$ ($|n|>2$). The 
{\it weak} singularity where $\mathscr{W}_R(R)= (2f-Rf_R)/f_{RR}$ diverges can be reached where $f_{RR}=0$, 
except if $2f-Rf_R$ vanishes at the same $R$. In fact for $|n|>2$ and $n\neq 3$ the quantity $\mathscr{W}_R(0)$ always diverges. 
In Figure~\ref{fig:HSU} (right panel), one can appreciate this divergence for $n=4$. For $n=3$ the quantity
 $\mathscr{W}_R(0)$ remains finite, but $\mathscr{W}_R(0) \neq 0$, so $R=0$ 
cannot be a possible asymptotic solution of Eq.~(\ref{TraceRsss}). Thus, the case $n=3$ is irrelevant for AF solutions. 
In consequence, for $|n|>2$ and $n\neq 3$, 
the {\it weak} singularity is approached as $R$ tries to reach its asymptotic value and at this point 
Eq.~(\ref{TraceRsss}) becomes singular. Therefore, in this range of $n$ we never find a well behaved solution regardless of 
the shooting parameter $R_h$ and $R$ systematically shows a divergent behavior in the numerical solutions.

\begin{figure}[h]
\includegraphics[width=5.9cm,height=5cm]{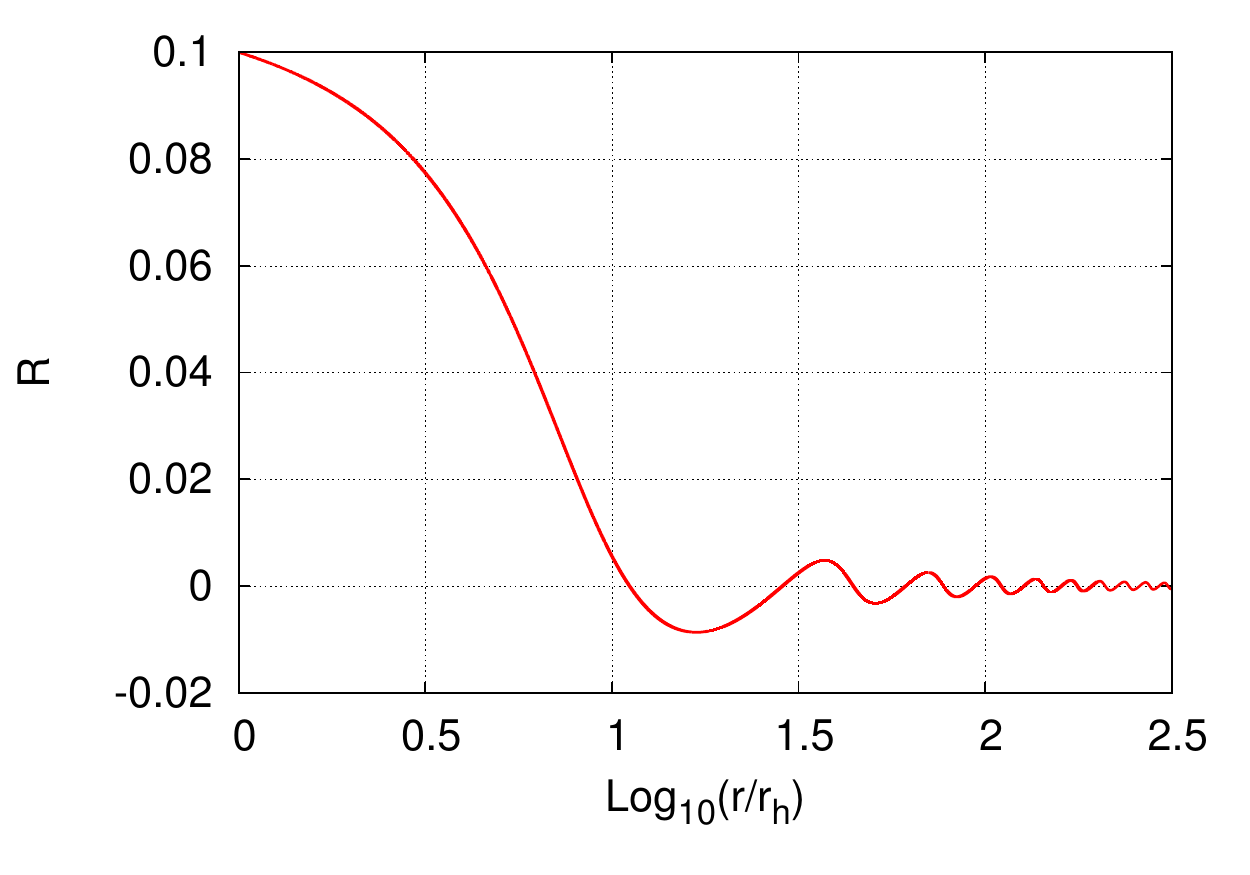}
\includegraphics[width=5.9cm,height=5cm]{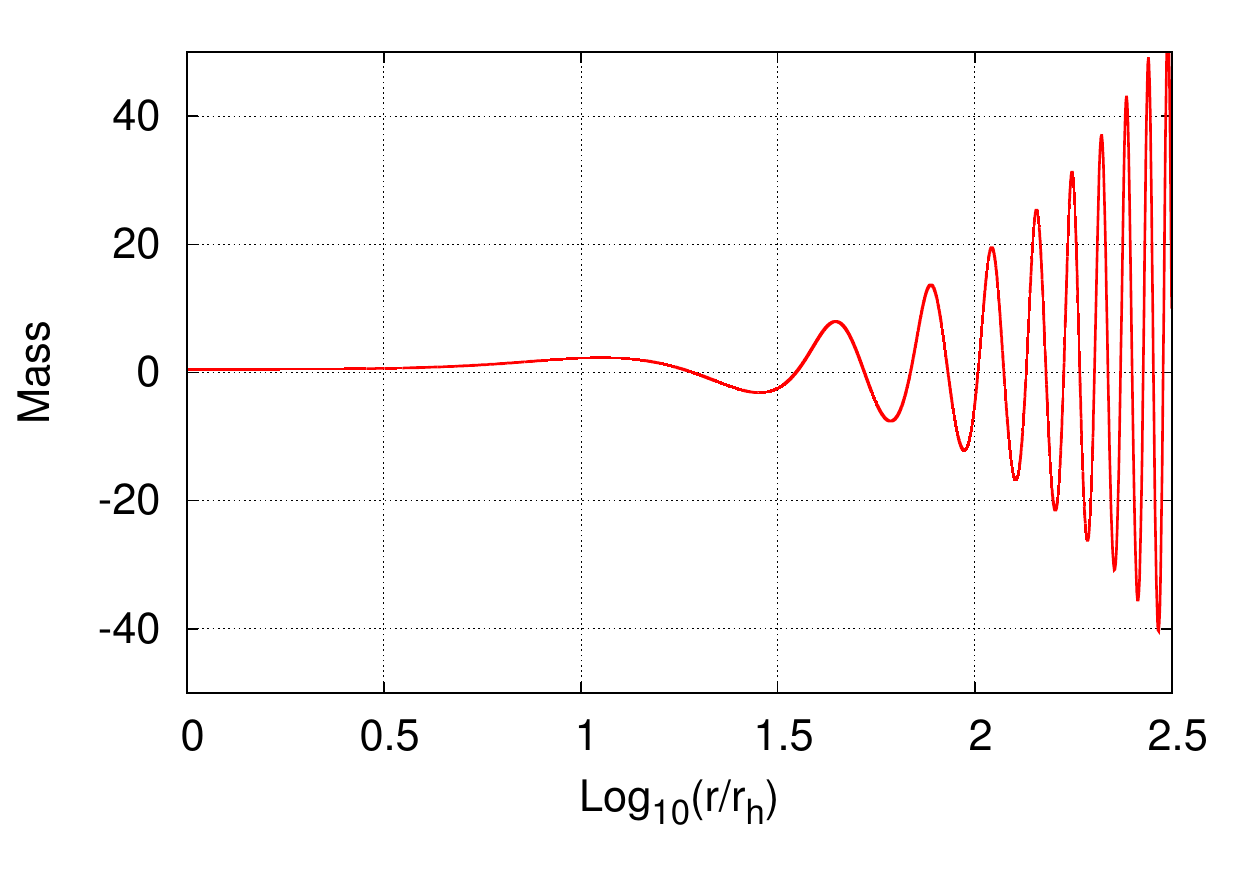}
\includegraphics[width=5.9cm,height=5cm]{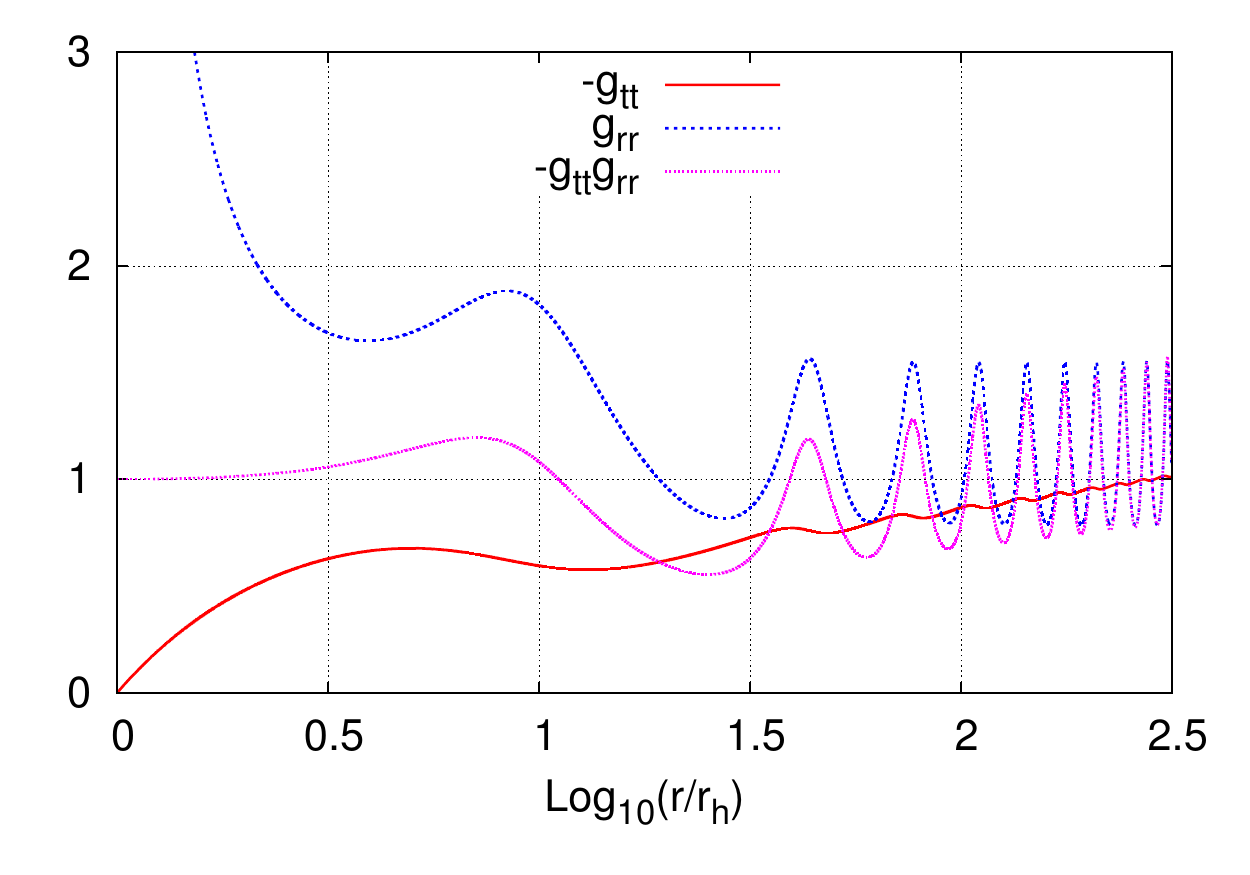}
\caption{(color online). Numerical solution for the Hu--Sawicky Model 6 with $n=1$ and 
and $c_1\approx 0.38$ and $c_2= 2.01\times 10^{-3}$. 
The units are like in Figure~\ref{fig:Starono-hair} but with $R_{\rm HS}$ instead of $R_{\rm ST}$. 
The description of panels is similar to Figure~\ref{fig:Starono-hairoscillating}.}
\label{fig:HS-nohairoscillating}
\end{figure}

In the interval $1 \leq n \leq 2$ the quantity $\mathscr{W}_R(0)$ vanishes and we do not find any pathologies as $R\rightarrow 0$. 
The numerical solutions are very similar to the oscillating solutions found in the Starobinsky Model 5 and the exponential Model 4. 
Figure~\ref{fig:HS-nohairoscillating} shows a prototype of such solutions. As we remarked before, for $n=2$ the Hu--Sawicky model 
is essentially the Starobinsky model with $q=1$, thus it not surprising to find, at least for these values of $n$, such an oscillating behavior 
for all of our trial values $R_h$. Our conclusions seem to be insensitive for several values of the constants $c_1$ and $c_2$. 

Hence, the numerical evidence indicates that AF solutions with non-trivial hair are absent as well in the Hu--Sawicky model.

\section{Conclusions}
\label{sec:concl}

We have argued that generically $f(R)$ gravity contains 
trivial solutions $R= const$. This property allows to mimic an effective cosmological constant that can produce the observed accelerated expansion 
in the universe. In the context of black-holes (stationary and axisymmetric, or static and spherically symmetric)  
the same property allows to find the same solutions known in GR (under the same symmetries) with or without a cosmological constant. 
The only difference between both type of solutions (i.e. the solutions found in one or the other theory) is that the two fundamental constants 
involved (the Newton's gravitational constant and the cosmological constant) are replaced by the effective ones, 
$G_{\rm eff}$ and $\Lambda_{\rm eff}$ in $f(R)$ gravity. Therefore, all solutions of this kind reported in the literature 
within the framework of $f(R)$ gravity do not provide any deeper knowledge than the ones we already know in GR. 

We then focused in the problem of finding non-trivial (hairy) AFSSS black hole solutions 
where the Ricci scalar is {\it not} constant in the domain of outer communication of the black hole but varies with the radial coordinate $r$. 
Within that aim  we provided the equations to study this scenario for an arbitrary $f(R)$ model and derived the conditions for a regular black 
hole. We then proceeded to analyze some specific models. Prior to a thorough numerical analysis, we studied 
the models under the scalar-tensor approach within the Einstein-frame, stressing that in vacuum $f(R)$ gravity takes the same form as in 
the Einstein-scalar-field system. Therefore, this method makes possible to check if the available no-hair theorems for AFSSS can be applied, 
and thus, sparing an unnecessary numerical effort. Thus, for the models where the NHT's apply, we concluded that {\it geometric} hair is absent. 
In those situations the numerically analysis simply confirms the NHT's.

For the cases where the resulting scalar-field potential $\mathscr{U}(\phi)$ does not satisfy the condition $\mathscr{U}(\phi) \geq 0$ 
required by the NHT's, we turned to a detailed numerical study. Our conclusion is that we did not find any such hair in any of the models 
where the NHT's do not apply. 

We also discussed some exact solutions that seem to represent AF hairy black holes, however, 
we showed that in fact the $r$ dependence in the Ricci scalar is due to the presence of a deficit angle. Therefore such solutions are not really 
AF. Similarly we report numerical solutions that are {\it not} genuinely AF as the 
the  mass function never converges to the Komar mass, a feature necessary to prove asymptotic flatness. Some of the solutions 
have an asymptotic oscillatory behavior but without encountering any singularity, while others produce singularities in the equations at a finite $r$.

It remains thus an open question to determine if hair exist in $f(R)$ gravity. In order to settle the question in the affirmative it is tantalizing to depart from the Einstein-$\phi$ system with a potential that has negative branches and that allow for hairy black 
holes~\cite{Nucamendi2003,Anabalon2012}, and then perform an ``inverse'' conformal transformation to obtain an $f(R)$ model. However, 
in practice this seems to be difficult in a closed explicit form. Moreover, even if this is possible, the resulting $f(R)$ model 
would be rather artificial and would need to be submitted to the usual test (i.e. cosmological,  Solar System, binary pulsar, etc.) 
to be better motivated physically, irrespective of the issue of hair. Some of the exact black hole solutions that have been 
reported in the literature (including in general relativity with exotic energy-momentum tensors) 
have been obtained using similar ``tricks'' or ad hoc confections, but presented afterwards in the more {\it logical} direction within the aim of enhance 
their merit. This is not the exception in $f(R)$ gravity, where an $f(R)$ model can be deduced by demanding 
the existence of some kind of exact solution, most of time, deprived of actual physical interest. For instance, 
one can impose $g_{rr}= -1/g_{tt}$ (in area coordinates), which is valid provided $T_{{\rm eff}\,\,\,t}^t= T_{{\rm eff}\,\,\,r}^r$, and then obtain a 
differential equation for $f(R)$ which can be solved if the assumed form for $g_{tt}(r)$ is simple enough. In fact, by using this method 
we were able to recover the model $f(R)= f(R) =  2a\sqrt{R - \alpha}$ given by (\ref{gfRSZ}). Notice however that such solution was not AF, and thus, 
it cannot be used as a counterexample to the no-hair conjecture in $f(R)$ gravity.

In this paper we limited ourselves to the case of AF spacetimes. In a future investigation we will analyze the case of hair in asymptotically De Sitter and 
Anti-De Sitter spacetimes. For such spacetimes the no-hair theorems for the AF scenario 
require amendments due the different asymptotic conditions, and the numerical treatment, 
although similar to the one presented in this work, 
is sufficiently different to require a detailed and separate analysis.

\section*{Acknowledgments}
M.S. is supported partially by DGAPA--PAPIIT UNAM grant IN107113 and SEP--CONACYT grants CB-166656 and CB-239639; L. G. J. acknowledges 
CONACYT postdoctoral fellowship 236937 and also L. Amendola and the Cosmology Group at the University of Heidelberg for their hospitality.

\newpage
\appendix


\section{Conservation of the energy-momentum tensor of matter in $f(R)$ gravity}
\label{sec:consEMT}

The simplest way to show the conservation equation $\nabla^a T_{ab}=0$ in $f(R)$ metric gravity is departing from Eq.~(\ref{fieldeq1}) 
written as:
\begin{equation}
\label{fieldeq1ap}
f_R R_{ab} -\frac{1}{2}fg_{ab} - 
\left(\nabla_b \nabla_a - g_{ab}\Box\right)f_R= \kappa T_{ab}\,\,.
\end{equation}
One can use the commutation relation $\nabla_a \nabla_b f_R = \nabla_b \nabla_a f_R$ since $f_R$ is a scalar function 
(i.e. in $f(R)$ metric gravity we assume the absence of torsion: the l.h.s of Eq.~(\ref{fieldeq1ap}) must be a symmetric 
tensor). So, applying the covariant derivative $\nabla^a$ to  Eq.~(\ref{fieldeq1ap}) we obtain
\begin{equation}
\label{Divfieldeq1}
f_R \nabla^a R_{ab} + R_{ab}\nabla^a f_R -\frac{1}{2}f_R g_{ab} \nabla^a R - 
\left(\nabla^a \nabla_b \nabla_a - \nabla_b \nabla^a \nabla_a \right)f_R= \kappa \nabla^a T_{ab}\,\,,
\end{equation}
where $\Box= \nabla^a\nabla_a$ and $f_R= df/dR$ were used. According to the Bianchi identities 
\begin{equation}
\nabla^a\Big(R_{ab}- \frac{1}{2}g_{ab}R\Big)=0\,\,\,,
\end{equation}
therefore the first and third terms of l.h.s of Eq.~(\ref{Divfieldeq1}) cancel out, yielding
\begin{equation}
\label{Divfieldeq2}
R_{ab}\nabla^a f_R - \left(\nabla^a \nabla_b \nabla_a - \nabla_b \nabla^a \nabla_a\right)f_R= \kappa \nabla^a T_{ab}\,\,.
\end{equation}
Furthermore
\begin{eqnarray}
\left(\nabla^a \nabla_b \nabla_a - \nabla_b \nabla^a \nabla_a\right)f_R &=& 
g^{ac}\left(\nabla_c \nabla_b  - \nabla_b \nabla_c \right) \nabla_a f_R = g^{ac} R_{cba}^{\hspace{0.4cm} d} \nabla_d f_R\nonumber \\
&=&  R_{bd}\nabla^d f_R \,\,\,.
\end{eqnarray} 
This result implies that the l.h.s of Eq.~(\ref{Divfieldeq2}) vanishes identically, leading then to
\begin{equation}
\label{Divfieldeq3}
\nabla^a T_{ab}= 0\,\,.
\end{equation}
We can summarize this result as a theorem: 
{\it in $f(R)$ metric gravity (under the same basic axioms assumed in GR concerning the manifold and the metric) the generalized tensor
\begin{equation}
 \mathscr{G}_{ab}:= f_R R_{ab} -\frac{1}{2}fg_{ab} - 
\left(\nabla_a \nabla_b - g_{ab}\Box\right)f_R
\end{equation}
obeys a generalized Bianchi identity}
\begin{equation}
\nabla^a \mathscr{G}_{ab}= 0 \,\,\,.
\end{equation}


\section{Supplementary regularity conditions}
\label{sec:regcond2}

In principle the regularity conditions at the horizon (\ref{Rprimereg}), (\ref{Mprimereg}) together with $\delta(r_{h})=\delta_{h}$, $R(r_{h})=R_{h}$, 
$M(r_{h})=r_{h}/2$, seem enough data to solve the system of ordinary differential Eqs.~(\ref{TraceRsss}), (\ref{Msss}), and (\ref{deltasss}). 
Nevertheless, in view of the method that we devised to solve these equations numerically, care must be taken when evaluating 
at the horizon the right-hand-side of Eqs.~(\ref{TraceRsss}) and (\ref{deltasss}). Therefore at this place we also need to know the values of 
$R''_H$ and $\delta'_H$ at the horizon when starting the numerical integration. If we do not impose these 
supplementary regularity conditions a mild numerical error is made at $r_H$, but we want to avoid this and impose also the right conditions on 
$R''_H$ and $\delta'_H$. As we mentioned in Sec.~\ref{sec:regcond}, in order to find these two conditions we require first to obtain an expression for 
$R'''$ and $\delta''$, then as before, develop the quantities around $r_H$, and finally impose that $R'''$ and $\delta''$ are finite at the horizon. 
We simply provide the final outcome of this process, and the regularity conditions turn out to be as follows: 
\begin{eqnarray}
\label{regdeltaprime}
\delta' \Big|_{r_H} &=& 4r^{3}\Bigg\{r^{2}ff_{RR}\Big[ f\Big( 10f -13R f_{R}\Big) + \frac{6f_{R}}{r^{2}}\Big(2f  - Rf_{R}  \Big)  \Big]  
+  r^{2}f f^{2}_{R}\Big[ 10f -R\Big( 13f_{R}  -4Rf_{RR}\Big)  \Big] \nonumber\\ &+&  2f^{3}_{R}\Big[ 6f + Rf_{R} \Big( 2r^{2}R - 3\Big) \Big]\Bigg\}
\Bigg\{f_{RR}\Big[ 2f_{R}\Big( 3 - r^{2}R\Big) + r^{2}f\Big]\Big[\Big(-41r^{2}f + 2f_{R} \Big( 13r^{2}R - 63\Big) \Big)r^{2}Rf_{R} \nonumber\\
 &+& 2r^{2}\Big(7r^{2}f^{2} + 18f_{R} \Big( \frac{4 f_{R}}{r^{2}}  + 3 f\Big)  \Big)  \Big]\Bigg\}^{-1} 
\left|\rule{0mm}{0.6cm}\right._{r=r_H} \,\,\,,
\end{eqnarray}
\begin{eqnarray}
\label{regRbiprime}
R'' \Big|_{r_h} &=&-4r^{2}f_{R}\Big(-Rf_{R} + 2f \Big)\Bigg\{ r^{2}f^{2}\Big[ -12r^{2}f f^{2}_{RR} + 3r^{2}f_{R}f_{RR}\Big( -4f_{R} 
+ f_{RR}\Big( 9R - \frac{20}{r^{2}}\Big)  \Big)\nonumber\\
&+& 4r^{2}f_{R}f_{RRR}\Big( 7f  -6 f_{R}\Big(  4R - \frac{9}{r^{2}}\Big) \Big) \Big] + f^{2}_{R}\Big[  r^{2}Rf_{R}f_{RRR}
\Big( 2f_{R} \Big( 63R  -\frac{72}{r^{2}} - 13 r^{2}R^{2}\Big)  -3f \Big( 120 - 31 r^{2}R\Big) \Big) \nonumber \\
&+& r^{2}f_{R}f_{RR}\Big( 3f_{R}\Big( 22R - 5r^{2}R^{2} - \frac{24}{r^{2}}\Big) -3f\Big( 20 - 9 r^{2}R\Big) \Big) + 288f f_{R}f_{RRR} 
+ 3r^{2}f f^{2}_{RR}\Big( 22R - \frac{24}{r^{2}}  - 5r^{2} R^{2}\Big)   \Big]\Bigg\}\nonumber \\
&\times& \Bigg\{ f^{3}_{RR}\Big[14r^{2}f  + f_{R} \Big( 24 - 13 r^{2}R\Big) \Big] \Big[r^{2}f +  2f_{R} \Big(3- r^{2}R \Big) \Big]^{3}  \Bigg\}^{-1}
\left|\rule{0mm}{0.6cm}\right._{r=r_H} \,\,.
\end{eqnarray}

Given that these two regularity conditions are quite involved, we performed minimal tests to prove its validity. Like in Sec.~\ref{sec:regcond} 
we used the exact solutions provided in Section~\ref{sec:SSS}. First we used the solution Eqs.~(\ref{Mrtriv})--(\ref{derpotcond}), notably for the model 
$f(R)= kR^2$, and then we considered the model (\ref{fRSZ}) with the solution provided by 
Eqs.~(\ref{Mrexact})--(\ref{deltafRSZ}). 
In both solutions $\delta(r) \equiv 0$. This means that the r.h.s of Eq.~(\ref{regdeltaprime}) must vanish in the two cases. 
We verified that indeed this happens.

On the other hand, the exact solution $R(r)=R_1=const$ obtained from the model $f(R)= kR^2$ 
leads simply to $R''_h= 0$, whereas $R''_h= 6/r_h^4$ for the model 
(\ref{fRSZ}). In both cases we checked that the 
r.h.s of Eq.~(\ref{regRbiprime}) gives respectively these two values 
at the horizon. We are therefore confident that our expressions are correct 
and so we enforced them in the numerical treatment presented in 
Section~\ref{sec:numerical}.

\section{Properties of the STT approach to $f(R)$ gravity}
\label{sec: BDidentification}

When introducing the action ~(\ref{Qframe}) in this way, we can recognize two things: 1) let us consider the functions 
$H(R,Q)= f'(Q)(R-Q) + f(Q)= Rf'(Q) - [Qf'(Q)- f(Q)]$, and ${\bar H}(R,Q,\chi)= Rf'(Q)- N(Q,\chi)$, where $N(Q,\chi):= Q\chi - f(Q)$, and 
in this Appendix a {\it prime} indicates differentiation with respect to the argument of the corresponding function. 
So, assuming $f(Q)$ to be a convex function, i.e., $f''(Q)>0$, then clearly $\partial_Q N=0$ if $\chi= f'(Q)$, and the inverse of this function 
$Q(\chi)$ allows to define $L(\chi):=  N(Q(\chi),\chi)= Q(\chi)\chi- f(Q(\chi))$. In this way we see that 
$L(\chi)$ is no other than the Legendre transformation of $f(Q)$, and $H(R,Q(\chi))= {\bar H}(R,Q(\chi),\chi)= 
R\chi- L(\chi)$. Moreover, $H(R,Q(\chi))$ defines in turn the Legendre transformation of $L(\chi)$ (at least in the region where $L''(\chi)=Q'(\chi)>0$)
 which in this case the condition $\partial_\chi H= R-L'(\chi)= R-Q(\chi)=0$, simply leads to $R=Q(\chi)$. 2) The condition for the second Legendre 
transformation can be imposed in the action by considering $\chi= f'(Q)$ as a Lagrange multiplier, so that the variation with respect to $\chi$ leads to 
$R=Q$. This is the formal construction when treating $f(R)$ theories as STT, and in practice it is achieved by taking the action ~(\ref{JordanF}).

In the following we perform explicitly the transformation between the original $f(R)$ theory and the special class of Brans--Dicke model 
$\omega_{\rm BD} = 0$ supplemented with a potential.

The Brans--Dicke action with a potential $W(\Phi)$ is given by\cite{Faraoni2004}
\begin{equation}
I_{\rm BD}[g_{ab},\Phi]= \frac{1}{2\kappa}\int \!\! d^4 x  \: \sqrt{-g} \Big[ \Phi R - \frac{\omega_{\rm BD}(\Phi)}{\Phi}g^{ab}(\nabla_a\Phi)(\nabla_b\Phi)
-W(\Phi)\Big] + I_{\rm matt}[g_{ab}, {\mbox{\boldmath{$\psi$}}}] \,\,\,.
\end{equation}
For comparison with the $f(R)$ model we shall focus only on the case $\omega_{\rm BD} = 0$. Thus, the field equations read~\cite{Faraoni2004}
\begin{eqnarray}
\label{EBDfield}
\Phi G_{ab} &=& \kappa T_{ab} + \left(\nabla_a \nabla_b - g_{ab}\Box\right) \Phi - \frac{W}{2} g_{ab}\,\,\,,\\
\label{PhiBDfield}
\Box \Phi &=& \frac{1}{3}\Big( \kappa T + \Phi \partial_\Phi W - 2W \Big)\,\,\,.
\end{eqnarray}
On the other hand when introducing $\chi=f_R$ Eq.~(\ref{fieldeq1}) simply reads
\begin{equation}
\label{fieldeq1JF}
\chi R_{ab} -\frac{1}{2} {\overline f}  g_{ab} - 
\left(\nabla_a \nabla_b - g_{ab}\Box\right) \chi = \kappa T_{ab}\,\,\,
\end{equation}
where ${\overline f} (\chi)= f(R(\chi))$. Moreover, this equation can be written as
\begin{equation}
\label{fieldeq1JF2}
\chi G_{ab}=  \kappa T_{ab} + \left(\nabla_a \nabla_b - g_{ab}\Box\right) \chi -\frac{1}{2} g_{ab}\Big[\chi R(\chi)  - f(R(\chi))\Big]\,\,\,,
\end{equation}
where we have made explicit the functional dependence $R(\chi)$, which means that if $f''(R)>0$, 
one can in principle invert the definition $\chi: = f_R(R)$, and obtain $R(\chi)$, and thus ${\overline f} (\chi)= f(R(\chi))$.
Therefore, if we choose the potential $W(\Phi)=  \Phi R(\Phi)  - f(R(\Phi)) \equiv U(\Phi)$ and 
identify $\Phi= \chi$, then Eq.~(\ref{EBDfield}) becomes exactly Eq.~(\ref{fieldeq1JF2}). Moreover, the trace of Eq.~(\ref{fieldeq1JF2}) reads
\begin{equation}
\label{chifield}
\Box \chi = \frac{1}{3}\Big( \kappa T + 2f- \chi R \Big)\,\,\,
\end{equation}
With the above identification of the fields $\Phi=\chi$ and the potential $W$ we appreciate that Eq.~(\ref{PhiBDfield}) also becomes 
Eq.~(\ref{chifield}) where one can easily verify that the expression $2f- \chi R$ coincides exactly with $\Phi \partial_\Phi W - 2W$. 
Henceforth, we conclude that $f(R)$ theory is equivalent to a Brans--Dicke theory with $\omega_{\rm BD} = 0$ and a potential $U(\Phi)$.


\section{Examples of AF spacetimes with a deficit angle}
\label{sec:AFdefang}
SSS spacetimes with zero charge that are asymptotically flat except for a deficit angle 
have the asymptotic form
\begin{equation}
\label{SSSD}
ds^2 \sim  - \Big(1 - \Delta - \frac{2M}{r}   \Big)
dt^{2} +  \frac{dr^{2}}{\Big(1 - \Delta - \frac{2M}{r} \Big) }  + r^{2}d\Omega^{2} \,\,\,.
\end{equation}
After a redefinition of coordinates $r= (1-\Delta)^{1/2} {\tilde r}, t = (1-\Delta)^{-1/2} {\tilde t}$, and $M= M_{ADM\Delta} (1-\Delta)^{3/2}$ 
the metric acquires the standard angle-deficit form
\begin{equation}
\label{SSSD2}
ds^2 \sim  - \Big(1 - \frac{2M_{ADM\Delta}}{\tilde r}   \Big)
d{\tilde t}^{2} +  \frac{d{\tilde r}^{2}}{\Big(1 - \frac{2M_{ADM\Delta}}{\tilde r} \Big) }  + (1-\Delta) {\tilde r}^{2}d\Omega^{2} \,\,\,.
\end{equation}
Under this parametrization the coefficient $M_{ADM\Delta}$ is then identified with the ADM mass associated with this kind of 
spacetimes~\cite{Nucamendi1997}. In the same way, an SSS metric which is ADS or AADS with a deficit angle
\begin{equation}
\label{SSSLambda}
ds^2 \sim  - \Big(1 - \Delta - \frac{2M}{r} -\frac{\Lambda r^2}{3}    \Big)
dt^{2} +  \frac{dr^{2}}{\Big(1 - \Delta - \frac{2M}{r} -\frac{\Lambda r^2}{3} \Big) }  + r^{2}d\Omega^{2} \,\,\,,
\end{equation}
can be transformed into 
\begin{equation}
\label{SSSLambda2}
ds^2 \sim  - \Big(1 - \frac{2M_{ADM\Delta}}{\tilde r} -\frac{\Lambda {\tilde r}^2}{3}  \Big)
d{\tilde t}^{2} +  \frac{d{\tilde r}^{2}}{\Big(1 - \frac{2M_{ADM\Delta}}{\tilde r} -\frac{\Lambda {\tilde r}^2}{3} \Big) }  + (1-\Delta) {\tilde r}^{2}d\Omega^{2} \,\,\,.
\end{equation}
Notice that the cosmological constant $\Lambda$ did not require to be redefined in order to obtain the standard metric (\ref{SSSLambda2}). 
We then need $M_{ADM\Delta}\equiv 0$, $\Delta=1/2$ 
and $\Lambda=\Lambda_\infty$ in order to recover the metric (\ref{gfRSZ}) in the standard form when $Q\equiv 0$.\\ 
Finally, the metric of an SSS that is ADS or AADS with a deficit angle and endowed with a charge $q$
\begin{equation}
\label{SSSLambdaQ}
ds^2 \sim  - \Big(1 - \Delta - \frac{2M}{r} -\frac{\Lambda r^2}{3} + \frac{q^2}{r^2}  \Big)
dt^{2} +  \frac{dr^{2}}{\Big(1 - \Delta - \frac{2M}{r} -\frac{\Lambda r^2}{3} + \frac{q^2}{r^2}\Big) }  + r^{2}d\Omega^{2} \,\,\,,
\end{equation}
is transformed into 
\begin{equation}
\label{SSSLambdaQ2}
ds^2 \sim  - \Big(1 - \frac{2M_{ADM\Delta}}{\tilde r} -\frac{\Lambda {\tilde r}^2}{3} + \frac{{\cal Q}^2}{{\tilde r}^2} \Big)
d{\tilde t}^{2} +  \frac{d{\tilde r}^{2}}{\Big(1 - \frac{2M_{ADM\Delta}}{\tilde r} -\frac{\Lambda {\tilde r}^2}{3} + \frac{{\cal Q}^2}{{\tilde r}^2} 
\Big) }  + (1-\Delta) {\tilde r}^{2}d\Omega^{2} \,\,\,,
\end{equation}
taking $q=  {\cal Q} (1-\Delta)$. The quantity ${\cal Q}$ is presumably the actual charge when $\Delta\neq 0$. Again, taking 
$M_{ADM\Delta}\equiv 0$, $\Delta=1/2$ and $\Lambda=\Lambda_\infty$ we recover the metric (\ref{gfRSZ}) written in the standard form but now with 
$Q\neq 0$ given by $Q=\pm q^2= \pm {\cal Q}^2/4$.

\newpage


\begin{thebibliography}{99}

\bibitem{f(R)}
S. Nojiri and S. D. Odintsov, Int. J. Geom. Meth. Mod. Phys. 4, 115 (2007); \Journal{\PR}{505}{59}{2011}; 
S. Capozziello, M. De Laurentis, and V. Faraoni, arXiv: 0909.4672; 
S. Capozziello, and M. De Laurentis, arXiv: 1108.6266; T. Clifton, P. G. Ferreira, A. Padilla, and C. Skordis, \Journal{\PR}{513}{1}{2011}; 
S. Capozziello, and M. Francaviglia, \Journal{\GRG}{40}{357}{2008}; T. P. Sotiriou and V. Faraoni, \Journal{\RMP}{82}{451}{2010}; 
A. De Felice, and S. Tsujikawa, \Journal{\LRR}{13}{3}{2010}.

\bibitem{Starobinsky1980}
A. A. Starobinsky, \Journal{\PLB}{91}{99}{1980}.

\bibitem{Schmidt2013}
H. J. Schmidt, \Journal{\GRG}{45}{395}{2013}.

\bibitem{BHs}
P. T. Chrusciel, Contemp. Math. {\bf 170}, 23 (1994); M. Heusler, {\it Black Holes Uniqueness Theorems}, 
Cambridge Univ. Press, Cambridge (UK), 1996; D. C. Robinson, in {\it Kerr Spacetime: Rotating Black Holes in General Relativity}, 
eds. D. L. Wiltshire, M. Visser, and S. M. Scott, Cambridge Univ. Press, Cambridge (UK), 2009, p.115-143.

\bibitem{Wald1984}  
R. M. Wald, {\it General Relativity}, Chicago Univ. Press,
Chicago, 1984.

\bibitem{Bekenstein2000}
J. D. Bekenstein, in {\it Cosmology and Gravitation}, M. Novello (ed.) (Atlantisciences, France, 2000) p. 1-85 (arXiv: 9808028)

\bibitem{Herdeiro2015}
C. A. R. Herdeiro, and E. Radu, \Journal{\IJMPD}{24}{1542014}{2015}.

\bibitem{Bizon1990}
P. Bizon, \Journal{\PRL}{64}{2844}{1990}.

\bibitem{Nucamendi2003}
U. Nucamendi, and M. Salgado, \Journal{\PRD}{68}{044026}{2003}

\bibitem{Anabalon2012}
A. Anabal\'on, and J. Oliva, \Journal{\PRD}{86}{107501}{2012}.

\bibitem{Radu2014}
C. A. R. Herdeiro, and E. Radu, \Journal{\PRL}{112}{221101}{2014}; C. A. R. Herdeiro, and E. Radu, arXiv: 1501.04319

\bibitem{BTaltth}
V. Faraoni, \Journal{\PRD}{81}{044002}{2010}; S. Capozziello, and D. S\'aez--G\'omez, arXiv: 1107.0948; 
S. Capozziello, A. Stabile, and A. Troisi, \Journal{\PRD}{76}{104019}{2007}.

\bibitem{Nzioki2010}
A. M. Nzioki, S. Carloni, R. Goswami, and P. K. S. Dunsby, \Journal{\PRD}{81}{084028}{2010}.

\bibitem{Carloni2013}
 S. Carloni, and P. K. S. Dunsby, arXiv: 1306.2473
 
\bibitem{f(R)stars}
 T. Multam\"aki, and I. Vilja, \Journal{\PRD}{74}{064022}{2006}; \Journal{\PLB}{659}{843}{2008}; 
 K. Kainulainen, J. Piilonen, V. Reijonen, and D. Sunhede, \Journal{\PRD}{76}{024020}{2007}; K. Kainulainen, and D. Sunhede, \Journal{\PRD}{78}{063511}{2008}; 
 K. Henttunen, T. Multam\"aki, and I. Vilja, \Journal{\PRD}{77}{024020}{2008}; T. Kobayashi, and K. Maeda, \Journal{\PRD}{78}{064019}{2008}; 
\Journal{\PRD}{79}{024009}{2009}; E. Babichev, and D. Langlois, \Journal{\PRD}{80}{121501(R)}{2009}; 
 \Journal{\PRD}{81}{124051}{2010}; A. Upadhye, and W. Hu, \Journal{\PRD}{80}{064002}{2009}; S. Capozziello, M. De Laurentis, S. D. Odintsov, and A. Stabile, 
 \Journal{\PRD}{83}{064004}{2011}; S. S. Yazadjiev, D. D. Doneva, K. D. Kokkotas, and Kalin V. Staykov, arXiv:1402.4469

\bibitem{Miranda2009}
V. Miranda, S. E. Jor\'as, I. Waga, and M. Quartin, \Journal{\PRL}{102}{221101}{2009}.

\bibitem{Jaime2011}
L. G. Jaime, L. Pati\~no, and M. Salgado, \Journal{\PRD}{83}{024039}{2011}.

\bibitem{spontscal}
G. Esposito--Far\`ese, and T. Damour, \Journal{\PRL}{70}{2220}{1993}; \Journal{\PRD}{54}{1474}{1996}; 
J. Novak, \Journal{\PRD}{57}{4789}{1998}; \Journal{\PRD}{58}{064019}{1998}; M. Salgado, U. Nucamendi, and D. Sudarsky, 
\Journal{\PRD}{58}{121403}{1998}.

\bibitem{Sudarsky1995}
D. Sudarsky, \Journal{\CQG}{12}{579}{1995}.

\bibitem{Bekenstein1995}
J. Bekenstein, \Journal{\PRD}{51}{R6608}{1995}.

\bibitem{Sotiriou2012}
T. P. Sotiriou, and V. Faraoni, \Journal{\PRL}{108}{081103}{2012}.

\bibitem{Bekenstein1972}
J. Bekenstein, \Journal{\PRL}{28}{452}{1972}; J. Bekenstein, \Journal{\PRD}{5}{1239}{1972}.

\bibitem{Lovelock}
D. Lovelock, \Journal{\JMP}{12}{498}{1971}; \Journal{\JMP}{13}{874}{1972}; A. Navarro, and J. Navarro, arXiv: 1005.2386.

\bibitem{Clifton}
T. Clifton, and J. Barrow, \Journal{\PRD}{72}{103005}{2005}; T. Clifton, \Journal{\CQG}{23}{7445}{2006}.

\bibitem{Sebastiani2011}
L. Sebastiani, and S. Zerbini, Eur. Phys. J. C {\bf 71}, 1591 (2011).

\bibitem{nonvacuum}
S. Habib Mazharimousavi, M. Halilsoy, and T. Tahamtan, \Journal{\EPJC}{72}{1851}{2012}; 
S. Habib Mazharimousavi, and M. Halilsoy, \Journal{\PRD}{86}{088501}{2012}.

\bibitem{Jaime2012e}
L. G. Jaime, L. Pati\~no, and M. Salgado, in Relativity and Gravitation, 100 Years After Einstein in Prague, Springer, Heidelberg, 2014; arXiv:1211.0015

\bibitem{Jaime2013}
L. G. Jaime, L. Pati\~no, and M. Salgado, \Journal{\PRD}{87}{024029}{2013}.

\bibitem{Jaime2014}
L. G. Jaime, L. Pati\~no, and M. Salgado, \Journal{\PRD}{89}{084010}{2014}.

\bibitem{exactsolsL}
B. Carter, \Journal{\CMM}{10}{280}{1968}; M. Demianski, Act. Astron. {\bf 23}, 211 (1973).

\bibitem{Davis1989}
P. C. W. Davis, \Journal{\CQG}{6}{1909}{1989}.

\bibitem{trivsols}
S. Capozziello, A. Stabile, and A. Troisi, \Journal{\CQG}{24}{2153}{2007}; \Journal{\CQG}{25}{085004}{2008}; 
A. de la Cruz--Dombriz, A. Dobado, and A. L. Maroto, \Journal{\PRD}{80}{124011}{2009}; \Journal{\PRD}{83}{029903(E)}{2011}; 
A. Larra\~naga, Pramana Journal of Physics, {\bf 78}, 697 (2012) [arXiv: 1108.6325]; J. A. R. Cembranos, 
A. de la Cruz--Dombriz, and P. Jimeno Romero, arXiv: 1109.4519; T. Moon, Y. S. Myung, and E. J. Son, \Journal{\CQG}{43}{3079}{2011}; 
A. Sheykhi, \Journal{\CQG}{43}{3079}{2011}; S. Habib Mazharimousavi, M. Halilsoy, and T. Tahamtan, \Journal{\EPJC}{72}{1958}{2012}.

\bibitem{Briscese2008}
F. Briscese, and E. Elizalde, \Journal{\PRD}{77}{044009}{2008}; F. Hammad, arXiv: 1508.05126.

\bibitem{Kehagias2015}
A. Kehagias, C. Kounnas, D. L\"ust, and A. Riotto, arXiv: 1502.04192

\bibitem{Bergliaffa2011} 
S. E. Perez Bergliaffa, and Y. E. Chiafarelli de Oliveira Nunes, \Journal{\PRD}{84}{084006}{2011}.

\bibitem{Halilsoy2012}
S. Habib Mazharimousavi, and M. Halilsoy, \Journal{\PRD}{86}{088501}{2012}.

\bibitem{singularsols}
N. Bocharova, K. Bronnikov, and V. Melnikov, Vestn. Mosk. Univ. Fiz. Astron. {\bf 6}, 706 (1970); 
J. D. Bekenstein, Ann. Phys. (NY) {\bf 82}, 535 (1974).

\bibitem{gtt=g^rr}
V. V. Kiselev, \Journal{\CQG}{20}{1183}{2003}; M. Salgado, \Journal{\CQG}{20}{4551}{2003}; 
T. Jacobson, \Journal{\CQG}{24}{5717}{2007}.

\bibitem{adsmass}
L. F. Abbott, and S. Deser, \Journal{\NPB}{195}{76}{1982}; G. W. Gibbons, S. W. Hawking, G. T. Horowitz, and M. J. Perry, 
\Journal{\CMM}{88}{295}{1983}; A. Ashtekhar, and A. Magnon, \Journal{\CQG}{1}{L39}{1984}; W. Boucher, G. W. Gibbons, and G. T. Horowitz, 
\Journal{\PRD}{30}{2447}{1984}; M. Henneaux, and C. Teitelboim, \Journal{\CMM}{98}{391}{1985}; 
V. Balasubramanian, and P. Krauss, \Journal{\CMM}{208}{413}{1999}; G. W. Gibbons, \Journal{\CQG}{16}{1677}{1999}; 
A. Ashtekhar, and S. Das, \Journal{\CQG}{17}{L17}{2000}; P. Chru\'sciel, and G. Nagy, \Journal{\CQG}{18}{L61}{2001}; 
Adv. Theor. Math. Phys. {\bf 5}, 697 (2002).

\bibitem{Magnon1985}
A. Magnon, \Journal{\JMP}{26}{3112}{1985}.

\bibitem{Vilenkin1994}
A. Vilenkin, and E. P. S. Shellard, {\it Cosmic Strings and other Topological Defects}, Cambridge University Press, Cambridge UK, 1994.

\bibitem{Nucamendi1997}
U. Nucamendi, and D. Sudarsky, \Journal{\CQG}{14}{1309}{1997}.

\bibitem{Nucamendi2000}
U. Nucamendi, and D. Sudarsky, \Journal{\CQG}{17}{4051}{2000}.

\bibitem{Ayon-Beato1998}
E. Ay\'on-Beato, and A. Garc\'ia, \Journal{\PRL}{80}{5056}{1998}.

\bibitem{Bartnik1988}
R. Bartnik, and J. McKinnon, \Journal{\PRL}{61}{141}{1988}.

\bibitem{Bertotti2003}
B. Bertotti, L. Iess, and P. Tortora, \Journal{\NAT}{425}{374}{2003}.

\bibitem{Hu2007}
W. Hu, and I. Sawicky, \Journal{\PRD}{76}{064004}{2007}.

\bibitem{Lombriser2015}
L. Lombriser, F. Simpson, and A. Mead, arXiv: 1501.04961

\bibitem{Mirandadebate}
A. de la Cruz-Dombriz, A. Dobado, and A. L. Maroto, \Journal{\PRL}{103}{179001}{2009}; 
I. Thongkool, M. Sami, R. Gannouji, and S. Jhingan, \Journal{\PRD}{80}{043523}{2009}

\bibitem{exponential}
G. Cognola, E. Elizalde, S. Nojiri, S. D. Odintsov, L. Sebastiani, S. Zerbini, \Journal{\PRD}{77}{046009}{2008}; 
E. Linder, \Journal{\PRD}{80}{123528}{2009}; L. Yang, C. C. Lee, L. W. Luo, C. Q. Geng, \Journal{\PRD}{82}{103515}{2010}; 
K. Bamba, C. Q. Geng, C. C. Lee, \Journal{\JCAP}{08}{021}{2010}; 
E. Elizalde, S. Nojiri, S. D. Odintsov, L. Sebastiani, S. Zerbini, \Journal{\PRD}{83}{086006}{2008}; 
E. Elizalde, S. D. Odintsov, L. Sebastiani, S. Zerbini, \Journal{\EPJC}{72}{1843}{2012}.

\bibitem{Starobinsky2007}
A. A. Starobinsky, JETP Lett. {\bf 86}, 157 (2007).

\bibitem{Dolgov2003}
A. D. Dolgov and M. Kawasaki, Phys. Lett. B 573, 1 (2003).

\bibitem{Press1990}
W. H. Press, B. P. Flannery, S. A. Teukoslky, and W. T. Vetterling, {\it Numerical Recipes} (FORTRAN Version), Cambridge University Press (1990).

\bibitem{Maeda2003}
Y. Fujii, and K. Maeda, {\it The Scalar-Tensor Theory of Gravitation}, Cambridge University Press, Cambridge UK, 2003.

\bibitem{Damour1992}
T. Damour, and G. Esposito-Far\`erse, \Journal{\CQG}{9}{2093}{1992}.

\bibitem{Jaime2012}
L. G. Jaime, L. Pati\~no, and M. Salgado, arXiv:1206.1642; \Journal{\PRD}{89}{084010}{2014}.

\bibitem{Faraoni2004}
V. Faraoni, {\it Cosmology in Scalar-Tensor Gravity}, Kluwer (Dordrecht), The Netherlands. 

\end{thebibliography}
\end{document}